\makeatletter
\def\cl@chapter{}
\makeatother

\makeatletter
\newcommand\newtag[2]{#1\phantomsection\def\@currentlabel{#1}\label{#2}}
\makeatother

 \documentclass[twocolumn,english]{svjour3}        
\usepackage[style = authoryear, maxbibnames=9, labelnumber, defernumbers = true, backend=bibtex]{biblatex}
\makeatletter
\if@twocolumn
  \renewcommand\normalsize{%
    \@setfontsize\normalsize\@xpt{12.5pt}%
    \abovedisplayskip=3 mm plus6pt minus 4pt
    \belowdisplayskip=3 mm plus6pt minus 4pt
    \abovedisplayshortskip=0.0 mm plus6pt
    \belowdisplayshortskip=2 mm plus4pt minus 4pt
    \let\@listi\@listI}%

  \renewcommand\small{%
    \@setfontsize\small{8.5pt}\@xpt
    \abovedisplayskip 8.5\p@ \@plus3\p@ \@minus4\p@
    \abovedisplayshortskip \z@ \@plus2\p@
    \belowdisplayshortskip 4\p@ \@plus2\p@ \@minus2\p@
    \def\@listi{\leftmargin\leftmargini
      \parsep 0\p@ \@plus1\p@ \@minus\p@
      \topsep 4\p@ \@plus2\p@ \@minus4\p@
      \itemsep0\p@}%
    \belowdisplayskip \abovedisplayskip}
\else
  \if@smallext
    \renewcommand\normalsize{%
      \@setfontsize\normalsize\@xpt\@xiipt
      \abovedisplayskip=3 mm plus6pt minus 4pt
      \belowdisplayskip=3 mm plus6pt minus 4pt
      \abovedisplayshortskip=0.0 mm plus6pt
      \belowdisplayshortskip=2 mm plus4pt minus 4pt
      \let\@listi\@listI}%

    \renewcommand\small{%
      \@setfontsize\small\@viiipt{9.5pt}%
      \abovedisplayskip 8.5\p@ \@plus3\p@ \@minus4\p@
      \abovedisplayshortskip \z@ \@plus2\p@
      \belowdisplayshortskip 4\p@ \@plus2\p@ \@minus2\p@
      \def\@listi{\leftmargin\leftmargini
        \parsep 0\p@ \@plus1\p@ \@minus\p@
        \topsep 4\p@ \@plus2\p@ \@minus4\p@
        \itemsep0\p@}%
      \belowdisplayskip \abovedisplayskip}
  \else
    \renewcommand\normalsize{%
      \@setfontsize\normalsize{9.5pt}{11.5pt}%
      \abovedisplayskip=3 mm plus6pt minus 4pt
      \belowdisplayskip=3 mm plus6pt minus 4pt
      \abovedisplayshortskip=0.0 mm plus6pt
      \belowdisplayshortskip=2 mm plus4pt minus 4pt
      \let\@listi\@listI}%

    \renewcommand\small{%
      \@setfontsize\small\@viiipt{9.25pt}%
      \abovedisplayskip 8.5\p@ \@plus3\p@ \@minus4\p@
      \abovedisplayshortskip \z@ \@plus2\p@
      \belowdisplayshortskip 4\p@ \@plus2\p@ \@minus2\p@
      \def\@listi{\leftmargin\leftmargini
        \parsep 0\p@ \@plus1\p@ \@minus\p@
        \topsep 4\p@ \@plus2\p@ \@minus4\p@
        \itemsep0\p@}%
      \belowdisplayskip \abovedisplayskip}
  \fi
\fi
\let\footnotesize\small
\makeatother

\usepackage{cmap}

\usepackage[ngerman,main=english]{babel}
\addto\extrasenglish{\languageshorthands{ngerman}\useshorthands{"}}

\usepackage{ifluatex}
\ifluatex
  \usepackage{fontspec}
  \usepackage[english]{selnolig}
\fi

\ifluatex
\else
  \usepackage[%
    rm={oldstyle=false,proportional=true},%
    sf={oldstyle=false,proportional=true},%
    tt={oldstyle=false,proportional=true,variable=false},%
    qt=false%
  ]{cfr-lm}
\fi

\ifluatex
\else
  \usepackage[T1]{fontenc}
  \usepackage[utf8]{inputenc} 
\fi

\smartqed  

\usepackage{graphicx}

\usepackage{upquote}

\usepackage{booktabs}

\usepackage{paralist}

\usepackage{csquotes}

\usepackage{textcmds}

\usepackage{tabularx}
\newcolumntype{Y}{>{\centering\arraybackslash}X}

\usepackage{xltabular}
\usepackage{multirow}
\usepackage{makecell}
\usepackage{caption}

\usepackage{url}

\usepackage{amsmath} 
\usepackage{bm}

\makeatletter
\g@addto@macro{\UrlBreaks}{\UrlOrds}
\makeatother

\usepackage[dvipsnames]{xcolor}

\usepackage[tikz]{ocgx2}
\usetikzlibrary{positioning,fit,arrows}

\usepackage{listings}
\renewcommand{\lstlistingname}{List.}
\usepackage{chngcntr}
\AtBeginDocument{\counterwithout{lstlisting}{section}}

\usepackage{stfloats}
\fnbelowfloat

\usepackage{hyperref}
\usepackage[all]{hypcap}

\usepackage[acronym,indexonlyfirst,nomain]{glossaries}
\glsdisablehyper

\usepackage[capitalise,nameinlink]{cleveref}
\crefname{section}{Sect.}{Sect.}
\Crefname{section}{Section}{Sections}
\crefname{listing}{\lstlistingname}{\lstlistingname}
\Crefname{listing}{Listing}{Listings}

\usepackage{xspace}

\DeclareFontFamily{U}{MnSymbolC}{}
\DeclareSymbolFont{MnSyC}{U}{MnSymbolC}{m}{n}
\DeclareFontShape{U}{MnSymbolC}{m}{n}{
  <-6>    MnSymbolC5
  <6-7>   MnSymbolC6
  <7-8>   MnSymbolC7
  <8-9>   MnSymbolC8
  <9-10>  MnSymbolC9
  <10-12> MnSymbolC10
  <12->   MnSymbolC12%
}{}
\DeclareMathSymbol{\powerset}{\mathord}{MnSyC}{180}

\usepackage{ifthen}
\usepackage{amssymb}
\usepackage[normalem]{ulem} 

\newboolean{showcomments}
\setboolean{showcomments}{false} 

\newboolean{showreview}
\setboolean{showreview}{false} 

\ifthenelse{\boolean{showcomments}}
{
	\usepackage{mparhack}

	\newcommand{\nbb}[3]{
		\marginpar[\hspace*{0.75cm}\parbox{35pt}{\tiny#1}]{\parbox{35pt}{\tiny#1}}
		{#2}
	}
	\newcommand{\modified}[1]{{\color{orange!80!black}#1}}
	\newcommand{\removed}[1]{{\color{red!90!black}\sout{#1}}}
	\newcommand{\added}[1]{{\color{green!49!black}#1}}

	\newcommand{\rmodified}[2]{\nbb{#1}{\color{orange!80!black} #2}{orange!80!black}}
	\newcommand{\rremoved}[2]{\nbb{#1}{\color{red!90!black} \sout{#2}}{red!90!black}}
	\newcommand{\radded}[2]{\nbb{#1}{\color{green!49!black} #2}{green!49!black}}

}
{
	\newcommand{\nbb}[3]{}
	\newcommand{\modified}[1]{#1}
	\newcommand{\added}[1]{#1}
	\newcommand{\removed}[1]{}

	\newcommand{\rmodified}[2]{#2}
	\newcommand{\rremoved}[2]{}
	\newcommand{\radded}[2]{#2}
	
}

\usepackage{reviews-style}
\usepackage{pdfpages}

\hypersetup{hidelinks,
  colorlinks=true,
  allcolors=black,
  pdfstartview=Fit,
  breaklinks=true}

\journalname{JOURNALNAME}

\usepackage[math]{blindtext}
\usepackage{mwe}

\lstdefinelanguage{ATL}
{
	morekeywords={
		uses,
		if,
		module,
		for,
		create,
		from,
		refining,
		helper,
		context,
		def,
		rule,
		to,
		using,
		in,
		do,
		not,
		true,
		false,
		endif,
		else,
		then,
		Boolean,
		unique,
		lazy
	},
	sensitive=false, 
	morecomment=[l]{//}, 
	morecomment=[s]{/*}{*/}, 
	morestring=[b]" 
}
\lstdefinelanguage{QVTR}
{
	morekeywords={
		top,
		relation,
		domain
		String
	},
	sensitive=false, 
	morecomment=[l]{//}, 
	morecomment=[s]{/*}{*/}, 
	morestring=[b]" 
}

\usepackage{color}
\definecolor{eclipseBlue}{RGB}{42,0.0,255}
\definecolor{eclipseGreen}{RGB}{63,127,95}
\definecolor{eclipsePurple}{RGB}{127,0,85}

\definecolor{pblue}{rgb}{0.13,0.13,1}
\definecolor{pgreen}{rgb}{0,0.5,0}
\definecolor{pred}{rgb}{0.9,0,0}
\definecolor{pgrey}{rgb}{0.46,0.45,0.48}

\begin{document}

\title{Advantages and Disadvantages of (Dedicated) Model Transformation Languages 
}
\subtitle{A Qualitative Interview Study}

\titlerunning{A Qualitative Interview Study on the Advantages and Disadvantages of Model Transformation Languages}        

%

\author{Stefan Höppner \and
	    Yves Haas \and
        Matthias Tichy \and
        Katharina Juhnke
}


\institute{S. Höppner \at
			  Ulm University\\
              James-Franck-Ring 1, 89081 Ulm \\
              Phone: +49-0731-5024208\\
              \email{stefan.hoeppner@uni-ulm.de}           
          \and
              Y. Haas \at
              Ulm University\\
              James-Franck-Ring 1, 80901 Ulm\\
              \email{yves.haas@uni-ulm.de}           
           \and
           M. Tichy \at
           Ulm University\\
              James-Franck-Ring 1, 80901 Ulm\\
              \email{matthias.tichy@uni-ulm.de}           
           \and
           K. Juhnke \at
           Ulm University\\
              James-Franck-Ring 1, 80901 Ulm\\
              \email{katharina.juhnke@uni-ulm.de}           
}

\date{Received: date / Accepted: date}

\maketitle

\begin{abstract}\hfill\break
\textbf{Context}\\
Model driven development envisages the use of model transformations to evolve models.
Model transformation languages, developed for this task, are touted with many benefits over general purpose programming languages.
However, a large number of these claims have not yet been substantiated.
They are also made without the context necessary to be able to critically assess their merit or built meaningful empirical studies around them.

\noindent
\textbf{Objective}\\
The objective of our work is to elicit the reasoning, influences and background knowledge that lead people to assume benefits or drawbacks of model transformation languages.

\noindent
\textbf{Method}\\
We conducted a large-scale interview study involving 56 participants from research and industry.
Interviewees were presented with claims about model transformation languages and were asked to provide reasons for their assessment thereof.
We qualitatively analysed the responses to find factors that influence the properties of model transformation languages as well as explanations as to how exactly they do so.

\noindent
\textbf{Results}\\
Our interviews show, that general purpose expressiveness of GPLs, domain specific capabilities of MTLs as well as tooling all have strong influences on how people view properties of model transformation languages.
Moreover, the \textit{Choice of MTL}, the \textit{Use Case} for which a transformation should be developed as well as the \textit{Skill}s of involved stakeholders have a moderating effect on the influences, by changing the context to consider.

\noindent
\textbf{Conclusion}\\
There is a broad body of experience, that suggests positive and negative influences for properties of MTLs.
Our data suggests, that much needs to be done in order to convey the viability of model transformation languages.
Efforts to provide more empirical substance need to be undergone and lacklustre language capabilities and tooling need to be improved upon.
We suggest several approaches for this that can be based on the results of the presented study.

\end{abstract}

\section{Introduction}
\label{sec:intro}

Model transformations are at the heart of model-driven engineering (MDE)~\parencite{Sendall2003,metzger2005systematic}.
They provide a way to consistently and automatically derive a multitude of artefacts such as source code, simulation inputs or different views from system models~\parencite{Schmidt2006}.
Model transformations also allow to analyse system aspects on the basis of models~\parencite{Schmidt2006} and can provide interoperability between different modelling languages, e.g. architecture description languages like those described by \textcite{Malavolta2010}.
Since the emergence of the MDE paradigm at the beginning of the century numerous dedicated model transformation languages (MTLs) have been developed to support engineers in their endeavours~\parencite{Jouault2006,Arendt2010,QvT2016}.
Their appeal is driven by the promise of many advantages such as increased productivity, comprehensibility and domain specificity associated with using domain specific languages~\parencite{Hermans2009,Johannes2009}.

A recent literature study of us revealed, that, while a large number of such advantages and also disadvantages are claimed in literature, there exist only a few studies investigating to what extend these claims actually hold~\parencite{Goetz2020}.
The study presents 15 properties of MTLs for which literature claims advantages or disadvantages.
In this context, a claimed positive effect on one of the properties means an advantage whereas a negative influence means a disadvantage.
The properties identified in the study are:
\textit{Analysability}, \textit{Comprehensibility}, \textit{Conciseness}, \textit{Debugging}, \textit{Ease of Writing} (a transformation), \textit{Expressiveness}, \textit{Extendability}, (being) \textit{Just Better}, \textit{Learnability}, \textit{Performance}, \textit{Productivity}, \textit{Reuse \& Maintainability}, \textit{Tool Support}, \textit{Semantics \& Verification} and \textit{Versatility}.

Our study also revealed, that most claims in literature are made broadly and without much explanation as to where the claim originates from~\parencite{Goetz2020}.
Claims such as \textit{``Model transformation languages make it easy to work with models.''}~\parencite{Liepins2012}, \textit{``Declarative MTLs increase programmer productivity''}~\parencite{Lawley2007} or \textit{``Model transformation languages are more concise''}~\parencite{Hinkel2019a} illustrate this.
We assume that authors make such claims while having certain context and background in mind, but choose to omit it for unspecified reasons.
Some likely reason for omission of the context are, that authors believe it to not be worth mentioning or to preserve space which is often sparse in publications.

Regardless of the concrete reasons, a result of this practice is a lack of cause and effect relations in the context of model transformation languages that explain both why and when certain advantages or disadvantages hold.
Claims are thus easily dismissed based on anecdotal evidence.
Furthermore, setting up proper evaluation is also difficult because the claims do not provide the necessary background to do so.

To close this gap, we executed a large-scale empirical study using semi-structured interviews.
It involved a total of 56 researchers and practitioners in the field of model transformations.
The \textbf{goal} of our study was to compile a comprehensive list of influences on properties of model transformation languages guided by the following research questions:\\

\textbf{RQ 1:} What are the factors that influence properties of model transformation languages?\\

\textbf{RQ 2:} How do the identified factors influence MTL properties?\\

To concentrate our efforts and best utilize all available resources, we decided to focus on 6 of the 15 properties of model transformation languages identified by us in the preceding SLR~\parencite{Goetz2020}.
The 6 properties investigated in this study are: \textit{Comprehensibility}, \textit{Ease of Writing}, \textit{(practical) Expressiveness}, \textit{Productivity}, \textit{Reuse and Maintainability} and \textit{Tool support}.
We have chosen these six because they all play a major role in providing reasons for the adoption of model transformation languages.

Interviewees were presented with a number of claims about MTLs from literature and asked to reveal their views on the matter, as well as assumptions and reasons that lead them to agree or disagree with the presented claims.
We qualitatively analysed the interviews to understand the participants perceived influence factors and reasons for the advantages or disadvantages stated in the claims. 
The extracted data was then analysed to find commonalities between interviewees.
This was done for single claims as well as for overarching factors and reasons that influence a variety of aspects of MTLs.

We present a comprehensive explanation of factors that, according to experts, play an essential role in the discussion of advantages and disadvantages of model transformation languages for the investigated properties.
This is accompanied by a detailed exposition of \textbf{how} factors are relevant for the properties given above.
Lastly, we discuss the most salient factors and argue actionable results for the community and further research.

As the first study of this type, we make the following contributions:
\begin{enumerate}
	\item A comprehensive categorisation and listing of factors resulting in advantages or disadvantages of MTLs in the 6 properties studied.
	\item A detailed description of why and how each identified factor exerts an influence on different properties.
	\item Suggestions for how the presented information can be utilised to empirically investigate MTL properties.
	\item Procedural proposals for improving current model transformation languages based on the presented data.
\end{enumerate}

The results of our study show, that there is a large number of factors that influence properties of model transformation languages.
There is also a number of factors on which this influence depends on, i.e. factors that have a moderation effect on the influence of other factors.
These factors provide a solid basis that allows further studies to be conducted with more focus.
They also enable precise decisions on where improvements and adjustments in or for model transformation languages can be made.

The remainder of this paper is structured as follows: \Cref{sec:background} introduces model-driven engineering and model transformation languages, the context in which our study integrates.
\Cref{sec:methodology} will detail our methodology for preparing and conducting the interviews and the procedures used to analyse the data accumulated through the interviews.
Afterwards \cref{sec:demographics} gives an overview over demographic data of our interview participants while \cref{sec:results} presents our code system and details the findings for each code based on the interviews and analysis thereof.
In \cref{sec:findings} we present overarching findings and in \cref{sec:discussion}, we discuss actionable results that can be drawn from our stud that indicate avenues to focus on for the research community.
\Cref{sec:threats} contains a detailed discussion of the validity threats of this research, and in \cref{sec:rw} related efforts are presented.
Lastly, \cref{sec:conclusion} draws a conclusion for our research and proposes future work.
\section{Background}
\label{sec:background}

This section will provide the necessary background for the context in which our study is integrated in.

\subsection{Model-driven engineering}
\label{sec:background:mde}

The \textit{Model-Driven Architecture} (MDA) paradigm was first introduced by the Object Management Group in 2001~\parencite{OMG2001}.
It forms the basis for an approach commonly referred to as \textit{Model-driven development} (MDD) \parencite{brown2005introduction}, introduced as means to cope with the ever growing complexity associated with software development.
At the core of it lies the notion of using models as the central artefact for development.
In essence this means, that models are used both to describe and reason about the problem domain as well as to develop solutions~\parencite{brown2005introduction}.
An advantage ascribed to this approach that arises from the use of models in this way, is that they can be expressed with concepts closer to the related domain than when using regular programming languages~\parencite{Selic2003}.

When fully utilized, MDD envisions automatic generation of executable solutions specialized from abstract models~\parencite{Selic2003,Schmidt2006}.
To be able to achieve this, the structure of models needs to be known.
This is achieved through so called meta-models which define the structure of models.
The structure of meta-models themselves is then defined through meta-models of their own.
For this setup, the OMG developed a modelling standard called \textit{Meta-object Facility} (MOF)~\parencite{OMG2016} on the basis of which a number of modelling frameworks such as the \textit{Eclipse Modelling Framework} (EMF)~\parencite{steinberg2008emf} and the \textit{.NET Modelling Framework}~\parencite{hinkel2016nmf} have been developed.

\subsection{Domain-specific languages}

Domain-specific languages (DSLs) are languages designed with a notation that is tailored for a specific domain by focusing on relevant features of the domain~\parencite{vanDeuersen2002}.
In doing so DSLs aim to provide domain specific language constructs, that let developers feel like working directly with domain concepts thus increasing speed and ease of development~\parencite{Sprinkle2009}.
Because of these potential advantages, a well defined DSL can provide a promising alternative to using general purpose tools for solving problems in a specific domain.
Examples of this include languages such as \textit{shell scripts} in Unix operating systems~\parencite{Kernighan1984}, \textit{HTML}~\parencite{raggett1999html} for designing web pages or AADL an architecture design language~\parencite{SAEMobilus2004}.

\subsection{Model transformation languages}
\label{sec:background:MTL}

The process of (automatically) transforming one model into another model of the same or different meta-model is called \textit{model transformation} (MT).
They are regarded as being at the heart of Model Driven Software Development~\parencite{Sendall2003,metzger2005systematic}, thus making the process of developing them an integral part of MDD.
Since the introduction of MDE at the beginning of the century, a plethora of domain specific languages for developing model transformations, so called model transformation languages (MTLs), have been developed~\parencite{Arendt2010,Balogh2006,Jouault2006,Kolovos2008,horn2013model,George2012,Hinkel2019}.
Model transformation languages are DSLs designed to support developers in writing model transformations.
For this purpose, they provide explicit language constructs for tasks involved in model transformations such as model matching.
There are various features, such as directionality or rule organization~\parencite{Czarnecki2006}, by which model transformation languages can be distinguished.
For the purpose of this paper, we will only be explaining those features that are relevant to our study and discussion in \Cref{sec:background:MTL:exvsin,sec:background:MTL:rules,sec:background:MTL:direction,sec:background:MTL:increment,sec:background:MTL:tracing,sec:background:MTL:rac,sec:background:MTL:navigation}.
\Cref{tbl:MTL_features} provides an overview over the presented features.

Please refer to \textcite{Czarnecki2006,kahani2019survey,Mens2006} for complete classification.

\begin{table*}
	\caption{MTL feature overview}
	\label{tbl:MTL_features}
	\begin{tabularx}{\textwidth}{l|l|X}
		\toprule
		\textbf{Feature} & \textbf{Characteristic} & \textbf{Representative Language}\\
		\midrule
		\midrule
		\multirow{2}{7em}{Embeddedness} & Internal & FunnyQT (Clojure), RubyTL (Ruby), NMF Synchronizations (C\#)\\
		\cmidrule{2-3}
		& External & ATL, Henshin, QVT\\
		\midrule
		\multirow{2}{7em}{Rules} & Explicit Syntax Construct & ATL, Henshin, QVT\\
		\cmidrule{2-3}
		& Repurposed Syntax Construct & NMF Synchronizations (Classes), FunnyQT (Macros)\\
		\midrule
		\multirow{2}{7em}{Location Determination} & Automatic Traversal & ATL, QVT\\
		\cmidrule{2-3}
		& Pattern Matching & Henshin\\
		\midrule
		\multirow{2}{7em}{Directionality} & Unidirectional & ATL, QVT-O\\
		\cmidrule{2-3}
		& Bidirectional & QVT-R, NMF Synchronisations\\
		\midrule
		\multirow{2}{7em}{Incrementality} & Yes & NMF Synchronizations\\
		\cmidrule{2-3}
		& No & QVT-O\\
		\midrule
		\multirow{2}{7em}{Tracing} & Automatic & ATL, QVT\\
		\cmidrule{2-3}
		& Manual & NMF Synchronizations\\
		\midrule
		\multirow{2}{8em}{Dedicated Model Navigation Syntax} & Yes & ATL (OCL), QVT (OCL), Henshin (implicit in rules)\\
		\cmidrule{2-3}
		& No & NMF Synchronizations, FunnyQT, RubyTL\\
	\end{tabularx}
\end{table*}

\subsubsection{External and Internal transformation languages}
\label{sec:background:MTL:exvsin}

Domain specific languages, and MTLs by extension, can be distinguished on whether they are embedded into another language, the so called host language, or whether they are fully independent languages that come with their own compiler or virtual machine.

Languages embedded in a host language are called \textit{internal} languages.
Prominent representatives among model transformation languages are \textit{FunnyQT}~\parencite{horn2013model} a language embedded in Clojure, \textit{NMF Synchronizations} and the \textit{.NET transformation language}~\parencite{Hinkel2019} embedded in C\#, and \textit{RubyTL}~\parencite{Cuadrado2006} embedded in Ruby.

Fully independent languages are called \textit{external} languages.
Examples of external model transformation languages include one of the most widely known languages such as the \textit{Atlas transformation language} (ATL)~\parencite{Jouault2006}, the graphical transformation language Henshin~\parencite{Arendt2010} as well as a complete model transformation framework called VIATRA~\parencite{Balogh2006}.

\subsubsection{Transformation Rules}
\label{sec:background:MTL:rules}

\textcite{Czarnecki2006} describe rules as being \textit{``understood as a broad term that describes the smallest units of [a] transformation [definition]''}.
Examples for transformation rules are the rules that make up transformation modules in ATL, but also functions, methods or procedures that implement a transformation from input elements to output elements.

The fundamental difference between model transformation languages and general-purpose languages that originates in this definition, lies in dedicated constructs that represent rules.
The difference between a transformation rule and any other function, method or procedure is not clear cut when looking at GPLs.
It can only be made based on the contents thereof.
An example of this can be seen in \Cref{lst:Java:example}, which contains exemplary Java methods.
Without detailed inspection of the two methods it is not apparent which method does some form of transformation and which does not.

In a MTL on the other hand transformation rules tend to be dedicated constructs within the language that allow a definition of a \textit{mapping} between input and output (elements).
The example rules written in the model transformation language ATL in \Cref{lst:ATL:example} make this apparent.
They define mappings between model elements of type \texttt{Member} and model elements of type \texttt{Male} as well as between \texttt{Member} and \texttt{Female} using \textit{rules}, a dedicated language construct for defining transformation mappings.
The transformation is a modified version of the well known Families2Persons transformation case~\parencite{fam2per}.

\begin{lstlisting}[float=t,language=Java,caption=Example Java methods,label=lst:Java:example]
public void methodExample(Member m) {
  System.out.println(m.getFirstName());
}
public void methodExample2(Member m) {
  Male target = new Male();
  target.setFullName(m.getFirstName() + " Smith");
  REGISTRY.register(target);
}
\end{lstlisting}

\begin{lstlisting}[float=t,language=ATL,caption=Example ATL rules,label=lst:ATL:example]
rule Member2Male {
  from
    s : Member (not s.isFemale())
  to
    t : Male (
    	fullName <- s.firstName + ' Smith'
    )
}
	
rule Member2Female {
  from
    s : Member (s.isFemale())
  to
    t : Female (
      fullName = s.firstName + ' Smith'
      partner = s.companion
    )
}
\end{lstlisting}

\subsubsection{Rule Application Control: Location Determination}
\label{sec:background:MTL:rac}

Location determination describes the strategy that is applied for determining the elements within a model onto which a transformation rule should be applied \parencite{Czarnecki2006}.
Most model transformation languages such as ATL, Henshin, VIATRA or QVT~\parencite{QvT2016}, rely on some form of \textit{automatic traversal} strategy to determine where to apply rules.

We differentiate two forms of location determination, based on the kind of matching that takes place during traversal.
There is the basic \textit{automatic traversal} in languages such as ATL or QVT, where single elements are matched to which transformation rules are applied.
The other form of location determination, used in languages like Henshin, is based on \textit{pattern matching}, meaning a model- or graph-\textit{pattern} is matched to which rules are applied.
This does allow developers to define sub-graphs consisting of several model elements and references between them which are then manipulated by a rule.

The \textit{automatic traversal} of ATL applied to the example from \Cref{lst:ATL:example} will result in the transformation engine automatically executing the \texttt{Member2Male} on all model elements of type \texttt{Member} where the function \texttt{isFemale()} returns \texttt{false} and the \texttt{Member2Female} on all other model elements of type \texttt{Member}.

\begin{figure}
	\includegraphics[width=\linewidth]{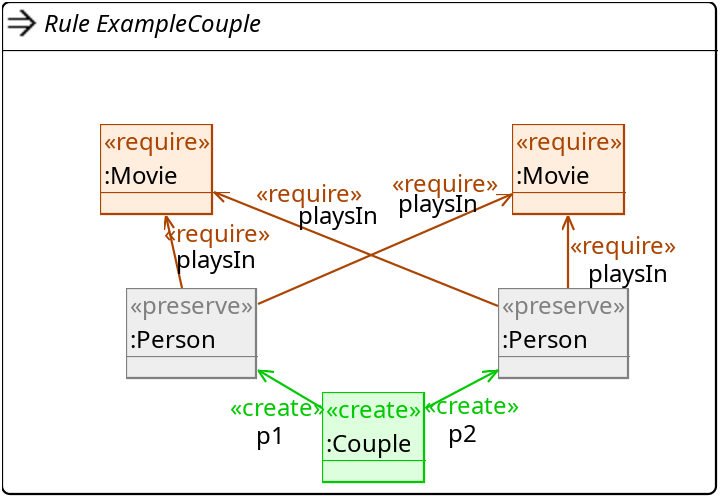}
	\caption{Example Henshin transformation}
	\label{fig:Henshin:example}
\end{figure}

The \textit{pattern matching} of Henshin can be demonstrated using \Cref{fig:Henshin:example}, a modified version of the transformation examples by \textcite{Krause2014}.
It describes a transformation that creates a couple connection between two actors that play in two films together.
When the transformation is executed the transformation engine will try and find instances of the defined graph pattern and apply the changes on the found matches.

This highlights the main difference between \textit{automatic traversal} and \textit{pattern matching} as the engine will search for a sub graph within the model instead of applying a rule to single elements within the model.

\subsubsection{Directionality}
\label{sec:background:MTL:direction}

The directionality of a model transformation describes whether it can be executed in one direction, called a unidirectional transformation or in multiple directions, called a multidirectional transformation~\parencite{Czarnecki2006}.

For the purpose of our study the distinction between unidirectional and bidirectional transformations is relevant.
Some languages allow dedicated support for executing a transformation both ways based on only one transformation definition, while other require users to define transformation rules for both directions.
General-purpose languages can not provide bidirectional support and also require both directions to be implemented explicitly.

The ATL transformation from \Cref{lst:ATL:example} defines a unidirectional transformation.
Input and output are defined and the transformation can only be executed in that direction.

The QVT-R relation defined in \Cref{lst:QVTR:example} is an example of a bidirectional transformation definition (For simplicity reasons the transformation omits the condition that males are only created from members that are not female).
Instead of a declaration of input and output, it defines how two elements from different domains relate to one another.
As a result given a \texttt{Member} element its corresponding \texttt{Male} elements can be inferred, and vice versa.

\begin{lstlisting}[float=t,language=QVTR,caption=Example QVT-R relation,label=lst:QVTR:example]
top relation Member2Male {
  n, fullName : String;
  domain Families s:Member {
    firstName = n };
  domain Persons t:Male {
    fullName = fullName};
  where {
  	fullName = n + ' Smith'; };
}
\end{lstlisting}

\subsubsection{Incrementality}
\label{sec:background:MTL:increment}

Incrementality of a transformation describes whether existing models can be updated based on changes in the source models without rerunning the complete transformation \parencite{Czarnecki2006}.
This feature is sometimes also called model synchronisation.

Providing incrementality for transformations requires active monitoring of input and/or output models as well as information which rules affect what parts of the models\modified{.
W}hen a change is detected the corresponding rules can then be executed.
It can also require additional management tasks to be executed to keep models valid and consistent.

\subsubsection{Tracing}
\label{sec:background:MTL:tracing}

According to \textcite{Czarnecki2006} tracing \textit{``is concerned with the mechanisms for recording different aspects of transformation execution, such as creating and maintaining trace links between source and target model elements''}.

Several model transformation languages, such as ATL and QVT have automated mechanisms for trace management.
This means that traces are automatically created during runtime.
Some of the trace information can be accessed through special syntax constructs while some of it is automatically resolved to provide seamless access to the target elements based on their sources.

An example of tracing in action can be seen in \texttt{line 16} of \Cref{lst:ATL:example}.
Here the \texttt{partner} attribute of a \texttt{Female} element that is being created, is assigned to \texttt{s.companion}.
The \texttt{s.companion} reference points towards a element of type \texttt{Member} within the input model.
When creating a \texttt{Female} or \texttt{Male} element from a \texttt{Member} element, the ATL engine will resolve this reference into the corresponding element, that was created from the referred \texttt{Member} element via either the \texttt{Member2Male} or \texttt{Member2Female} rule.
ATL achieves this by automatically tracing which target model elements are created from which source model elements.

\subsubsection{Dedicated Model Navigation Syntax}
\label{sec:background:MTL:navigation}

Languages or syntax constructs for navigating models is not part of any feature classification for model transformation languages.
However, it was often discussed in our interviews and thus requires an explanation as to what interviewees refer to.

Languages such as OCL~\parencite{OCL2014}, which is used in transformation languages like ATL, provide dedicated syntax for querying and navigating models.
As such they provide syntactical constructs that aid users in navigation tasks.
Different model transformation languages provide different syntax for this purpose.
The aim is to provide specific syntax so users do not have to manually implement queries using loops or other general purpose constructs.
OCL provides a functional approach for accumulating and querying data based on collections while Henshin uses graph patterns for expressing the relationship of sought-after model elements.

\section{Methodology}
\label{sec:methodology}

To collect data for our research question, we decided on using semi-structured interviews and a subsequent qualitative content analysis that follows the guidelines detailed by \textcite{Kuckartz2014}.
Semi-structured interviews were chosen as a data collection method because they present a versatile approach to eliciting information from experts.
They provide a framework that allows to get insights into opinions, thoughts and knowledge of experts~\parencite{Meyer1990,Hove2005,Kallio2016}.
The qualitative content analysis guidelines by \textcite{Kuckartz2014} were chosen because of their detailed descriptions for all steps of the analysis process.
As such they provide a more detailed and modern framework compared to the procedures introduced by \textcite{mayring1994qualitative}, which have long been a gold standard in qualitative content analysis.

\begin{figure*}[ht]
	\includegraphics[width=\textwidth]{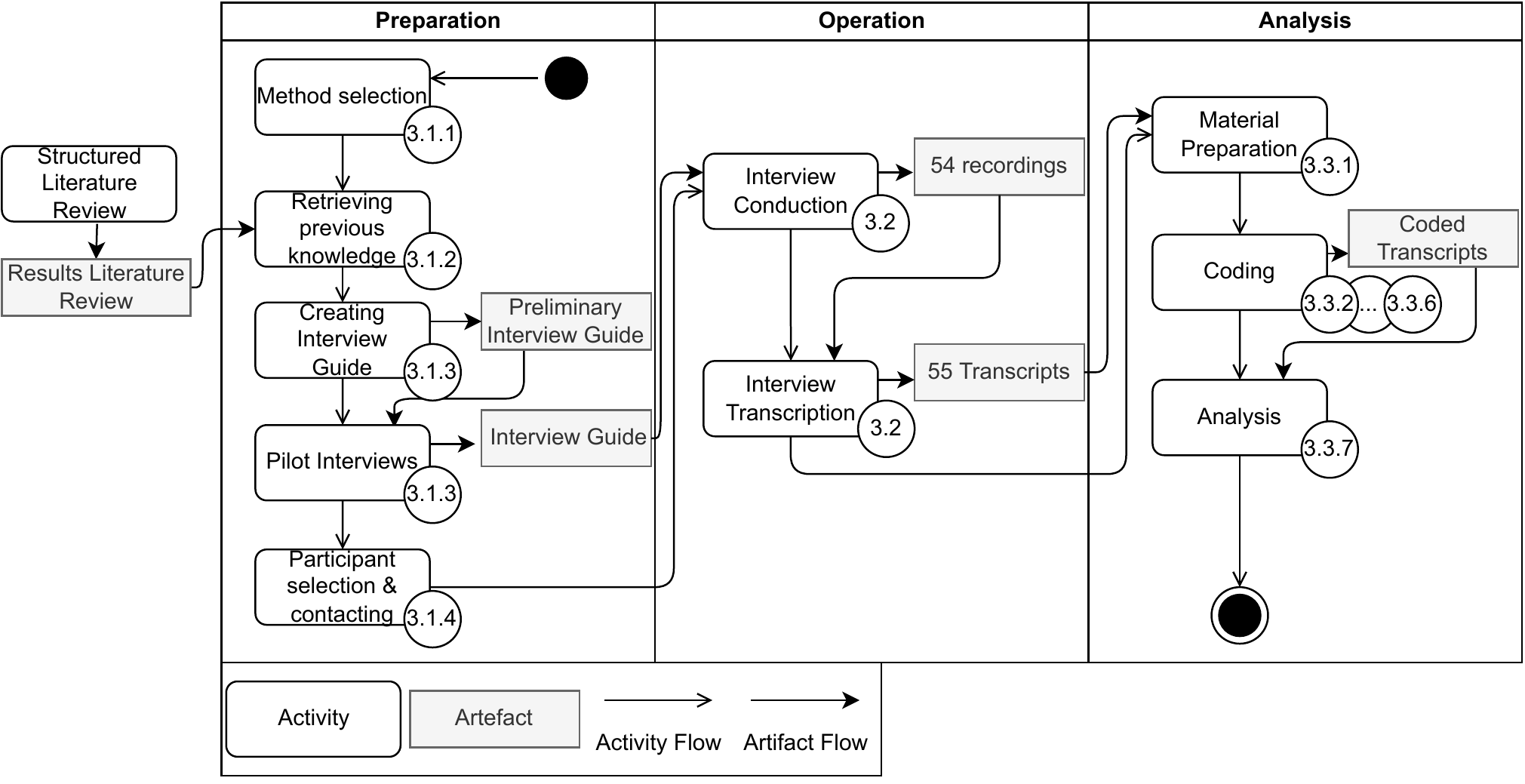}
	\caption{Overview over the study design}
	\label{fig:study_design}
\end{figure*}

An overview over our complete study design can be found in~\cref{fig:study_design}.
It shows the order of activities that were planned and executed as well as the artefacts produced and used throughout the study.
Each activity is annotated with the section number in which we detail the activity.
We split our approach into three main-phases: \textit{Preparation} (detailed in \Cref{sec:methodology:preparation}), \textit{Operation} (detailed in \Cref{sec:methodology:interview_conduction}) and \textit{Coding \& Analysis} (detailed in \Cref{sec:methodology:coding_analysis}).

For the preparation phase, we used a subset of the claimed properties of model transformation languages identified by us~\parencite{Goetz2020} to develop an interview guide.
The guide focuses around asking participants \textbf{whether} they agree with a claim from one of the properties and then envisages the usage of \textbf{why} questions to gain a deeper understanding of their opinions on the matter.
After identifying and contacting participants based on the publications considered during our previous literature review~\parencite{Goetz2020}, we conducted 54 interviews with 55 interviewees (at the request of two participants, one interview was conducted with both of them together) and collected one additional written response.
During the \textit{Coding \& Analysis} phase, we coded and analysed all 54 transcripts, as well as the written response, guided by the framework detailed by Kuckartz.
In doing so, we focused first on factors and reasons for the individual properties and then on common factors and reasons between them.

The remainder of this section will describe in detail how each of the three phases of our study was conducted.

\subsection{Interview Preparation}
\label{sec:methodology:preparation}
Our interview preparation phase consists of the creation of an interview guide plus selecting and contacting appropriate interview participants.
We use the guidelines by \textcite{Kallio2016} for the creation of our interview guide and expand the steps detailed there with steps for selecting and contacting participants.
In addition, we use the guidance from \textcite{newcomer2015handbook} to construct our study in the best possible way.

According to Kallio et al.~\parencite{Kallio2016} the creation of an interview guide consists of five consecutive steps.
First, researchers are urged to evaluate how appropriate semi-structured interviews are as a data collection method for the posed research questions.
Then, existing knowledge about the topic should be retrieved by means of a literature review.
Based on the knowledge from the knowledge retrieval phase, a preliminary interview guide can then be formulated and in another step be pilot tested.
Lastly the complete interview guide can then be presented and used.
As previously stated, we enhance these steps with two additional steps for selecting and contacting potential interview participants.

In the following, we detail how the presented steps were executed and the results thereof.

\subsubsection{Identifying the appropriateness of semi-structured interviews}

The goal of our study, outlined in our research questions, is to collect and analyse reasons and background information of why people believe claims about model transformation languages to be true.
Data such as this is qualitative by nature and hence requires a research method capable of producing qualitative data.
According to \textcite{Hove2005,Meyer1990} expert interviews are one of the most widely used research methodologies in the technical field for this purpose.
They allow to ascertain qualitative data such as opinions and estimates.
Interviews also enable qualitative interpretation of already available data~\parencite{Meyer1990} which perfectly aligns with our goal.
Moreover the opportunity to ask further questions about specific statements made by the participants~\parencite{newcomer2015handbook} fits the open ended nature of our research question.
For these reasons, we believe semi-structured interviews to be a well suited to answer our research questions.

%
%

\subsubsection{Retrieving previous knowledge}

In our previous publication~\parencite{Goetz2020}, we detailed the preparation, execution and results of an extensive structured literature review on the topic of claims about model transformation languages.
The literature review resulted in a categorization of 127 claims into 15 different categories (i.e. properties of MTLs) namely \textit{Analysability}, \textit{Comprehensibility}, \textit{Conciseness}, \textit{Debugging}, \textit{Ease of Writing a transformation}, \textit{Expressiveness}, \textit{Extendability}, \textit{Just better}, \textit{Learnability}, \textit{Performance}, \textit{Productivity}, \textit{Reuse \& Maintainability}, \textit{Tool Support}, \textit{Semantics and Verification} and lastly \textit{Versatility}.
These properties and the claims about them serve as the basis for the design of our interview study presented here.


\subsubsection{Interview guide}
\label{sec:methodology:interview_guide}

The interview guide involves presenting each interview participant with several claims on model transformation languages.
We use claims from literature instead of formulating our own statements, to make them more accessible.
This also prevents any bias from the authors to be introduced at this step.
Participants are first asked to assess their agreement with a claim before transitioning into a discussion on what the reasons for their decision are based on an open-ended question.
This style of using close-ended questions as a prerequisite for open-ended or probe questions has been suggested by multiple guides~\parencite{newcomer2015handbook, Hove2005}.

We focus on a subset of six properties.
This is due to the aim of keeping the length of interviews within an acceptable range for participants.
According to \textcite{newcomer2015handbook} semi-structured interviews should not exceed a maximum length of one hour.
As a result, only a number of properties can be discussed per interview.
In order to still talk with enough participants about each property, the number of properties examined must be reduced.
The properties we discuss in the interviews and the reasons why they are relevant are as follows:

\begin{itemize}
	\item \textit{Comprehensibility}: Is an important property when transformations are being developed as part of a team effort or evolve over time.
	\item \textit{Ease of Writing}: Is a decisive property that influences whether developers want to use a languages to write transformations in.
	\item \textit{Expressiveness}: Is one of the most cited properties in literature~\parencite{Goetz2020} and main selling point of domain specific languages in general.
	\item \textit{Productivity}: Is a property that is highly relevant for industrial adoption.
	\item \textit{Reuse \& Maintainability}: Is another property that enables wider adoption of model transformation languages in project settings.
	\item \textit{Tools}: High-quality tools can provide huge improvements to the development.
\end{itemize}

The list consists of the 5 most claimed properties form the previous literature review~\parencite{Goetz2020} and is supplemented with \textit{Productivity}, because we believe this attribute to be the most relevant for industry adoption.

To maximize the response rate of contacted persons, we aim for an interview length of 30 minutes.
This decision is based on experiences from previous interview studies conducted at our research group~\parencite{Groner2020, Juhnke2020} and fits within the maximum interview length suggested by \textcite{newcomer2015handbook}.

To best utilize the limited time per interview, the six properties are split into three sets of two properties each.
In each interview one of the three sets is discussed.

For each property, one non-specific, one specific and one negative claim is used to structure all interviews involving this property around.
A complete overview over all selected claims can be found in~\cref{tbl:catsclaims}.

We consider non-specific claims to be those that do not provide any rationale as to why the claimed property holds, e.g. \textit{``Model transformation languages ease the writing of model transformations.''}.
The non-specific claims chosen simply reflect the property itself.
They serve the purpose of getting participants to state their assumptions and beliefs for the property without any influence exerted by the discussed claim.

We consider those claims as specific, that provide a rationale or reason for why the claimed property holds, e.g. \textit{``Model transformation languages, being DSLs, improve the productivity.''}.
And we consider negative claims to be those, that state a negative property of model transformation languages, e.g. \textit{``Model transformation languages lack sophisticated reuse mechanisms.''}.
Generally, we use claims where we believe the discussions about the reasons to provide useful insights.

There exist several reasons why we believe this setup of using the same three none-specific, specific and negative claims for each property to be appropriate.
First, the non-specific claim allows participants to provide any and all factors or reasons that they believe influence a claimed property.
The specific claim then allows us to introduce a reason, that participants might not have thought about.
It also prompts a discussion about a particular reason or factor that is shared between all participants.
This ensures at least one area for cross comparison between answers.
The negative claim forces participants to also deliberate negative aspects, providing a counterbalance that counteracts bias.
Furthermore, the non-specific claim provides an easy introduction into the discussion about a specific MTL property that can present the interviewer with an overview of the participants thoughts on the matter.
It also allows participants to provide other influence factors not specifically covered through the discussed claims or even new factors and reasons not present in the collection of claims from our literature review~\parencite{Goetz2020}.

\begin{table*}
	\caption{Properties and Claims}
	\label{tbl:catsclaims}
	\begin{tabularx}{\textwidth}{l|X}
		\toprule
		\textbf{Property} & \textbf{Claim}\\
		\midrule
		\midrule
		
		\multirow{3}{7em}{Comprehensibility} & The use of model transformation languages increases the comprehensibility of model transformations.\\
		\cmidrule{2-2}
		& Model transformation languages incorporate high-level abstractions that make them more understandable than general purpose languages.\\
		\cmidrule{2-2}
		& Most model transformation languages lack convenient facilities for understanding the transformation logic.\\
		\midrule
		
		\multirow{3}{7em}{Ease of Writing} & The use of model transformation languages increases the ease of writing model transformations.\\
		\cmidrule{2-2}
		& Model transformation languages ease development efforts by offering succinct syntax to query from and map model elements between different modelling domains.\\
		\cmidrule{2-2}
		& Model transformation languages require specific skills to be able to write model transformations.\\
		\midrule
		
		\multirow{3}{7em}{Expressiveness} & The use of model transformation languages increases the expressiveness of model transformations.\\
		\cmidrule{2-2}
		& Model transformation languages hide transformation complexity and burden from the user.\\
		\cmidrule{2-2}
		& Having written several transformations we have identified that current model transformation languages are too low a level of abstraction for succinctly expressing transformations between DSLs because they demonstrate several recurring patterns that have to be reimplemented each time.\\
		\midrule
		
		\multirow{3}{7em}{Productivity} & The use of model transformation languages increases the productivity of writing model transformations.\\
		\cmidrule{2-2}
		& Model transformation languages, being DSLs, improve the productivity.\\
		\cmidrule{2-2}
		& Productivity of GPL development might be higher since expert users for general purpose languages are easier to hire.\\
		\midrule
		
		\multirow{3}{7em}{Reuse \& Maintainability} & The use of model transformation languages increases the reusability and maintainability of model transformations.\\
		\cmidrule{2-2}
		& Bidirectional model transformations have an advantage in maintainability.\\
		\cmidrule{2-2}
		& Model transformation languages lack sophisticated reuse mechanisms.\\
		\midrule

		\multirow{3}{7em}{Tool Support} & There is sufficient tool support for the use of model transformation languages for writing model transformations.\\
		\cmidrule{2-2}
		& Tool support for external transformation languages is potentially more powerful than for internal MTL or GPL because it can be tailored to the DSL.\\
		\cmidrule{2-2}
		& Model transformation languages lack tool support.\\

		\bottomrule
	\end{tabularx}
\end{table*}

\begin{figure}[ht]
	\centering
	\includegraphics[width=\linewidth]{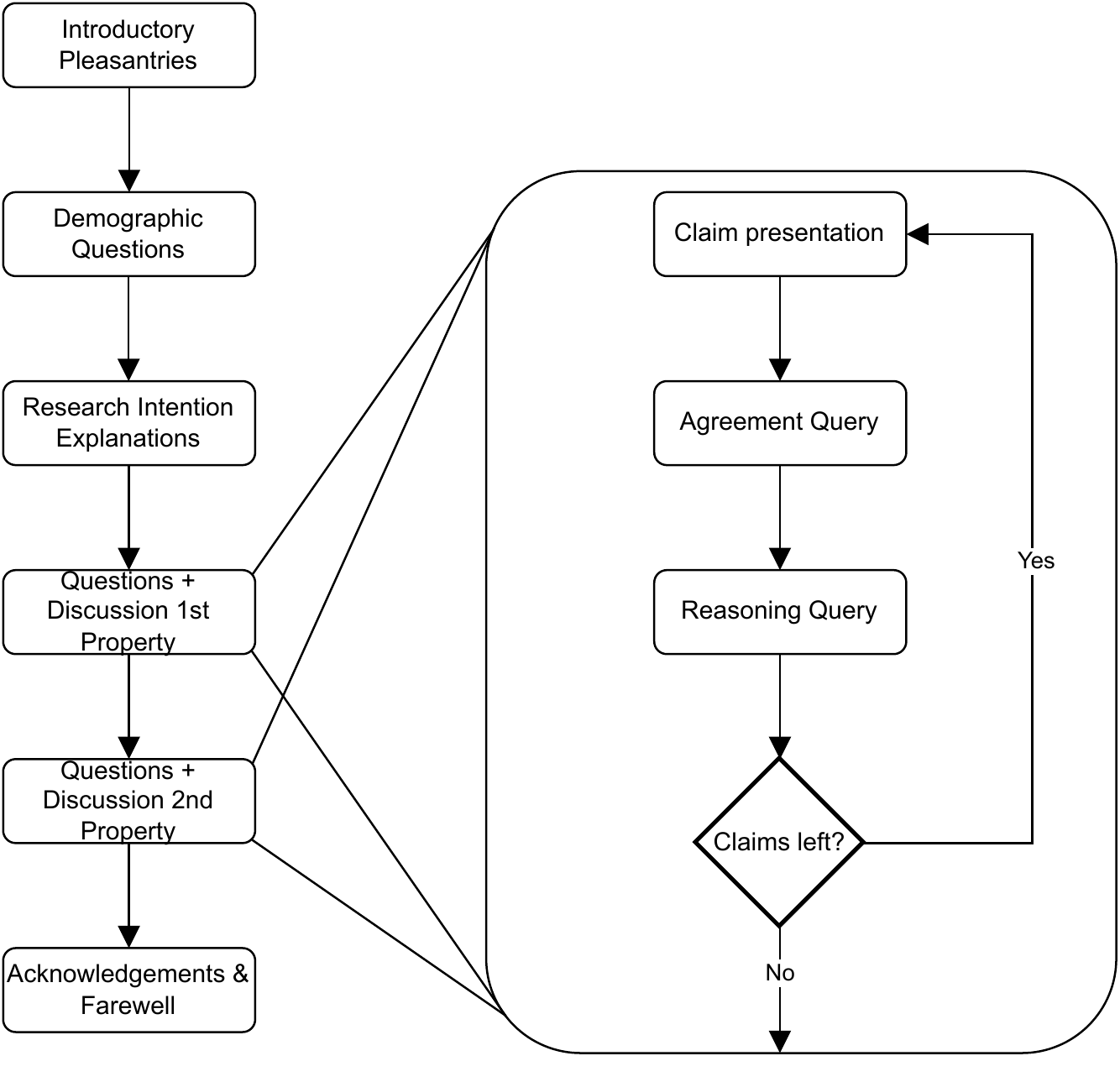}
	\caption{Interview guide}
	\label{fig:interview_guide}
\end{figure}

The complete interview guide resulting from the aforementioned considerations can be seen in~\cref{fig:interview_guide}.
After introductory pleasantries we start all interviews of with demographic questions.
Although some sources discourage asking demographic questions early in the interview due to their sensitive nature~\parencite{newcomer2015handbook}, we use them to break the ice between the interviewer and interviewee because our demographic questions do not probe any sensitive information.

After this initial get-to-know each other phase, the interviewer then proceeds to explain the research intentions, goals and the procedure of the remaining interview.
Depending on the property-set selected for the interview, participants are then presented with a claim about a property.
They are asked to rate their agreement with the claim based on a 5-point likert scale (5: completely agree, 4: agree, 3: neither agree nor disagree, 2: disagree, 1: completely disagree).
The likert scale is used to allow the interviewer to better assess the participants tendency compared to a simple yes or no question.
This part of the interview is intended solely to get a first impression of the view of the participant and not for a quantitative analysis.
It also creates a casual point of entry for the interviewee to think about the topic under consideration.
We communicate this to all participants to reduce any pressure they might feel to answer the question correctly.
Afterwards an open-ended question inquiring about the reasons for the interviewees assessment is asked.

Some terms used within the discussed claims have ambiguous definitions.
We tried to ask participants to explain their understanding of such terms, to prevent errors in analysis due to interviewees having different interpretations thereof.
This allows for better assessment during analysis.
The terms we have deemed to be ambiguous are: \textit{`succinct syntax'}, \textit{`mapping'}, \textit{`specific skills'}, \textit{`high-level abstractions'}, \textit{`convenient facilities'}, \textit{`sufficient tool support'}, \textit{`powerful tool support'}, \textit{`sophisticated reuse mechanisms'} and \textit{`expressiveness'}.
We provide a definition for the term \textit{expressiveness}.
This is, because we are only interested in a specific type of \textit{expressiveness}, i.e. how concisely and readily developers can express something.
We are not interested in expressiveness in a theoretical sense, i.e. the closeness to Turing completeness.

This process of presenting a claim, querying the participants agreement before discussing their reasons for the assessment is repeated for all 3 claims about both properties.
After discussing all claims, it is explained to the participants that the formal part of the interview is finished and that they are allowed to make final remarks about all discussed topics or other properties they want to address.
After this phase of the interview acknowledgements on the part of the interviewer are expressed before saying goodbye.
The complete question catalogue for the interviews can be found in~\cref{apdx:questions}.

The interview guide was tested in a pilot study by the main author with one co-author that was not involved in its creation.
After pilot testing, we changed the question about agreement with a claim from a yes-no question to one that uses a likert scale.
We also extended the question sets with non-specific claims that do not contain any reasoning.
Before adding the non specific claim, discussions focused too much on the narrow view within the presented claims.


\subsubsection{Selecting \& contacting participants}
\label{sec:method:selection}

The target population for our study consists of all users of model transformation languages.
To select potential participants for our study we rely on data from our previous literature review~\parencite{Goetz2020}.
The literature review produced a list of publications that address the topic of model transformations and model transformation languages.
Because search terms such as `model to text' and the like were not used in the study, using this list limits our results to model to model transformation languages.
We discuss this limitation more thoroughly in \Cref{sec:threats:external}.

All authors of the resulting publications are deemed to be potential interview participants.
We assume, that people using MTLs in industry do have some research background and thus have published work in the field.
There is also no other systematic way to find industry users.
We also assume that people who are still active in the field have published within the last 5 years.
This limits outreach but makes the set of potential participants more manageable.
For this reason, the list was shortened to publications more recent than 2015 before the authors of all publications was compiled.
This resulted in a total of 645 potential participants.

After selection, the authors were contacted via mail.
First, everyone was contacted once and then, after a week, everyone who had not responded by then was contacted again.
The texts we use for both mails can be found in~\cref{apdx:mails}.
Ten potential participants, from the list of potential participants, were not contacted through this channel but via personalised emails, as they are personal contacts of the authors.

Within the contact mails, potential participants are asked to select a suitable date for the interview and fill out a data consent form allowing us to record and transcribe the interviews.

Overall of the 645 contacted authors, 55 agreed to participate in our interview study resulting in a response rate of 8.53\%\footnote{when including the written response in this statistic, the resulting response rate is 8.68\%.}.

\subsection{Interview Conduction and Transcription}
\label{sec:methodology:interview_conduction}

All but one interview were conducted by the first author using the online conferencing tool WebEx and lasted between 20 and 80 minutes.
Due to scheduling issues, one interview had to be conducted by the second author, who had a preparatory mock interview with the main interviewer.
Additionally, at the request of two participants, one interview was conducted with both of them together.
Since our main focus for all interviews is on discussions, we do not believe this to have any effect on its results.
WebEx is the chosen conferencing tool, due to its availability to the authors and its integrated recording tool which is used to record all interviews.
For data privacy reasons and for easier in-depth analysis later on, all recordings are transcribed by two authors.
To increase the readability of heavily fragmented sentences they are shortened to only contain the actual content without interruptions.
In case of audibility issues the transcribing authors consulted with each other to try and resolve the issue.
Altogether the interviews produced just over 32 hours of audio and about 162.100 words of transcriptions.

Each day, the main author decided on which question sets to use for all participants that had agreed to partake in the interviews.
The question sets had to be chosen daily, as many participants only responded to the invitation after interviews had already taken place.

The goal of the decision process was, to ensure an even spread of participants over the question sets based on relevant demographic backgrounds, namely \textit{research}, \textit{industry}, \textit{MTL developer} and \textit{MTL user}.
We consider those relevant because each group has a different view point on model transformation languages and their usage for writing transformations.
It is therefore important to have answers from each group for each set of questions, to reduce the risk of missing relevant opinions.

We were able to ensure that at least one representative for each demographic group provided answers for each question set.
A complete uniform distribution was not possible due to overlaps in the demographic groups.

\subsection{Coding \& Analysis}
\label{sec:methodology:coding_analysis}

\begin{figure}[ht]
	\includegraphics[width=\linewidth]{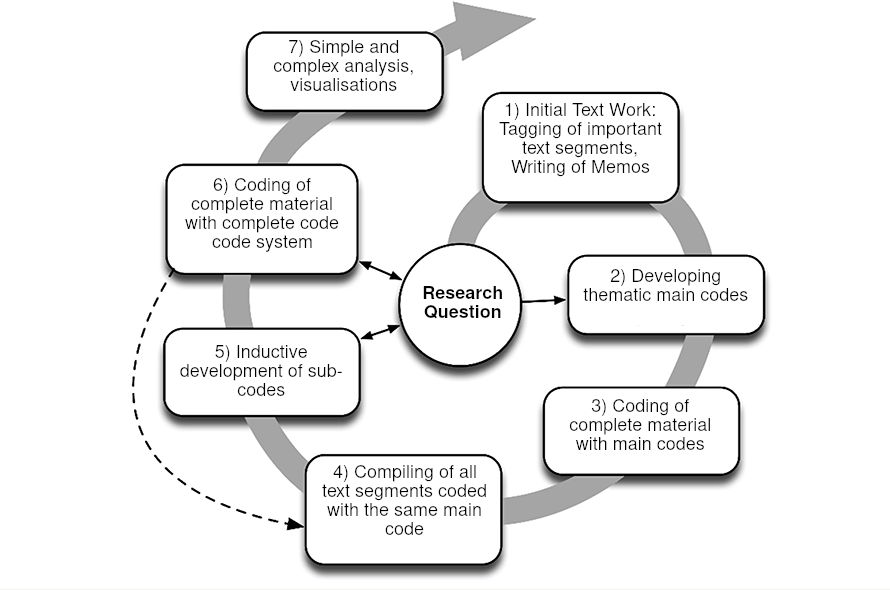}
	\caption{Process of a content structuring content analysis as presented by \textcite{Kuckartz2014}}~\label{fig:process_qualitative_content_analysis}
\end{figure}

Coding and analysing the interview transcripts is done in accordance with the guidelines for \textit{content structuring content analysis} suggested by \textcite{Kuckartz2014}.
The guideline recommends a seven step process (depicted in \Cref{fig:process_qualitative_content_analysis}) for coding and analysing qualitative data.
All steps are carried out with tool support provided by the MAXQDA\footnote{\url{https://www.maxqda.com/}} software.
In the following, we explain how each process step is conducted in detail.
We will use the following statement as a running example to show how codes and sub-codes are assigned and how the coding of text segments evolved throughout the process:
\textit{``Of course some MTLs use explicit traceability for instance. But even then you have a mechanism to access it. And if you have a MTL with implicit traceability where the trace links are created automatically then of course you gain a lot of expressivity because you don’t have to write something that you would otherwise have to write for almost every rule.''} (\textbf{P30})

\subsubsection{Initial Text Work}

The initial text work step initiates our qualitative analysis.
Kuckartz~\parencite{Kuckartz2014} suggests to read through all the material and highlight important segments as well as to write notes for the transcripts using memos.
Following these suggestions, we apply \textit{initial} coding from constructivist grounded theory~\parencite{Charmaz2014,Vollstedt2019,10.1145/2884781.2884833} to mark and summarize all text segments where interviewees reason about their beliefs on influence factors about the discussed properties.
To do so, the two authors, which conducted and transcribed the interviews, read through all transcripts and mark all relevant text segments with codes that preferably represented the segment word for word.
The codes allow for easier reference in later steps and, due to tooling, we are still able to quickly read the underlying text segment if necessary.

During this step, the example statement was labelled with the code \texttt{automatic tracing increases expressiveness because no manual overhead}.

\subsubsection{Developing thematic main codes}
\label{sec:method:coding_analysis:developing_main_codes}

For developing the thematic main codes for our study we follow the common practice of inferring them from our research questions as suggested by \textcite{Kuckartz2014}.
Since the goal of our research is to investigate implicit assumptions, and factors that influence the assessment of experts about properties of model transformation languages three main codes arise:
\begin{itemize}
	\item \textbf{Properties}: Denoting which property is being discussed (e.g. \textit{Comprehensibility}).
	\item \textbf{Factors}: Denoting what influences a discussed property according to an interviewee (e.g. \textit{Bidirectional functionality of a MTL}).
	\item \textbf{Factor assessment}: Denoting an evaluation of how a factor influences a property (e.g. \textit{positive} or \textit{negative} or \textit{mixed depending on other factors}).
\end{itemize}

The sub-codes for the property code can be directly defined based on the six properties from our previous literature review~\parencite{Goetz2020}.
As such they are deductive (a-priori) codes that are intended to mark text segments based on the properties that are being discussed in them.

\subsubsection{Coding of all the material with main codes}

In order to code of all the material with the main codes one author analyses all interview transcripts.
While doing so, the conversations about a discussed claim are marked with the code that is based on the property stated in the claim.
To exemplify this, all discussions on the claim \textit{"The use of MTLs increases the comprehensibility of model transformations."} are coded with the main code \textit{comprehensibility}.

This realisation of the process step breaks with Kuckartz's specifications in multiple ways.
First, we do not code the material with the main codes \textbf{Factors} and \textbf{Factor assessment}, because all factors and factor assessments are already coded with the summarising \textit{initial} codes.
These will be refined into actual sub-codes of \textbf{Factors} and \textbf{Factor assessment} in a later step.
Second, we directly code segments with the sub-codes for the \textbf{Property} main code, because the differentiation comes naturally with the structure of the interviews and delaying this refinement makes no sense.
And third, this way of coding makes it possible that unimportant segments are also coded, something that Kuckartz suggests not to do.
However, we actively decided in favour of this, because it accelerates the coding process enormously.
Furthermore, only overlaps of the property codes with the other codes are considered, in later steps, thus automatically excluding unimportant text segments from consideration.

During this step, the coding for the example text segment was extended with the code \texttt{Expressiveness}.
While this does not look like much of an enhancement on the surface, it is paramount to allow for systematic analysis in later steps.

After this step the example segment had its \textit{initial code}, summarising the essence of the statement, and the explicit \textit{property sub-code} \texttt{Expressiveness}, providing the first systematic categorisation of the segment.

\subsubsection{Compilation of all text passages coded with the same main code}

This step forms the basis for the subsequent iterative process of inductively developing sub-codes for each main code.
Due to the use of the MAXQDA tool, this step is purely technical and does not require any special procedure outside of the selection of the main code that is being considered in the tool.

\subsubsection{Inductive development of sub-codes}

The inductive development of sub-codes forms the most important coding step in our study.
Inductive development here means that the sub-codes are developed based on the transcripts contents.

\textcite{Kuckartz2014} suggests to read through all segments coded with a main code to iteratively refine the code into several sub-codes that define the main category more precisely.
We optimize this step by analysing all the \textit{initial} codes from the \textit{Initial Text Work} step, to construct concise and comprehensive codes for similar \textit{initial} codes that could be used as sub-codes for the \textbf{Factor} or \textbf{Factor assessment} main codes.
In doing so we follow the \textit{focused} coding procedure of constructivist grounded theory to refine the initial code system.

All sub-codes of the \textbf{Factor} main code, that are refined using this process, are thematic codes, meaning they denote a specific topic or argument made within the transcripts.
As a result, the sub-codes represent factors explicitly named by interviewees that influence the different properties.
In contrast, all sub-codes of the \textbf{Factor assessment} main code, that are refined using this process, are evaluative codes, meaning they represent an evaluation, made by the authors, about an effect.
More specifically, the codes represent an evaluation of how participants believe factors influence various properties.

Because of the importance of this coding step, the sub code refinement is created in a joint effort by three of the authors.
First, over a period of three meetings, the authors develop comprehensive codes based on the \textit{initial} codes of 18 interviews through discussions.
Then the main author complements the resulting code system by analysing the remaining set of interview transcripts, while the two other authors each analyse half of them.
In a final meeting any new sub code, devised by one of the authors, is discussed and a consensus for the complete code system is established.

During this step no code segment is extended with additional codes.
Instead new codes derived from the \textit{initial codes} are saved for usage in the following steps.

From the example code segment and its \textit{initial code}, a sub-code \textit{automatic tracing} for the \textbf{Factors} code was derived.
The finalised sub-code \texttt{Traceability} was decided upon based on the combination with other derived codes of similar meaning, like \textit{traces}.

\subsubsection{Coding of all the material with complete code system}

After the final code system is established, the main author processes all transcripts to replace the \textit{initial} codes with codes from the final code system.
For this, each coded statement is re-coded with codes indicating the influence factors expressed by the interviewees as well as a factor assessment, if possible.
This final coding step is done by the main author while all three co-authors each check 10 coded transcripts to validate the correct and consistent use of all codes and to make sure all relevant statements are considered.
The results from the reviews are discussed in pairwise meetings between the main author and the reviewing co-author before being incorporated in a final coding approved by all authors.

During this step, the \textit{initial code} for the example segment was dropped and replaced by the codes \texttt{MTL advantage} and \texttt{Traceability}.

The final codes assigned to the example text segment thus were: \textit{Expressiveness}, \textit{Traceability} and \textit{MTL advantage}.
The reasoning given within the statement as to why automatic tracing provides an expressiveness advantage, are manually extracted during analysis using tooling provided by MAXQDA.

\subsubsection{Simple and complex analysis and visualisation}

\begin{figure}
	\includegraphics[width=\linewidth]{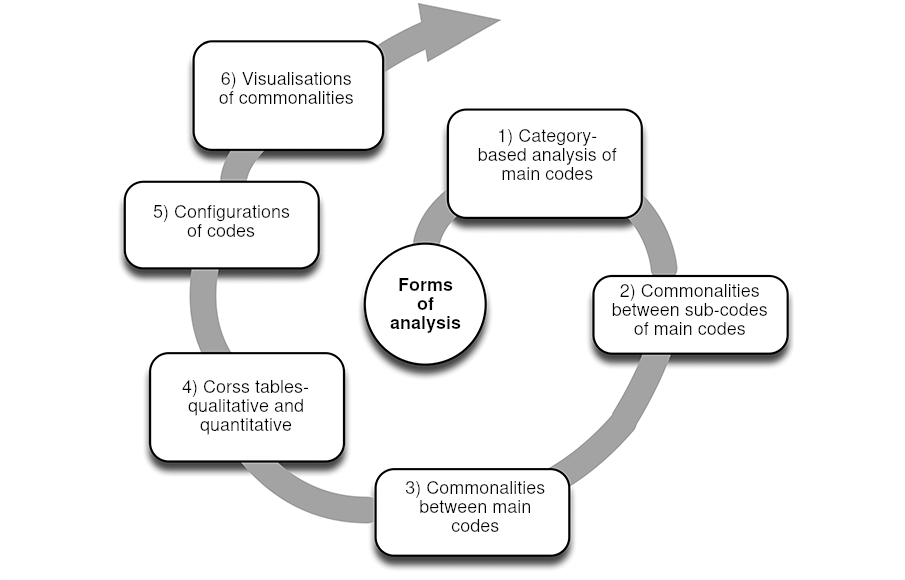}
	\caption{Analysis forms in a content structuring content analysis as presented by \textcite{Kuckartz2014}}~\label{fig:process_analysis}
\end{figure}

The resulting coding and the coded text segments are then used as the basis for our analysis which, in accordance with our research question, focuses on identifying and evaluating factors that influence the properties of MTLs.
As recommended by \textcite{Kuckartz2014}, this \modified{is} first done for each \textbf{Property} individually before analysis across all properties is conducted (as shown in \Cref{fig:process_analysis}).

For analysing the influence factors of an individual property, we use the MAXQDA tooling to find segments coded with both a factor and the considered property.
Using this approach we first compile a list of all factors relevant for a property, before then doing an in-depth analysis of all the gathered statements for each factor.
Here the goal is to elicit commonalities and contradictions between the opinions of our interviewees that can be used to establish a theory on how each factor influences each property individually.

In terms of our example text segment, the segment and all other segments coded with \texttt{Expressiveness} and \texttt{Traceability} were read and analysed.
The goal was to see if reduced overhead from implicit trace capabilities played a role in the argumentation of other participants and to gather all the other mentioned reasons. 

For the analysis over all properties combined we apply the \textit{theoretical} coding process of constructivist grounded theory~\parencite{Charmaz2014,10.1145/2884781.2884833} to develop a model of influences.
To do so, the \textbf{Factor assessment}s are used to examine how the factors influence the respective properties, what the commonalities between properties are and where the differences lie.
The goal here is to develop a cohesive theory which explains the influences of factors on the individual properties but also on the properties as a whole and potential influences between the factors themselves.

In terms of our example text segment, the results from analysing \texttt{Expressiveness} and \texttt{Traceability} segments were compared to results from analysing segments coded with other property codes and \texttt{Traceability}.
The goal was to find commonalities and differences between the analysed groups.

\subsubsection{Privacy and Ethical concerns}
\label{sec:method:privacy}

All interview participants were informed of the data collection procedure, the handling of the data and their rights surrounding the process, prior to the interview.

During selection of potential participants the following data was collected and processed.
\begin{itemize}
	\item First \& last name.
	\item E-Mail address.
\end{itemize}

For participants that agreed to the partake in the interview study the following additional data was collected and processed during the course of the study.
\begin{itemize}
	\item Non anonymised audio recording of the interview.
	\item Transcripts of the audio recordings.
\end{itemize}

All data collected during the study was not shared with any person outside of the group of authors.
Audio recordings were handled only by the first and second author.

The complete information and consent form can be found in \Cref{apdx:consent_form}.
All participants have consented to having their interview recorded, transcribed and analysed based on this information.
All interview recordings were stored on a single device with hardware encryption and deleted as soon as transcriptions were finalised.
The interview transcripts were processed to prevent identification of participants.
For this, identifying statements and names were removed.

Apart from the voice recordings and names, no sensitive information about the interviewees was collected.

The study design was not presented to an ethical board.
The basis for this decision are the rules of the German Research Foundation (DFG) on when to use a ethical board in humanities and social sciences\footnote{\url{https://www.dfg.de/foerderung/faq/geistes_sozialwissenschaften/}}.
We refer to these guidelines because there are none specifically for software engineering research and humanities and social sciences are the closest related branch of science for our research.
\section{Demographics}
\label{sec:demographics}

We interviewed a total of 55 experts from 16 different countries with varied backgrounds and experience levels and collected one comprehensive written response.
\Cref{tbl:demographicsOverview} in \Cref{apdx:demographics} presents an overview of the demographic data about all interview participants.
Experts and their statements are distinguished via an anonymous ID (\textbf{P1} to \textbf{P56}).

\subsection{Background}
\label{sec:demographics:background}

\begin{figure}[ht]
	\centering
	\includegraphics[width=\linewidth]{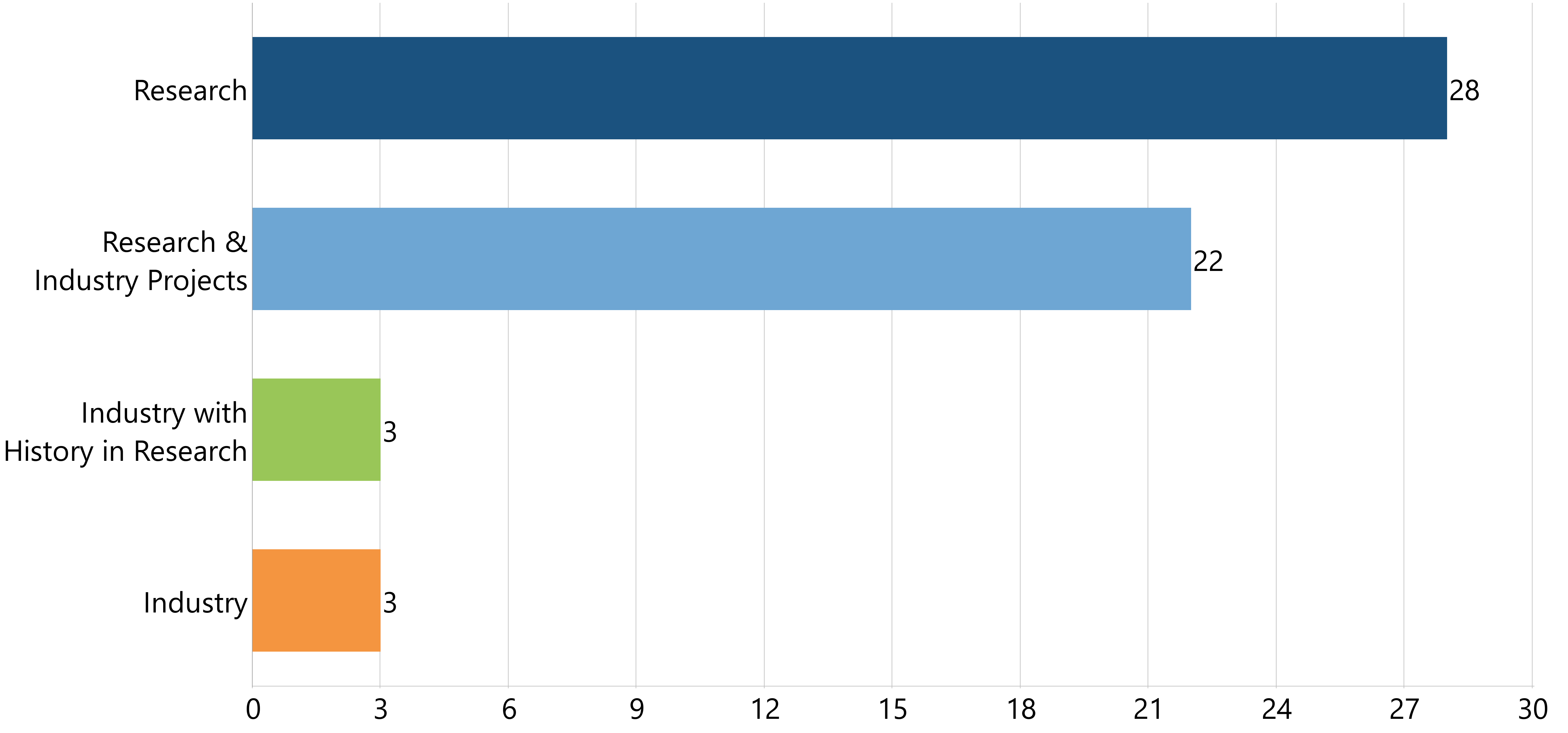}
	\caption{Distribution of participants background}~\label{fig:participant_types}
\end{figure}

As evident from \Cref{fig:participant_types} participants with a research background constitute the largest portion of our interviewees.
Overall there is an even split between participants solely from research and those that have at least some degree of industrial contact (either through industry projects or by working in industry).
Only 3 participants stated to have used model transformations solely in an industrial context.
This is in part offset by the fact that 25 of interviewees have executed research projects in cooperation with industry or have worked both in research and industry (22 and 3 respectively).
While there is a definitive lack of industry practitioners present in our study, a large portion of interviewees are still able to provide insights into model transformations and model transformation languages with an industry view.

Lastly, 10 of our participants are, in some capacity, involved in the development of model transformation languages.
They can provide a different angle on advantages or disadvantages of MTLs compared to the 46 participants that use them solely for transformation purposes.

\subsection{Experience}

50 interviewees expressed to have 5 or more years of experience in using model transformations.
Moreover, 24 of the participants have over 10 years of experience in the field.
Lastly there was a single participant that had only used model transformations for a brief amount of time during their masters thesis.


\subsection{Used languages for transformation development}
\label{sec:demographics:langs}

\begin{figure}[ht]
	\centering
	\includegraphics[width=\linewidth]{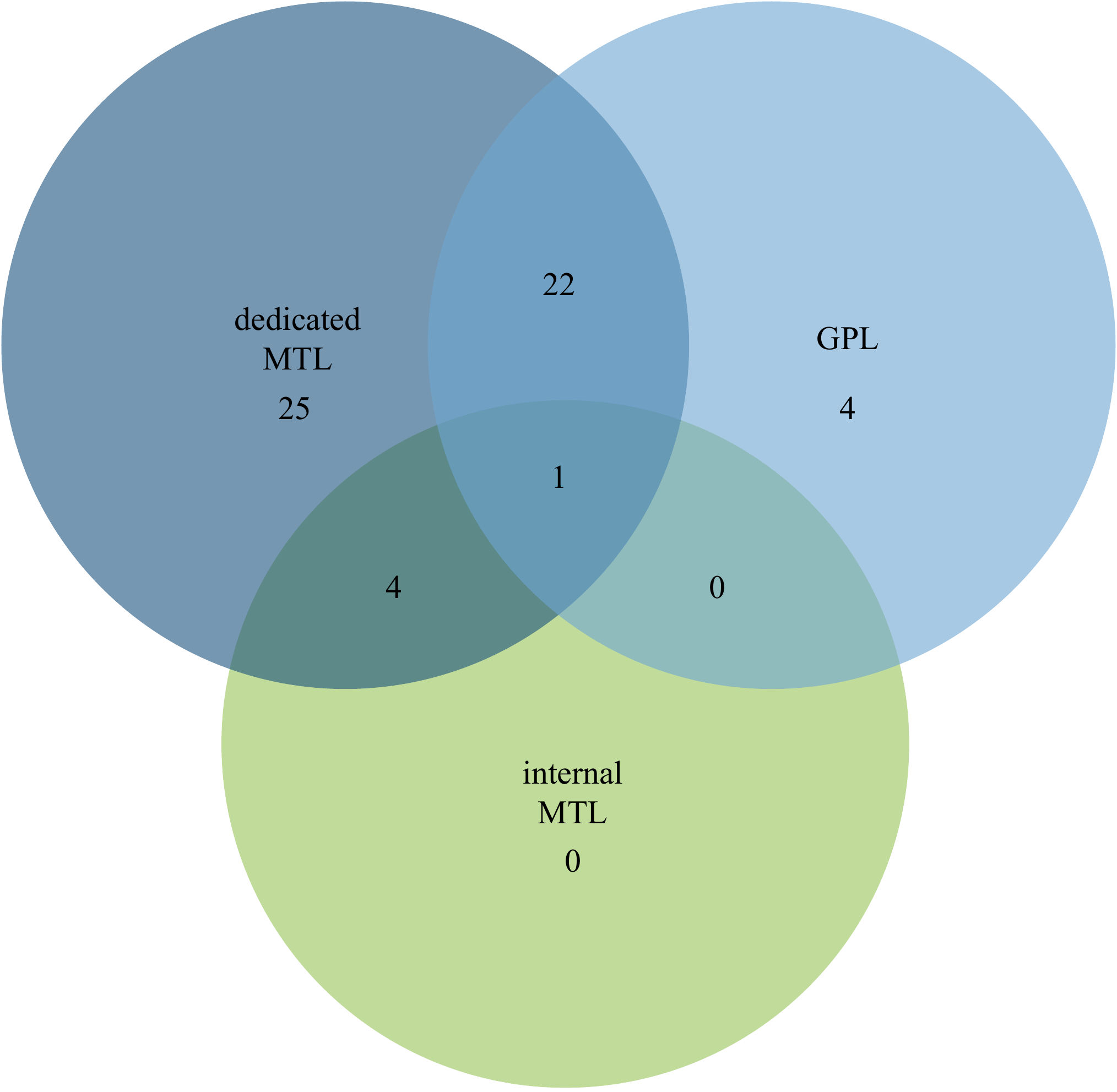}
	\caption{Venn diagram depicting the language usage of participants}~\label{fig:lang_types_used}
\end{figure}

To better assess our participants and to qualify their answers with respect to their background we asked all interviewees to list languages they used to develop model transformations.
\Cref{fig:lang_types_used} summarises the answers given by participants while categorizing languages in one of three categories namely \textit{dedicated MTL}, \textit{internal MTL} and \textit{GPL}.
This differentiation is based on the classifications from \textcite{Czarnecki2006,kahani2019survey}.

The distinction between GPL and dedicated/internal MTL is made, to gain an overview over how large the portion of users of general purpose languages for the development is, compared to the users of model transformation languages.
Furthermore, it also allows for comprehending the viewpoint participants will take when answering questions throughout the interview, i.e. do they compare general purpose languages with model transformation languages based on their experience with both or do they give specific insights into their experiences with one of the two approaches.
Internal MTL is separated from dedicated MTL because one claim within the interview protocol specifically explores the topic of internal model transformation languages.

52 participants have used dedicated model transformation languages such as ATL, Henshin or Viatra for transforming models.
Only half as many (27) stated to have used general purpose languages for this goal.
Lastly, only 5 indicated the use of internal MTLs.

\begin{figure}[ht]
	\centering
	\includegraphics[width=\linewidth]{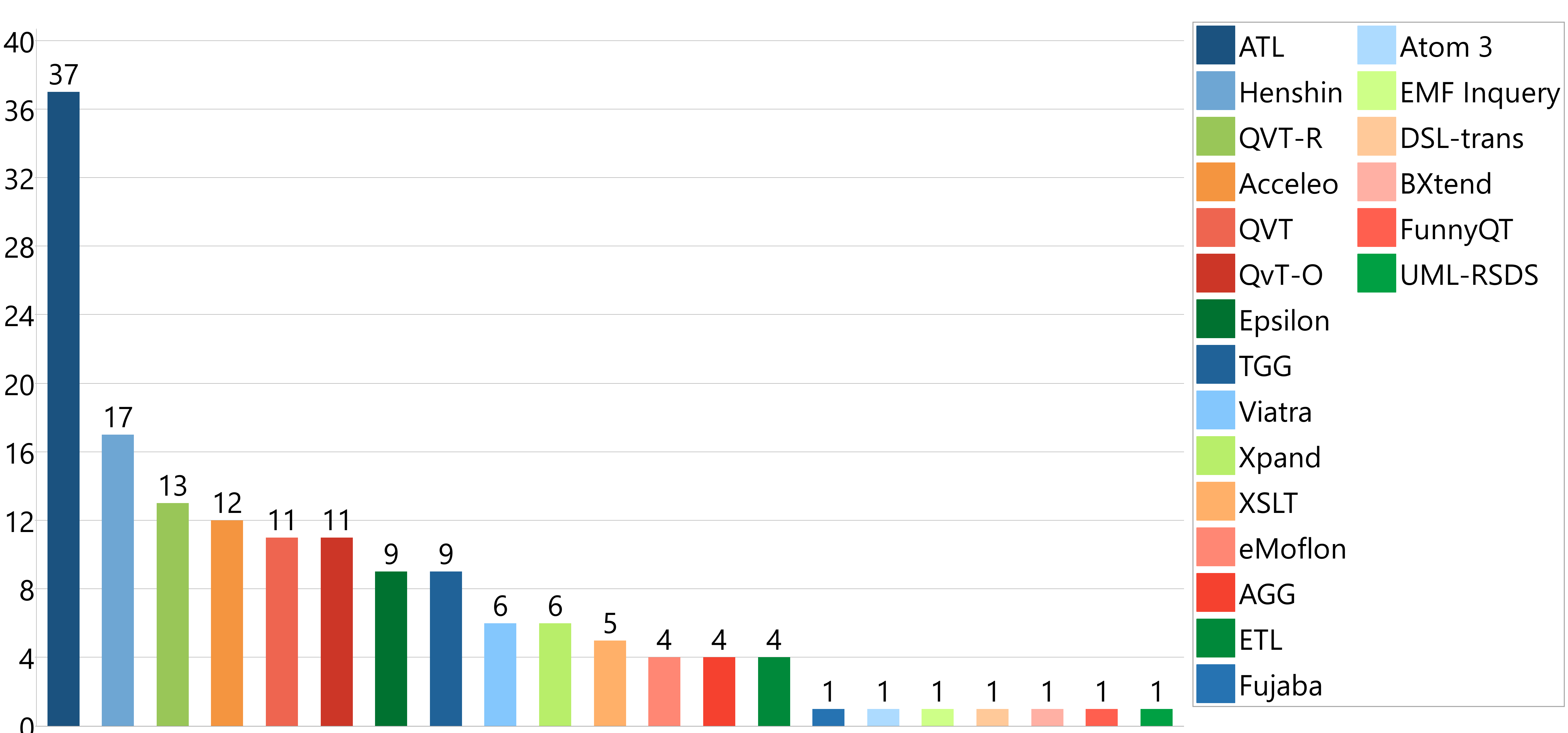}
	\caption{Number of participants using a specific dedicated MTL}~\label{fig:dedicated_MTLs_used}
\end{figure}

When looking at the specific dedicated MTLs used ATL is by far the most prominent one used by interviewees.
A total of 37 participants mention having used ATL.
This is more than double the amount of the second most used language namely Henshin which is only mentioned by 17 interviewees.
The QVT family then follows in third place with QvT-R having been used by 13 participants, QvT-O by 11.
A complete overview over all dedicated model transformation languages used by our interviewees can be found in~\Cref{fig:dedicated_MTLs_used}.
Note that several interviewees mentioned using more than one language, making the total number of data points in this figure larger than 52.

\begin{figure}[ht]
	\centering
	\includegraphics[width=\linewidth]{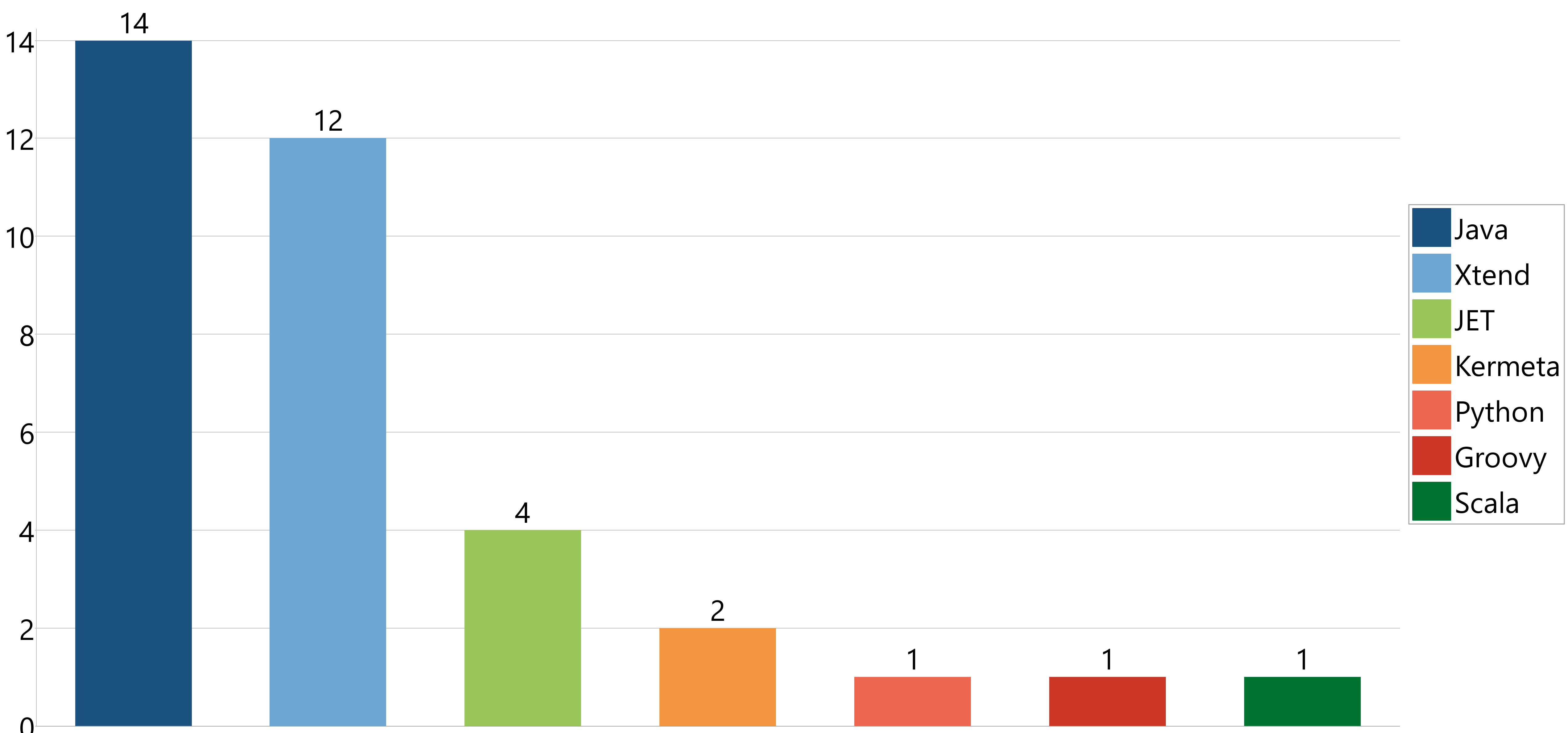}
	\caption{Number of participants using a specific GPL}~\label{fig:GPLs_used}
\end{figure}

In the group of GPL languages used for model transformation (summarised in~\Cref{fig:GPLs_used}), Java is the most used language with 14 participants stating so.
Note that several interviewees mentioned using more than one language, making the total number of data points in this figure larger than 27.
Java is closely followed by Xtend which is mentioned by 12 interviewees.
Then follows a steep drop of in popularity with Java Emitter Templates having been used by only four participants.

Lastly, only four internal model transformation languages, namely RubyTL, NTL, NMF Synchronizations and FunnyQT, are mentioned.
This shows a lack of prominence thereof.
Moreover none of the languages is used by more than two interviewees.
\section{Findings}
\label{sec:results}

Based on the responses of our interviewees and our analysis, we developed a framework to classify influence factors.
It allows us to categorize how factors influence properties of MLTs and each other according to our interviewees.
Note that we split the property \textit{Reuse \& Maintainability} into two properties for the purpose of reporting.
This is done because interviewees chose to consider them separately.
Thus reporting on them separately allows for presenting more nuanced results.

The factors themselves are split into six top-level factors namely \textit{GPL Capabilities}, \textit{MTL Capabilities}, \textit{Tooling}, \textit{Choice of MTL}, \textit{Skills} and \textit{Use Case}.
The first factor, \textit{GPL Capabilities}, encompasses sub-factors related to writing model transformations in general purpose languages.
\textit{MTL Capabilities} encompasses sub-factors that originate from transformation specific features of model transformation languages.
\textit{Tooling} contains factors surrounding tool support for MTLs.
\textit{Choice of MTL} details how the choice of language asserts its influence.
The factor \textit{Skills} encompasses sub-factors associated with skills.
Lastly, the \textit{Use Cases} factor contains sub-factors that relate to the involved use case an its influences.

Within the framework we differentiate between two kinds of factors.
The first kind are factors, that have a positive or negative impact on properties of MTLs. These include the factors \textit{GPL Capabilities}, \textit{MTL Capabilities} and \textit{Tooling} as well as their sub-factors.
The second kind are factors that, depending on their characteristic, moderate how other factors influence properties, e.g. depending on the language, its syntax might have a positive or negative influence on the comprehensibility of written code.
We call such factors \textit{moderating} factors.
These include the factors \textit{Choice of MTL}, \textit{Skills} and \textit{Use Case} and their sub-factors.

\Cref{tbl:overview_influence_factors} provides an overview over the answers given by our interviewees.
The table shows factors on its rows and MTL properties on its columns.
A \texttt{+} in a cell denotes, that interviewees mentioned the factor to have a positive effect on their view of the MTL property.
A \texttt{-} means interviewees saw a negative influence and \texttt{+/-} describes that there have been mentions of both positive and negative influences.
Lastly, a \texttt{M} in a cell denotes, that the factor represents a moderating factor for the MTL property, according to some interviewees.
The detailed extent of the influence of each factor is described throughout \Cref{sec:results,sec:findings}.

\begin{table*}[ht]
	\caption{Overview over quality attribute influences per factor}~\label{tbl:overview_influence_factors}
	\begin{tabularx}{{\textwidth}}{@{}X|l|Y|Y|Y|Y|Y|Y|Y|@{}}
		\toprule
		\textbf{Top-level Factor} & \textbf{Sub-Factor} & \textbf{Compre-hensibil-ity} & \textbf{Ease of Writing} & \textbf{Expres-siveness} &  \textbf{Maintain-ability}  & \textbf{Produc-tivity} & \textbf{Reuse-ability} & \textbf{Tool Support}\\
		\midrule
		\midrule
		GPL Capabilities  &   & +/- & +/- & - & - & +/- & + & +/- \\
		\midrule
		\multirow{11}{4em}{MTL Ca-pabilities} &  Domain Focus & + & + & +/- & +  & +/- &   & + \\
		&  Bidirectionality & +/- & +/- & + &  +/-  &  +  &   &  \\
		&  Incrementality & + & +/- & + &   &  &  &  \\
		&  Mappings & + & +/- & + & +  &   & +/- &  \\
		&  Traceability & + & +/- & +/-  &    & + &   &   \\
		&  Model Traversal & + & + & +  &    & + &   &   \\
		&  Pattern Matching & + &  & + &    & + &   &   \\
		&  Model Navigation & + & + & + &    &   &   &   \\
		&  Model Management & + & + & + &    &   &   &   \\
		&  Reuse Mechanisms &   &   &   &    &   & +/- &   \\
		&  Learnability &  & -  &   &    &   &   &   \\
		\midrule
		\multirow{11}{4em}{Tooling} &  Analysis Tooling & +  &   &  &   & + &   & +/-  \\
		&  Code Repositories &   &   &   &   &   & - &  - \\
		&  Debugging Tooling &  +/-  &   &  &   &   &   & +/-  \\
		&  Ecosystem &   &   &   &  -  & - &   & - \\
		&  IDE Tooling &  & +/- &   & -  &   &   & - \\
		&  Interoperability &   &   &   &   &   &   &  - \\
		&  Tooling Awareness &   &   &   &  &   &   &  - \\
		&  Tool Creation Effort & & & &  & & & - \\
		&  Tool Learnability &  & -  &   &  &  &   &  - \\
		&  Tool Usability &  & -  &   &   & - &   & - \\
		&  Tool Maturity &   &   &   &  &   &   &  - \\
		&  Validation Tooling &   &   &   &   &   &   & -  \\
		\midrule
		\midrule
		Choice of MTL &  & M & M & M & M  & M & M & M  \\
		\midrule
		\multirow{2}{4em}{Skills} &  Language Skills & M &  M &  &  M  &   & M &  \\
		&  User Experience/Knowledge &  & M &   &  M  & M &   &  \\
		\midrule
		\multirow{3}{4em}{Use Case} &  (Meta-) Models &  & M &   &    &   &   &   \\
		&  I/O Semantic gap & M & M &  &    & M &   &   \\
		&  Size & M & M &  &    &   &   &   \\
		\bottomrule
	\end{tabularx}
\end{table*}

In the following we present all top-level factors and their sub-factors and describe their place within our framework.
For each factor we detail its influence on properties of model transformation languages or on other factors, based on the statements made by our interviewees.

\subsection{GPL Capabilities}
Using general purpose languages for developing model transformations, as an alternative to using dedicated languages was extensively discussed in our interviews.
Interviewees mentioned both advantages and disadvantages that GPLs have compared to MTLs that made them view MTLs more or less favourably.

The disadvantages of GPLs compared to MTLs stem from additional features and abstractions that MTLs bring with them and will be discussed later in \Cref{subsec:mtl_capabilities}.
The advantages of GPLs on the other hand can not be placed within the \textit{MTL Capability} factors.
These will instead be presented separately in this section.


According to our interviewees, advantages of GPLs are a relevant factor for all properties of MTLs.

General purpose languages are better suited for \textit{writing} transformations that require lots of computations.
This is because they were streamlined for these kinds of activities and designed for this task, with language features like streams, generics and lambdas.
As a result, general purpose languages are far more advanced for such situations compared to model transformation languages\added{,} which sacrifice this for more domain expressiveness [\ref{que:1}].

Much like the language design for GPLs, their tools and ecosystems are mature and designed to integrate well with each other.
Moreover, according to several interviewees, their tools are of high quality making developers feel more \textit{Productive} [\ref{que:2}].

Lastly, multiple participants noted, that there are much more GPL developers readily available for companies to hire, thus making GPLs more attractive for them.
This helps the \textit{Maintainability} of existing code as such experts are more likely to \textit{Comprehend} GPL code [\ref{que:3}].
Whether this aspect also improves the overall \textit{Productivity} of transformation development in a GPL was disagreed upon, because it might be that developers trained in a MTL could produce similar results with less resources.

It was also mentioned, that much more training resources are available for GPL development, making it easier to start learning and using a new GPL compared to a MTL.

\subsection{MTL Capabilities}
\label{subsec:mtl_capabilities}
The capabilities that model transformation languages provide that are not present in GPLs, are important factors that influence properties of the languages.
This view is shared by our interviewees that raised many different aspects and abstractions present in model transformation languages.

The influence of capabilities specifically introduced in MTLs is diverse and depends on the concrete implementation in a specific language, the skills of the developers using the MTL and the use case in which the MTL is to be applied.
We will discuss all the implications raised by our interviewees regarding the transformation specific capabilities of MTLs for the properties attributed to MTLs in detail, in this section.

\subsubsection{Domain Focus}
\label{sec:results:domain_focus}

\textit{Domain Focus} describes the fact that model transformation languages provide transformation specific constructs, abstractions or workflows.
Interviewees remarked the domain focus, provided by MTLs, as influencing \textit{Comprehensibility}, \textit{Ease of Writing}, \textit{Expressiveness}, \textit{Maintainability}, \textit{Productivity} and \textit{Tool Support}.
But the effects can differ depending on the specific MTL in question.


There exists a consensus that MTLs can provide better domain specificity than GPLs by introducing domain specific language constructs and abstractions.
This increases \textit{Expressiveness} by lifting the problem onto the same level as the involved models allowing developers to express more while writing less.
MTLs allow developers to treat the transformation problem on the same abstraction level as the involved modelling languages [\ref{que:4}].
This also improves the \textit{ease of development}.

Several interviewees argued, that when moving towards domain specific concepts the \textit{Comprehensibility} of written code is greatly increased.
The reason for this is, that because transformation logic is written in terms of domain elements, unnecessary parts are omitted (compared to GPLs) and one can focus solely on the transformation aspect [\ref{que:5}].

Having domain specific constructs was also raised as facilitating better \textit{Maintainability}.
Co-evolving transformations written in MTLs together with hardware-, technology-, platform- or model changes is said to be easier than in GPLs because \textit{``Once you have things like rules and helpers and things like left hand side and right hand side and all these patterns then [it is] easier to create things like meta-rules to take rules from one version to another version [...]''} (\textbf{P23}).

Domain focus also enforces a stricter code structure on model transformations.
This reduces the amount of variety in which they can be expressed in MTLs.
As a result, developing \textit{Tool Support} for analysing transformation scripts gets easier.
Achieving similarly powerful tool support for general purpose languages, and even for internal MTLs, can be a lot harder or even impossible because much less is known solely based on the structure of the code.
Analysis of GPL transformations has to deal with the complete array of functionality of general purpose language constructs [\ref{que:6}].
While MTLs can be Turing complete too, they tend to limit this capability to specific sections of the transformation code.
They also make more information about the transformation explicit compared to GPLs.
This allows for easier analysis of properties of the transformation scripts which reduces the amount of work required to develop analysis tooling.

The influence of domain abstractions on \textit{Productivity} was heavily discussed in our interviews.
Interviewees agreed that, depending on the used language, \textit{Productivity} gains are likely, due to their domain focus.
However, one interviewee explained that precisely because of \textit{Productivity} concerns companies in the industry might use general purpose languages.
The reason for this boils down to the \textit{Use Case} and project context.
Infrastructure for general purpose languages might already be set up and developers do not need to be trained in other languages [\ref{que:7}].
Moreover, different tasks might require different model transformation languages to fully utilise their benefits, which, from an organisational standpoint, does not make sense for a company.
So instead one GPL is used for all tasks.

\subsubsection{Bidirectionality}
\label{sec:results:bx}

According to our interviewees bidirectional functionality in a model transformation language influences its \textit{Comprehensibility}, \textit{Ease of Writing}, \textit{Expressiveness} and \textit{Maintainability} and \textit{Productivity}.
Its effects on these properties then depends on the concrete implementation of the functionality in a \textit{MTL}.
It also depends on the \textit{Skills} of the developers and the concrete \textit{Use Case}.


Our interviewees mentioned that the problem of bidirectional transformations is inherently difficult and that high level formalisms are required to concisely define all aspects of such transformations.
Many believe that because of this solutions using general purpose languages can never be sufficient.
Statements in the vein of \textit{``in a general purpose programming language you would have to add a bit of clutter, a bit of distraction, from the real heart of the matter''} (\textbf{P42}) were made several times.
This, combined with having less optimal querying syntax, then shifts focus away from the actual transformation and decreases both the \textit{Comprehensibility} and \textit{Maintainability} of the written code.

\textit{Maintainability} is also hampered because GPL solutions scatter the implementation throughout two or more rules (or methods or files) that have to be adapted in case of changes [\ref{que:9}].
Expressive and high level syntax in MTLs helps alleviate these problems and increases the \textit{ease} at which developers can \textit{write} bidirectional model transformations.

Interviewees also commented on the fact that, thanks to bidirectional functionalities, consistency definitions and synchronisations between both sides of the transformation can be achieved easier.
This improves the \textit{Maintainability} of modelling projects as a whole and allows for more \textit{Productive} workflows.
Manual attempts to do so have been stated to be error-prone and labour-intensive.

It was also pointed out that the inherent complexity of bidirectionality leads to several problems that have to be considered.
MTLs that offer syntax for defining bidirectional transformations are mentioned to be more complex to use as their unidirectional counterparts.
They should thus only be used in \textit{cases} where bidirectionality is a requirement.
Moreover, one interviewee mentioned that developers are not generally used to thinking in bidirectional way [\ref{que:10}].

Lastly, the models involved in bidirectional transformations also play a role regardless of the language used to define the transformation.
Often the models are not equally powerful making it hard to actually achieve bidirectionality between them, because of information loss from one side to the other [\ref{que:11}].

\subsubsection{Incrementality}

Dedicated functionality in MTLs for executing incremental transformations has been discussed as influencing \textit{Comprehensibility}, \textit{Ease of Writing} and \textit{Expressiveness}.
Similar to bidirectionality its influence is again heavily dependent on the \textit{Use Case} in which incremental languages are applied as well as the \textit{Skills} of the involved developers.


Declarative languages have been mentioned to facilitate incrementality because the execution semantics are already hidden and thus completely up to the execution engine.
This increases the \textit{Expressiveness} of language constructs.
It can, however, hamper the \textit{Comprehensibility} of transformation scripts for developers inexperienced with the language because there is no direct way of knowing in which order transformation steps are executed [\ref{que:12}].

On the other hand interviewees also explained that writing incremental transformations in a GPL is unfeasible.
Manual implementations are error-prone because too many kinds of changes have to be considered and chances are high that developers miss some specific kind.
Due to the high level of complexity that the problem of incrementality inherently posses interviewees argued that \textit{writing} such transformations in MTLs is much \textit{easier} [\ref{que:13}].

The same argumentation also applied for the \textit{Comprehensibility} of transformations.
All the additional code required to introduce incrementality to GPL transformations is argued to clutter the code so much that developers ``\textit{[will be] in way over their head[s]}'' (\textbf{P13}).

As with bidirectionality interviewees agreed, that the \textit{Use Case} needs to be carefully considered when arguing over incremental functionality.
Only when ad-hoc incrementality is really needed should developers consider using incremental languages.
In cases where transformations are executed in batches, maybe even over night, no actual incrementality is necessary and then ``\textit{general purpose programming languages are very much contenders for implementing model transformations}'' (\textbf{P42}).
It was also explained that using general purpose languages for simple transformations is common practice in industry as they are ``\textit{very good in expressing [the required] control flow}'' (\textbf{P42}) and because none of the aforementioned problems for GPLs have a strong impact in these cases.

\subsubsection{Mappings}

The ability of a MTL to define mappings influences that languages \textit{Comprehensibility}, \textit{Ease of Writing}, \textit{Expressiveness}, \textit{Maintainability} and \textit{Reuse} of model transformations.
Developer \textit{Skills}, the used \textit{Language} and concrete \textit{Use Case} also play an important role in the kind of influence.


Interviewees agreed, that the \textit{Expressiveness} of transformation languages utilising syntax for mapping is increased due to them hiding low level operations [\ref{que:14}].
However, as remarked by one participant, the semantic complexity of transformations can not be hidden by mappings, only the computational complexity.

According our interviewees mappings form a natural way of how people think about transformations.
They impose a strict structure on how transformations need to be defined, making it easy for developers to start of \textit{writing} transformations.
The structure also aids general development, because all elements of a transformation have a predetermined place within a mapping.
Being this restrictive has the advantage of directing ones thoughts and focus solely on the elements that should be transformed [\ref{que:15}].
To transform an element, developers only need to write down the element and what it should be mapped to.

The simple structure expressed by mappings also benefits the \textit{Comprehensibility} of transformations.
It allows to easily grasp which elements are involved in a transformation, even by people that are not experienced in the used language.
Trying to understand the same transformation in GPLs would be much harder because \textit{``[one] would not recognize [the involved elements] in Java code any more''} (\textbf{P32}).
Instead, they are hidden in between all the other instructions necessary to perform transformations in the language.
Interviewees also mentioned that, due to the natural fit of mappings for transformations, it is much easier to find entry points from where to start and understand a transformation and to reconstruct relationships between input and output.
This is aided by the fact that the order of mappings within a transformation does not need to conform with its execution sequence and thus enables developers to order them in a comprehensible way [\ref{que:16}].

One interviewee explained that, from their experience, mappings lead to less code being written which makes the transformations both easier to \textit{comprehend} and to \textit{maintain}.
However, they conceded that the competence of the involved developers is a crucial factor as well.
According to them, language features alone do not make code maintainable.
Developers need to have language engineering skills and intricate domain knowledge to be able to design well maintainable transformations [\ref{que:17}].
Both are skills that too little developers posses.

Moreover, several interviewees raised the concern, that complex \textit{Use Cases} can hamper the \textit{Comprehensibility} of transformations.
Understanding which elements are being mapped can be hard to grasp if several auxiliary functions are used for selecting the elements.
Here one interviewee suggested that a standardized way of annotating such selections could help alleviate the problem.

It was also mentioned that, while mappings and other MTL features increase the \textit{Expressiveness} of the language, they might make it harder for developers to start learning the languages.
Because a lot of semantics are hidden behind keywords, developers need to first understand the hidden concepts to be able to utilise them correctly [\ref{que:18}].

Other features that highlight how much \textit{Expressiveness} is gained from mappings have also been mentioned.
Mappings hide how relations between input and output are defined.
This creates a formal and predictable correspondence between them and thus enables \textit{Tracing}.
Moreover, the correspondence between elements allows languages to provide functionality such as \textit{Bidirectionality} and \textit{Incrementality} [\ref{que:19}].


Because many languages that utilise mappings can forgo definitions of explicit control flow, mappings allow transformation engines to do execution optimisations.
However, one interviewee explained that they encountered \textit{Use Cases} where developers want to influence the execution order, forcing them to introduce imperative elements into their code effectively hampering this advantage.
It has also been mentioned that in \textit{complex cases} the code within mappings can get complicated to the point where non experts are unable to \textit{comprehend} the transformation again.
This problem also exists for \textit{writing} transformations as well.
According to one interviewee mappings are great for linear transformations and are thus very dependent on the involved (meta-)models.
Also in \textit{cases} where complex interactions needs to be defined mappings do not present any advantage over GPL syntax and sometimes it can even be easier to define such logic in GPLs [\ref{que:20}].

Lastly, mappings enable more modular code to be written.
This in turn facilitates \textit{reuse}, because reusing and changing code results in local changes instead of several changes throughout different parts of GPL code [\ref{que:21}].

\subsubsection{Traceability}

The ability in model transformation languages to automatically create and handle trace information about the transformation has been discussed by our interviewees to influence \textit{Comprehensibility}, \textit{Ease of writing}, \textit{Expressiveness} and \textit{Productivity}.
However, the concrete effect depends on the \textit{MTL} and the \textit{skill} of users.


All interviewees talking about automatic tracing agreed that it increases the \textit{Expressiveness} of the language utilising it.
In GPLs this functionality would need to be manually implemented using structures like hash maps.
Code to set up traces would then also need to be added to all transformation rules [\ref{que:22}].


However, interviewees disagreed on how much this actually impacts the overall transformation development.
Most interviewees felt like automatic trace handling \textit{Eases Writing} transformations and even increases \textit{Productivity} since no manual workarounds need to be implemented.
This is because manual implementation requires developers to think about when and in which cases traces need to be created and how to access them correctly.
It also enables languages that allow developers to define rules independent from the execution sequence.
One interviewee however felt like this was not as effort intensive as commonly claimed and thus automatic trace handling to them is more of a nice to have feature than a requirement for writing transformations effectively.
Moreover, for complex \textit{Use Cases} of tracing such as QvTs late resolve, the \textit{Users} are required to understand the principle of tracing [\ref{que:23}].
And according to another interviewee teaching how tracing and trace models work is hard.






\textit{Comprehending} written transformations can also be aided by automatic trace management.
Manual implementations introduce unnecessary clutter into transformation code that needs to be understood to be able to understand a whole transformation.
This is especially true if effort has been put into making tracing work efficiently, according to one interviewee.
Understanding a transformation is much more straight forward when only the established relationships between the input and output domains need to be considered, without any additional code to setup and use traces [\ref{que:24}].


Lastly, one interviewee raised the issue that manual trace handling might be necessary to write complex transformations involving multiple source and target models, as current engines are not intended for such \textit{Use Cases}.

\subsubsection{Automatic Model Traversal}

According to our interviewees, the automatic traversal of the input model to apply transformations influences \textit{Ease of Writing}, \textit{Expressiveness}, \textit{Comprehensibility} and \textit{Productivity}.
They also explain that depending on the implementation in a concrete \textit{MTL} the effects can differ.
\textit{Use Cases} are also mentioned to be relevant to the influence of automatic traversal.


Automatic model traversal hides the traversal of the input model and how and when transformations are applied to the input model elements.
Because of this many interviewees expressed that this feature in MTLs increases their \textit{Expressiveness}.
The reduced code clutter also helps with \textit{Comprehensibility}.


It also \textit{Eases the Writing} of transformations because developers do not need to worry about traversing the input and finding all relevant elements, a task that has been described as complicated by interviewees.
This can be of significant help to developers.
One interviewee explained, that they ran an experiment with several participants where they observed model traversal to be \textit{``one of the biggest problems for test persons''} (\textbf{P49}).








Not having to manually define traversal reduces the amount of code that needs to be written and thus increases the overall \textit{Productivity} of development, according to one interviewee.
However, there can also be drawbacks from this practice.
Hiding the traversal automatically leads to the context of model elements to be hidden from the developer.
In cases where the context contains relevant information this can be detrimental and even mask errors that are hard to track down [\ref{que:25}].

Lastly, automatic input traversal enables transformation engines to optimize the order of execution in declarative MTLs.
And MTLs where no automatic execution ordering can be performed have been described as being ``\textit{close to plain GPLs}'' (\textbf{P52}).

\subsubsection{Pattern-Matching}

Some model transformation languages, such as Henshin, allow developers to define sub-graphs of the model graph, often using a graphical syntax, to be matched and transformed.
This pattern-matching functionality influences the \textit{Comprehensibility}, \textit{Expressiveness} and \textit{Productivity}, according to our interviewees.
It is, however, strongly dependent on the specific \textit{language} and \textit{Use Case}\modified{.
The} feature is only present in a small portion of MTLs and brings with it its own set of restrictions depending on the concrete implementation in the language.


Pattern-matching functionality greatly increases the \textit{Expressiveness} of MTLs.
Similar to the basic model traversal no extra code has to be written to implement this semantic.
However, the complexity of the abstracted functionality is even higher, since it is required to perform sub-graph matching to find all the relevant elements in a model.
These patterns can also become arbitrarily complex and thus all interviewees talking about pattern-matching agreed that manual implementations are nearly impossible.
Nevertheless one interviewee mentioned, that all languages they used that provided pattern-matching functionality (Henshin and TGG) had the drawback of providing no abstractions for resolving traces which takes away from its overall usefulness for certain \textit{Use Cases} [\ref{que:26}].


Not having to implement complex pattern-matching algorithms manually is also mentioned to increase the \textit{Productivity} of writing transformations because this task is labour-intensive and error-prone.

Improvements for the \textit{Comprehensibility} of transformations have also been recognized by some interviewees.
They explained that the, often times graphical, syntax of languages with pattern-matching functionality allows to directly see the connection between involved elements.
In GPLs this would be hidden behind all the code required to find and collect the elements.
As such MTL code is \textit{``less polluted''} (\textbf{P52}) than GPL code.
Moreover, the \textit{Comprehensibility} is also promoted by the fact that in some languages the graphical syntax shows the involved elements as they would be represented in the abstract syntax of the model.





\subsubsection{Model Navigation}

Dedicated syntax for expressing model navigation has influence on the \textit{Comprehensibility} and \textit{Ease of Writing} of model transformations as well as on the \textit{Expressiveness} of the \textit{MTL} that utilises it.


Having dedicated syntax for model navigation helps to \textit{ease development} as it allows transformation engineers to simply express which elements or data they want to query from a model while the engine takes care of everything else.
Furthermore, it has been mentioned that this has a positive effect on transformation development because developers do not need to consider the efficiency of the query compared to when defining such queries using nested loops in general purpose languages [\ref{que:27}].

Because languages like OCL abstract from how a model is navigated to compute the results of a query, interviewees attributed a higher \textit{Expressiveness} to them than GPL solutions and described code written in these languages as more concise.
Several interviewees attribute a better \textit{Comprehensibility} to OCL as a result of this conciseness, arguing that well designed conditions and queries written in OCL are easy to read [\ref{que:28}].

OCL has however also been criticised by an interviewee.
According to them, the language is too practically oriented, misses a good theoretical foundation and lacks elegance to properly express ones intent.
They explain that because of this, the worth of learning such a language compared to using a more common language is uncertain.



%

\subsubsection{Model Management}
\label{sec:results:mtl_caps:model_management}

The impact of having to read and write models from and to files, i.e., model management, has been discussed by several interviewees.
Automatic model management was discussed in our interviews as influencing the \textit{Comprehensibility}, \textit{Ease of Writing} and \textit{Expressiveness} of model transformations in MTLs.


The argument for all three properties boils down to developers not having to write code for reading input models or writing output models, as well as the automatic creation of output elements and the order thereof.
Interviewees agreed that implicit language functionality for these aspects raised the \textit{Expressiveness} of languages.
It reduces clutter when reading a written transformation and thus improving the \textit{Comprehensibility}.
Finally, developers do not have to deal with model management tasks, e.g. using the right format, that are not relevant to the actual transformation which helps with \textit{writing} transformations [\ref{que:29}].

%
%
%

\subsubsection{Reuse Mechanism}
\label{sec:results:mtl_caps:reuse}

Mechanisms to reuse model transformations mostly influence the \textit{Reusability} of model transformations in MTLs.
Their concrete influence depends on the used \textit{Language} and how reuse is handled in it.
Interviewees also reported on cases where the users \textit{Skills} with the language was relevant because novices might not be familiar with how the provided facilities can be utilised to achieve reuse.


There exists discourse between the interviewees about reuse mechanisms and their usefulness in model transformation languages.
Several interviewees argued that MTLs do not have any reuse mechanisms that go beyond what is already present in general purpose languages.
They believe that most, if not all, the reuse mechanisms that exist in MTLs are already present in GPLs and as such MTLs do not provide any reuse advantages [\ref{que:30}].
According to them such reuse mechanisms include things like rule inheritance from languages like ATL or modules and libraries.

Other interviewees on the other hand suggested that while the aforementioned mechanisms stem from general purpose languages, they are still more transformation specific than their GPL counterpart.
This is, because the mechanisms work on transformation specific parts in MTLs rather than generic constructs in GPLs [\ref{que:31}, \ref{que:32}].
Because of this focus, interviewees argue that they are more straight forward to use and thus improve \textit{Reusability} in MTLs.

Interviewees also explained that there exist many languages that do not provide any useful reuse or modularisation mechanisms and that even in those that do it can be hard to achieve \textit{Reusability} in a useful manner.
However, one participant acknowledged that in their case, the reason for this might also relate to the inability of the \textit{Users} to properly utilize the available mechanisms.

It has also been mentioned that reuse in model transformations is an inherently complex problem to solve.
Transferring needs between two transformations which apply on different meta-models is difficult to do.
As such, model transformation are often tightly tied to the domain in which they are used which makes reuse hard to achieve and most reuse between domains is currently done via copy \& paste.
This argument can present a reason why, as criticised by several interviewees, no advanced reuse mechanisms are broadly available.

The desire for advanced mechanisms has been expressed several times.
One interviewee would like to see a mechanism that allows to define transformations to adapt to different input and output models to really feel like MTLs provide better reusability than GPLs.
Another mentioned, that all reuse mechanisms conferred from object orientation rely on the tree like structure present in class structures while models are often more graph like and cyclic in nature.
They believe that mechanisms that address this difference could be useful in MTLs.

Another disadvantage in some MTLs that was raised\modified{,} is the granularity on which reuse can be defined.
In languages like Henshin, for example, reuse is defined on a much coarser level than what is possible in GPLs.

Not having a central catalogue, similar to maven for Java, from which transformations or libraries can be reused, has also been critiqued as hindering reuse in model transformation languages.

\subsubsection{Learnability}
\label{sec:results:mtl_caps:learn}

The learnability of model transformation languages has been discussed as influencing the \textit{Ease of Writing} model transformations.


It has been criticised by several interviewees, that the learning curve for MTLs is steep.
This is, in part, due to the fact that users not only need to learn the languages themselves, but also accompanying concepts from MDE which are often required to be able to fully utilise model transformation languages.
The learning curve makes it difficult for users to get started and therefore hampers the \textit{Ease of Writing} transformations [\ref{que:33}].
This effect could be observed among computer science students at several of the universities of our interviewees.
The students were described to having difficulties adapting to the vastly different approach to development compared to more traditional methods.
A potential reason for this could be that people come into contact with MDE principles too late, as noted by an interviewee [\ref{que:34}].



\subsection{Tooling}
\label{sec:results:tooling}

While \textit{Tool Support} is a MTL property that was investigated in our study, the tooling provided for MTLs, as well as several functional properties thereof, have been raised many times as factors that influence other properties attributed to model transformation languages as well.
Most of the time this influence is negative, as tooling is seen as one of the greatest weak points of model transformation languages by our interviewees.

Many interviewees explained, that the most common languages do in fact have tools.
The problem, however, lies in the fact that some helpful tools only exist for one language while others only exist for another language.
As a result there is always some tool missing for any specific language.
This leads people to feel like \textit{Tool Support} for MTLs is bad compared to GPLs.
Though there was one interviewee that explained that for their \textit{Use Cases}, all tools required to be productive were present.

In the following, we will present several functional properties and tools that interviewees expressed as influential for \textit{Tool Support} as well as other properties of MTLs.

\subsubsection{Analysis Tooling}

Analysis tools are seen as a strong suit of MTLs.
Their existence in MTLs is said to impact \textit{Productivity}, \textit{Comprehensibility} and perceived \textit{Tool Support}.


According to the interviewees, some analyses can only be carried out on MTLs, as the abstraction in transformations in GPLs is not high enough and too much information is represented by the program logic and not in analysable constructs.
As one interviewee explained, this comes from the fact that for complex analysis, such as validating correctness, languages need to be more structured.
Nevertheless, participants mainly mentioned analyses they would like to see, which is an indication that, while the potential for analysis tools for MTLs is high, they do not yet see usable solutions for it, or are unaware of it.
This is highlighted by one interviewee that explained that they are missing ways to check properties of model transformations, even though such solutions exist for certain MTLs [\ref{que:35}].

A desired analysis tool mentioned in the interviews is rule dependency analysis and visualisation.
They believe that such a tool would provide valuable insights into the structure of written transformations and help to better \textit{comprehend} them and their intent.
\textit{``What I would need for QVT-R, for example, in more complex cases, would be a kind of dependency graph.''} (P32).
Moreover two interviewees expressed the desire for tools to verify that transformations uphold certain properties or preserve certain properties of the involved models.

\subsubsection{Code Repositories}

A gap in \textit{Tool Support} that has been brought up several times, is a central platform to share transformation libraries, much like maven-central for Java or npm for JavaScript.
This tool influences \textit{Tool Support} and the \textit{Reusability} of MTLs.


According to two interviewees, not having a central repository where transformations, written by other developers, can be browsed or downloaded, greatly hinders their view on the \textit{Reusability} of model transformation languages.
This is because it creates a barrier for reuse.
For one thing, it is difficult to find model transformations that can be reused.
Secondly, mechanisms that would simplify such reuse are then also missing.
\textit{``I think what is currently missing is a catalogue or a tool like maven for having repositories for transformations so you can possibly find transformations to reuse.''} (\textbf{P14})

\subsubsection{Debugging Tooling}

Debuggers have been raised as essential tools that help with the \textit{Comprehensibility} of written model transformations.
The existence of a debugger for a given language therefore influences its \textit{Tool Support} as well as its \textit{Comprehensibility}.


One interviewee explained that, especially for declarative languages, where the execution deviates greatly from the definition, debugging tools would be a tremendous help in understanding what is going on.
In this context, opinions were also expressed that more information is needed for debugging model transformations than for traditional programming and that the tools should therefore be able to do more.
Interviewees mentioned the desire to be able to see how matchings arise or why expected matches are not produced as well as the ability to challenge their transformations with assertions to see when and why expressions evaluate to certain values.
\textit{``Demonstrate to me that this is true, show me the section of the transformation in which this OCL constraint is true or false.''} (\textbf{P28}).

Valuable debugging of model transformations is mainly possible in dedicated MTLs, according to one interviewee.
They argue that debugging model transformations in GPLs is cumbersome because much of the code does not relate to the actual transformation thus hampering a developers ability to grasp problems in their code.

\subsubsection{Ecosystem}
\label{sec:results:tooling:eco}

The ecosystems around a language, as well as existing ecosystems, in which model transformations languages would have to be incorporated into, were remarked as mostly limiting factors for \textit{Productivity}, \textit{Maintainability} as well as the perceived amount of \textit{Tool Support}.


One interviewee explained, that for many companies, adopting a model transformation language for their modelling concerns is out of the question because it would negatively impact their \textit{Productivity}.
The reason for this are existing ecosystems, which are designed for GPL usage.
Moreover, it was noted that, to fully utilise the benefits of dedicated languages many companies would need to adopt several languages to properly utilise their benefits.
This is seen as hard to implement as \textit{``people from industry have a hard time when they are required to use multiple languages''} (\textbf{P49}) making it hard for them to \textit{maintain} code in such ecosystems.

Ecosystems surrounding MTLs have also been criticised in hampering \textit{Productivity} and perceived \textit{Tool Support}.
Several interviewees mentioned, that developers tend to favour ecosystems where many activities can be done in one place, something they see as lacking in MTL ecosystems.
One interviewee even referred to this problem as the reason why they turned away from using model transformation languages completely [\ref{que:37}].

This issue somewhat contrasts a concern raised by a different group of interviewees.
They felt that the coupling of much of MDE tooling to Eclipse is a problem that hampers the adoption of MTLs and MDE.
This coupling allows many tools to be available within the Eclipse IDE but, according to them, the problem lies in the fact that Eclipse is developed at a faster pace than what tool developers are able to keep up with, leaving much of the \textit{Tool Support} for MTLs in an outdated state, limiting their exposure and usability [\ref{que:38}].







\subsubsection{IDE Tooling}
\label{sec:results:tooling:ide}

One essential tool for \textit{Tool Support}, \textit{Ease of Writing} and \textit{Maintainability} of MTLs are language specific editors in IDEs.


Several interviewees mentioned, that languages without basic IDE support are likely to be unusable, because developers are used to all the quality-of-life improvements, with autocompletion and syntax highlighting being the two most important features offered by such tools.
Refactoring capabilities in IDEs, like renaming, have also been raised as crucial, especially for easing the \textit{Maintainability} of transformations.

\subsubsection{Interoperability}

How well tools can be integrated with each other has been raised as a concern for the \textit{Tool Support} of MTLs by several interviewees.


Interviewees see a clear advantage for GPLs when talking about interoperability between different MTL tools.
They believe, that due to the majority of tools being research projects, little effort is spent into standardizing those in a way that allows for interoperability on the level that is currently provided for GPLs.
\textit{``But the technologies, to combine them, it is difficult [...]''} (\textbf{P36}).
One interviewee described their first hand experience with this.
They could not get a MTL to process models they generated with a software architecture tool because it produced non standard UML models which could not be used by the MTL.
This problem has been echoed by another interviewee who explained that many MTLs do not work with non EMF compatible models.




\subsubsection{Tooling Awareness}
\label{sec:results:tooling:aware}

A few interviewees talked about the availability of information about tools and the general awareness of which tools for MTLs exist.
According to them, this strongly influences the perceived lack of \textit{Tool Support} for model transformation languages in general.


When starting out with model transformations it can be hard to find out which tools one should use or even which tools are available at all.
Two interview participants mention experiencing this first hand.
They further explain that there exists no central starting point when looking for tools and tools are generally not well communicated to potential users outside of research [\ref{que:40}].

Another interviewee suspected that the same problem also happens the other way around.
They believe that some well designed MTL tools are completely unknown outside of the companies that developed them for internal use.




\subsubsection{Tool Creation Effort}

The amount of effort, that is required to be put into the development of MTL tools, has been raised by many interviewees as a reason why \textit{Tool Support} for MTLs is seen as lacking.


All interviewees talking about the effort involved in creating tools for MTLs agree that there is a lot of effort involved in developing tools.
This is not a problem in and of itself but, when comparing tooling with GPLs interviewees felt like MTLs being at a disadvantage.
The disadvantage stems from the community for MTLs being much smaller and thus having less man power to develop tools which limits the amount of tools that can be developed.
Several interviewees noted, that the only solution they see for this problem is industrial backing or commercial tool vendors because \textit{``I am keenly aware of the cost to being able to develop a good programming language, the cost of maintaining it and the cost of adding debuggers and refactoring engines. It is enormous.''} (\textbf{P01}).

When comparing the actual effort for creating transformation specific tools, some interviewees explained that their experience suggests easier tool development for MTLs than for GPLs.
They explained that, extracting the transformation specific information necessary for such tools out of GPL code complicates the whole process, whereas dedicated MTLs with their small and focused language core provide much easier access to such information [\ref{que:42}].

\subsubsection{Tool Learnability}

The learning curve for someone starting off with MTLs and MTL tools is discussed as a heavy burden to the perceived effectiveness of \textit{Tools} and even influences \textit{Ease of Writing}.


Several interviewees criticised the fact that when starting off with a new MTL and its accompanying tools there is little support for users.
Many tools lack basic documentation on how to set them up properly and how to use them.
As a result users feel lost and find it difficult to start off \textit{writing} transformations [\ref{que:43}].

\subsubsection{Tool Usability}

Related to the topic of learnability, the usability of tools for model transformation languages is discussed as influencing the quality of \textit{Tool Support} for the languages as well as the \textit{Ease of Writing} and \textit{Productivity}.


To fully utilise the potential of MTLs useable tools are essential.
Due to their higher level of abstraction, high quality tools are necessary to properly work with them and \textit{Write} well rounded transformations [\ref{que:44}].

This is currently not the case when looking at the opinions of our interviewees talking about the topic of tool usability.
There are tools available for people to start off with developing transformations but they are not well rounded and thus not ready for professional use, according to one interviewee.
This is supported by several other interviewees opinions, many tools are faulty, which hinders the workflow and reduces \textit{Productivity} [\ref{que:45}].
It has also been stated that if there were high quality useable tools available, they would be used.
The reality for many users is, however, more in line with the experience of one interviewee who stated that they were unable to get many tools (for bidirectional languages) to even work at all.

\subsubsection{Tool Maturity}

A reason given for many of the criticised points surrounding MTL tools is their maturity.
It is said to be a pivotal factor for everything related to \textit{Tool Support}.


The maturity of tools for model transformation languages was commented on a lot.
Tools need to be refined more in order to raze many of their current faults.
The fact that this is not currently done relates back to the effort that is involved with it and the limited personnel available to do so.
This is highlighted in an argument made by one of the interviewees who feels, that the community should not be hiding behind the argument of maturity [\ref{que:46}].

\subsubsection{Validation Tooling}

Tools or frameworks to support the validation and testing of transformations written in MTLs have been discussed to influence the perceived \textit{Tool Support} for nearly all MTLs.


Too much of the available tool support focuses solely on the writing phase of transformation development.
There is little tool support for testing developed transformations, which has been raised as an area where much progress can, and has to be, made.
Especially when comparing the current state of the art with GPLs, MTLs are seen as lacking [\ref{que:47}].
Not only are there little to no tools like unit testing frameworks, there is also too few transformation specific support such as tools to specifically verify binding or helper code in ATL.

\subsection{Choice of MTL}

The choice of MTL is an obvious factor that influences how other factors, such as the \textit{MTL Capabilities}, influence the properties of model transformation languages.
However, it should be explicitly mentioned, because it has been brought up countless times by interviewees while not often being considered in literature.
Depending on the chosen model transformation language its capabilities and whole makeup changes, which has strong implications on all aspects of model transformation development.

A large number of the interviewees have commented on this.
They either directly raised the concern, by prefacing a discussion with a statement such as \textit{``[...] it depends on the MTL''}, or indirectly raised the concern, when comparing specific languages that do or do not exhibit certain capabilities and properties.


\subsection{Skills}

Skills of involved stakeholders is another group of factors that does not have a direct influence on how MTLs are perceived but instead plays a passive role.
Many interviewees cited skills as a limiting factor to other influence factors.
They argue that insufficient user skills could hinder advantages that MTLs can provide and might even create disadvantages compared to the more well-known and commonly used GPLs.

In this section we present the different types of skills mentioned by our interviewees as being relevant to the discussion of properties of model transformation languages.

\subsubsection{Language Skills}

The skill of developers in a specific model transformation language was raised by several interviewees as critical in facilitating many of the advantages provided through the languages capabilities.
So much so that, according to them, the ability of developers to use and read a language can make or break any and all advantages and disadvantages of MTLs related to \textit{Comprehensibility}, \textit{Ease of Writing}, \textit{Maintainability} and \textit{Reuseability}.

Basic skills in any language are a prerequisite to being able to use it.
They are also necessary to understand written code.
There is no difference between GPLs and MTLs.
It was however mentioned, that developers are generally more used to the development style in general purpose languages.
Thus users need to learn how to solve a problem with the functionality of the model transformation language to be able to successfully develop transformations in a MTL [\ref{que:48}].
This is especially relevant for complex transformations, where users are required to know of abstractions such as tracing or automatic traversal.
Following on on this, one interviewee explained, that while learning the language is a requirement, using a new library, e.g. one for developing model transformations, in a GPL also entails learning and as such this must not be regarded as a disadvantage.

For \textit{reuse} it is also paramount for users to know what elements of a transformation can be reused through language functionality.
As a result the \textit{Reuseability} is again limited by the knowledge of users in the specific language.

Lastly, being able to \textit{maintain} a transformation written in an MTL also requires users to know the language to be able to understand where changes need to be made [\ref{que:49}].

\subsubsection{User Experience/Knowledge}

Apart from mastering a used language, the amount of experience users have with said languages and techniques also play a vital role in bringing out the full potential of said languages.
Our interviewees discussed this for \textit{Ease of Writing}, \textit{Maintainability} and \textit{Productivity}.

One interviewee explained that, from their experience, the amount of experience developers have with a language greatly impacts their \textit{Productivity} when using said language.
The problem for MTLs that results from this is the fact that there is little incentive for a person that is trying to build up their CV to spend much time on dedicated languages such as MTLs [\ref{que:50}].
Developers are more inclined to learning and accumulating experience in languages that are commonly used in different companies to improve their chances of landing jobs.
As a result people tend to have little to no experience in using MTLs.
This in turn results in them having a harder time \textit{developing} transformations in these languages, and the final product being of lower quality than what they could achieve using a GPL in which they have more experience in.

The problem is further exacerbated in teaching.
\textit{``Many MDSE courses are just given too late, when people are too acquainted with GPLs, and then its really hard for students to see the point of using models, modelling and MTLs, because it's comparable with languages and stuff they have already learned and worked with.''} (\textbf{P06}).

%



\subsection{Use Case}

Similar as the \textit{MTL} itself and stakeholder \textit{Skills}, the concrete \textit{Use Case} in which model transformations are being developed is another factor that does not directly influence how properties of MTLs are being assessed.
Instead, interviewees often mention that, depending on the \textit{Use Case}, other influence factors could either have a positive or negative effect.

Use cases are distinguished along three dimensions.
The complexity of involved models based on their structure, the complexity of the transformation based on the semantic gap between source and target, and the size of the transformation based on the use case.
Depending on which differentiation is referred to by the interviewees, the considerations look differently.

\subsubsection{Involved (meta-) models}

The involved models and meta-models can have a large impact on the transformation and can hence heavily influence the advantages or disadvantages that MTLs exhibit.

Writing transformations for well behaved models, meaning models that are well structured and documented, can be immensely productive in a MTL while `badly' behaved models bring out problems that require well trained experts to properly solve in a MTL.
The UML meta-model was put forth as an example for such a badly behaved meta-model by one interviewee.
According to them, transformations involving UML models can be problematic due to templates, which are model elements that are parametrized by other model elements, and \texttt{eGenericType}.
The problem with these complex model elements is often worsened by low-quality documentation [\ref{que:52}].
In cases where these badly behaved models are involved, many of the advantages from advanced features of MTLs can not be properly utilised without powerful tooling.

\subsubsection{Semantic gap between input and output}

Many interviewees formulate considerations based on the differentiation between `simple' and `complex' transformations in terms of the semantic gap that needs to be overcome.
Transformations are considered simple when there is little semantic difference between the source and target models.
Common comparisons read like: \textit{``transforming boxes into circles''} (\textbf{P32}).

For simple transformations, model transformation languages are regarded as taking a lot of work off of the developers through the different language features discussed in \Cref{subsec:mtl_capabilities}.
In more complex cases, transformations will get more complex and the developers experience gets more and more relevant, as more advanced language features need to be utilised, which can favour GPLs [\ref{que:53}].

Others argue that the advantages of MTLs only really come into play in more complex cases or when high level features, such as bidirectionality or incrementality, are required.
The reasoning for this argument is, that in simple cases the overhead of GPLs is not that prominent.
Moreover, for writing complex transformations, dedicated query languages in MTLs are regarded by some to be much better than having to manually define complex conditions and loops in a GPL.

\subsubsection{Size}

The Size of the transformation based on the \textit{Use Case} is considered by some interviewees to be a relevant factor as well.
In cases with many rules that depend on each other, MTLs are seen as having advantages [\ref{que:54}].
The size of transformations has been said to be a limiting factor for the use of graphical languages as enormous transformations would make graphical notations confusing.
Modularisation mechanisms of languages also become a relevant feature in these cases.

\section{Cross-Factor Findings}
\label{sec:findings}

Based on interview responses, we developed a structure model from structural equation modeling~\parencite{Weiber2021} that models interactions between the presented influence factors and the properties of model transformation languages.

Structure models depict assumed relationships between variables~\parencite{Weiber2021}.
They divide their components into endogenous and exogenous variables.
The endogenous variables are explained by the causal influences assumed in the model.
The exogenous variables serve as explanatory variables, but are not themselves explained by the causal model.
Exogenous variables either directly influence a endogenous variable or they moderate an influence of another exogenous variable on an endogenous variable.

Structure models are therefore well suited to provide a theoretical framework for the findings of our work.
Factors identified during analysis constitute exogenous variables while MTL properties constitute endogenous variables.
Moderating factors also constitute exogenous variables, with the caveat of only having moderating influences on other influences.

\begin{figure}[ht]
	\centering
	\includegraphics[width=\linewidth]{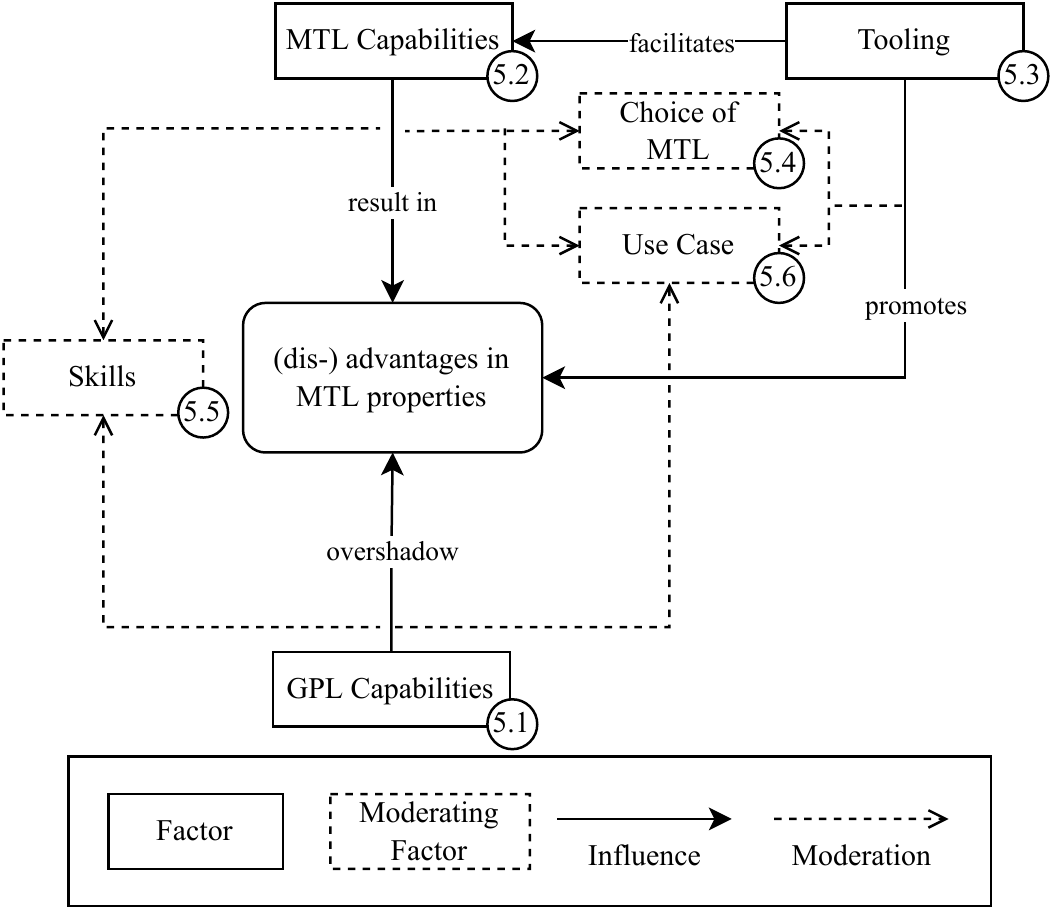}
	\caption{Graphical overview over factor influences and moderations}
	\label{fig:factor_theory}
\end{figure}

A graphical overview over the influences identified by us can be found in \Cref{fig:factor_theory}.
The detailed structure model is depicted in \Cref{fig:structure_model}.

The structure model depicts which MTL properties are influenced by which of the identified factors.
For each MTL property the model also illustrates which factors moderate the influence on the property.
Rectangles represent factors, rounded rectangles represent MTL properties.
Below each MTL property the moderating factors for the property are displayed.
Arrows between a factor and a MTL property represent the factor having an influence on the MTL property.
Each influence on a MTL property is moderated by its moderating factors.
The graphical representation deviates from standard presentation due to its size.

The capabilities of model transformation languages based on domain specific abstractions are aimed at providing advantages over general purpose languages.
Whether these advantages are realised or whether disadvantages emerge is moderated by the \textit{Skills} of the users, the concrete \textit{MTL chosen} as well as the \textit{Use Case} for which transformations are applied.
Depending on these organisational factors the versatility provided by general purpose languages may overshadow advantages provided by MTLs.

Tooling can aid the usage of advanced features of MTLs by supporting developers in their endeavours beyond simple syntax highlighting.
As a result, tools can further promote the advantages that stem from the domain specific abstractions.
The biggest problem that tools for MTLs face is their availability and quality.

In the following we will present thorough discussions of the most salient observations based on our interviews and the presented structure model.
Note that, as will be thoroughly discussed in \Cref{sec:threats:external}, the observations have a limited applicability for industry use cases due to the lack of interviewees that use MTLs in an industry setting.


\tikzstyle{factor} = [rectangle, draw, node distance=2cm, text width=8em, text centered, minimum height=3em]
\tikzstyle{property} = [rectangle, rounded corners, draw, node distance=3.25cm, text width=8em, text centered, minimum height=3em,fill=white]

\tikzstyle{mod-factor-box} = [rectangle, draw, node distance=4em, text width=8em, text centered, minimum height=5.65em, dashed, rounded corners]

\tikzstyle{l} = [draw, -latex']

\pgfdeclarelayer{background}
\pgfsetlayers{background,main}

\begin{figure*}
\begin{tikzpicture}
	\node[factor,actions ocg={}{reset-b df2comp df2eow df2exp df2mtb df2pro df2ts}{bx2comp bx2eow bx2exp bx2mtb bx2pro inc2comp inc2eow inc2exp map2comp map2eow map2exp map2mtb map2reu man2comp man2eow man2exp nav2comp nav2eow nav2exp gpl2comp gpl2eow gpl2exp gpl2mtb gpl2pro gpl2reu gpl2ts trv2comp trv2eow trv2exp trv2pro pm2comp pm2exp pm2pro rm2reu trc2comp trc2eow trc2exp trc2pro debug2comp debug2ts mat2ts lea2eow lea2ts val2ts use2eow use2pro use2ts ide2eow ide2mtb ide2ts int2ts cre2ts ana2comp ana2pro ana2ts eco2mtb eco2pro eco2ts rep2reu rep2ts awa2ts ler2eow}] (factor-df) at (0,0) {Domain Focus};
	\node[factor,below of=factor-df,actions ocg={}{reset-b bx2comp bx2eow bx2exp bx2mtb bx2pro}{df2comp df2eow df2exp df2mtb df2pro df2ts inc2comp inc2eow inc2exp map2comp map2eow map2exp map2mtb map2reu man2comp man2eow man2exp nav2comp nav2eow nav2exp gpl2comp gpl2eow gpl2exp gpl2mtb gpl2pro gpl2reu gpl2ts trv2comp trv2eow trv2exp trv2pro pm2comp pm2exp pm2pro rm2reu trc2comp trc2eow trc2exp trc2pro debug2comp debug2ts mat2ts lea2eow lea2ts val2ts use2eow use2pro use2ts ide2eow ide2mtb ide2ts int2ts cre2ts ana2comp ana2pro ana2ts eco2mtb eco2pro eco2ts rep2reu rep2ts awa2ts ler2eow}] (factor-bx) {Bidirectionality};
	\node[factor,below of=factor-bx,actions ocg={}{reset-b inc2comp inc2eow inc2exp}{df2comp df2eow df2exp df2mtb df2pro df2ts bx2comp bx2eow bx2exp bx2mtb bx2pro map2comp map2eow map2exp map2mtb map2reu man2comp man2eow man2exp nav2comp nav2eow nav2exp gpl2comp gpl2eow gpl2exp gpl2mtb gpl2pro gpl2reu gpl2ts trv2comp trv2eow trv2exp trv2pro pm2comp pm2exp pm2pro rm2reu trc2comp trc2eow trc2exp trc2pro debug2comp debug2ts mat2ts lea2eow lea2ts val2ts use2eow use2pro use2ts ide2eow ide2mtb ide2ts int2ts cre2ts ana2comp ana2pro ana2ts eco2mtb eco2pro eco2ts rep2reu rep2ts awa2ts ler2eow}] (factor-inc) {Incrementality};
	\node[factor,below of=factor-inc,actions ocg={}{reset-b map2comp map2eow map2exp map2mtb map2reu}{df2comp df2eow df2exp df2mtb df2pro df2ts bx2comp bx2eow bx2exp bx2mtb bx2pro inc2comp inc2eow inc2exp man2comp man2eow man2exp nav2comp nav2eow nav2exp gpl2comp gpl2eow gpl2exp gpl2mtb gpl2pro gpl2reu gpl2ts trv2comp trv2eow trv2exp trv2pro pm2comp pm2exp pm2pro rm2reu trc2comp trc2eow trc2exp trc2pro debug2comp debug2ts mat2ts lea2eow lea2ts val2ts use2eow use2pro use2ts ide2eow ide2mtb ide2ts int2ts cre2ts ana2comp ana2pro ana2ts eco2mtb eco2pro eco2ts rep2reu rep2ts awa2ts ler2eow}] (factor-map) {Mappings};
	\node[factor,below of=factor-map,actions ocg={}{reset-b man2comp man2eow man2exp}{df2comp df2eow df2exp df2mtb df2pro df2ts bx2comp bx2eow bx2exp bx2mtb bx2pro inc2comp inc2eow inc2exp map2comp map2eow map2exp map2mtb map2reu nav2comp nav2eow nav2exp gpl2comp gpl2eow gpl2exp gpl2mtb gpl2pro gpl2reu gpl2ts trv2comp trv2eow trv2exp trv2pro pm2comp pm2exp pm2pro rm2reu trc2comp trc2eow trc2exp trc2pro debug2comp debug2ts mat2ts lea2eow lea2ts val2ts use2eow use2pro use2ts ide2eow ide2mtb ide2ts int2ts cre2ts ana2comp ana2pro ana2ts eco2mtb eco2pro eco2ts rep2reu rep2ts awa2ts ler2eow}] (factor-man) {Model Management};
	\node[factor,below of=factor-man,actions ocg={}{reset-b nav2comp nav2eow nav2exp}{df2comp df2eow df2exp df2mtb df2pro df2ts bx2comp bx2eow bx2exp bx2mtb bx2pro inc2comp inc2eow inc2exp map2comp map2eow map2exp map2mtb map2reu man2comp man2eow man2exp gpl2comp gpl2eow gpl2exp gpl2mtb gpl2pro gpl2reu gpl2ts trv2comp trv2eow trv2exp trv2pro pm2comp pm2exp pm2pro rm2reu trc2comp trc2eow trc2exp trc2pro debug2comp debug2ts mat2ts lea2eow lea2ts val2ts use2eow use2pro use2ts ide2eow ide2mtb ide2ts int2ts cre2ts ana2comp ana2pro ana2ts eco2mtb eco2pro eco2ts rep2reu rep2ts awa2ts ler2eow}] (factor-nav) {Model Navigation};
	\node[factor,below of=factor-nav,actions ocg={}{reset-b gpl2comp gpl2eow gpl2exp gpl2mtb gpl2pro gpl2reu gpl2ts}{df2comp df2eow df2exp df2mtb df2pro df2ts bx2comp bx2eow bx2exp bx2mtb bx2pro inc2comp inc2eow inc2exp map2comp map2eow map2exp map2mtb map2reu man2comp man2eow man2exp nav2comp nav2eow nav2exp trv2comp trv2eow trv2exp trv2pro pm2comp pm2exp pm2pro rm2reu trc2comp trc2eow trc2exp trc2pro debug2comp debug2ts mat2ts lea2eow lea2ts val2ts use2eow use2pro use2ts ide2eow ide2mtb ide2ts int2ts cre2ts ana2comp ana2pro ana2ts eco2mtb eco2pro eco2ts rep2reu rep2ts awa2ts ler2eow}] (factor-gpl) {GPL Capabilities};
	\node[factor,below of=factor-gpl,actions ocg={}{reset-b trv2comp trv2eow trv2exp trv2pro}{df2comp df2eow df2exp df2mtb df2pro df2ts bx2comp bx2eow bx2exp bx2mtb bx2pro inc2comp inc2eow inc2exp map2comp map2eow map2exp map2mtb map2reu man2comp man2eow man2exp nav2comp nav2eow nav2exp gpl2comp gpl2eow gpl2exp gpl2mtb gpl2pro gpl2reu gpl2ts pm2comp pm2exp pm2pro rm2reu trc2comp trc2eow trc2exp trc2pro debug2comp debug2ts mat2ts lea2eow lea2ts val2ts use2eow use2pro use2ts ide2eow ide2mtb ide2ts int2ts cre2ts ana2comp ana2pro ana2ts eco2mtb eco2pro eco2ts rep2reu rep2ts awa2ts ler2eow}] (factor-trv) {Model Traversal};
	\node[factor,below of=factor-trv,actions ocg={}{reset-b pm2comp pm2exp pm2pro}{df2comp df2eow df2exp df2mtb df2pro df2ts bx2comp bx2eow bx2exp bx2mtb bx2pro inc2comp inc2eow inc2exp map2comp map2eow map2exp map2mtb map2reu man2comp man2eow man2exp nav2comp nav2eow nav2exp gpl2comp gpl2eow gpl2exp gpl2mtb gpl2pro gpl2reu gpl2ts trv2comp trv2eow trv2exp trv2pro rm2reu trc2comp trc2eow trc2exp trc2pro debug2comp debug2ts mat2ts lea2eow lea2ts val2ts use2eow use2pro use2ts ide2eow ide2mtb ide2ts int2ts cre2ts ana2comp ana2pro ana2ts eco2mtb eco2pro eco2ts rep2reu rep2ts awa2ts ler2eow}] (factor-pm) {Pattern Matching};
	\node[factor,below of=factor-pm,actions ocg={}{reset-b rm2reu}{df2comp df2eow df2exp df2mtb df2pro df2ts bx2comp bx2eow bx2exp bx2mtb bx2pro inc2comp inc2eow inc2exp map2comp map2eow map2exp map2mtb map2reu man2comp man2eow man2exp nav2comp nav2eow nav2exp gpl2comp gpl2eow gpl2exp gpl2mtb gpl2pro gpl2reu gpl2ts trv2comp trv2eow trv2exp trv2pro pm2comp pm2exp pm2pro trc2comp trc2eow trc2exp trc2pro debug2comp debug2ts mat2ts lea2eow lea2ts val2ts use2eow use2pro use2ts ide2eow ide2mtb ide2ts int2ts cre2ts ana2comp ana2pro ana2ts eco2mtb eco2pro eco2ts rep2reu rep2ts awa2ts ler2eow}] (factor-rm) {Reuse Mechanisms};
	\node[factor,below of=factor-rm,actions ocg={}{reset-b trc2comp trc2eow trc2exp trc2pro}{df2comp df2eow df2exp df2mtb df2pro df2ts bx2comp bx2eow bx2exp bx2mtb bx2pro inc2comp inc2eow inc2exp map2comp map2eow map2exp map2mtb map2reu man2comp man2eow man2exp nav2comp nav2eow nav2exp gpl2comp gpl2eow gpl2exp gpl2mtb gpl2pro gpl2reu gpl2ts trv2comp trv2eow trv2exp trv2pro pm2comp pm2exp pm2pro rm2reu debug2comp debug2ts mat2ts lea2eow lea2ts val2ts use2eow use2pro use2ts ide2eow ide2mtb ide2ts int2ts cre2ts ana2comp ana2pro ana2ts eco2mtb eco2pro eco2ts rep2reu rep2ts awa2ts ler2eow}] (factor-trc) {Traceability};
	\node[factor,below of=factor-trc,actions ocg={}{reset-b ler2eow}{df2comp df2eow df2exp df2mtb df2pro df2ts bx2comp bx2eow bx2exp bx2mtb bx2pro inc2comp inc2eow inc2exp map2comp map2eow map2exp map2mtb map2reu man2comp man2eow man2exp nav2comp nav2eow nav2exp gpl2comp gpl2eow gpl2exp gpl2mtb gpl2pro gpl2reu gpl2ts trv2comp trv2eow trv2exp trv2pro pm2comp pm2exp pm2pro rm2reu debug2comp debug2ts mat2ts lea2eow lea2ts val2ts use2eow use2pro use2ts ide2eow ide2mtb ide2ts int2ts cre2ts ana2comp ana2pro ana2ts eco2mtb eco2pro eco2ts rep2reu rep2ts awa2ts trc2comp trc2eow trc2exp trc2pro}] (factor-ler) {Learnability};
	
	\node[factor,actions ocg={}{reset-b debug2comp debug2ts}{df2comp df2eow df2exp df2mtb df2pro df2ts bx2comp bx2eow bx2exp bx2mtb bx2pro inc2comp inc2eow inc2exp map2comp map2eow map2exp map2mtb map2reu man2comp man2eow man2exp nav2comp nav2eow nav2exp gpl2comp gpl2eow gpl2exp gpl2mtb gpl2pro gpl2reu gpl2ts trv2comp trv2eow trv2exp trv2pro pm2comp pm2exp pm2pro rm2reu trc2comp trc2eow trc2exp trc2pro mat2ts lea2eow lea2ts val2ts use2eow use2pro use2ts ide2eow ide2mtb ide2ts int2ts cre2ts ana2comp ana2pro ana2ts eco2mtb eco2pro eco2ts rep2reu rep2ts awa2ts ler2eow}] (factor-debug) at (14.25,0) {Debugging Tooling};
	\node[factor,below of=factor-debug,actions ocg={}{reset-b mat2ts}{df2comp df2eow df2exp df2mtb df2pro df2ts bx2comp bx2eow bx2exp bx2mtb bx2pro inc2comp inc2eow inc2exp map2comp map2eow map2exp map2mtb map2reu man2comp man2eow man2exp nav2comp nav2eow nav2exp gpl2comp gpl2eow gpl2exp gpl2mtb gpl2pro gpl2reu gpl2ts trv2comp trv2eow trv2exp trv2pro pm2comp pm2exp pm2pro rm2reu trc2comp trc2eow trc2exp trc2pro debug2comp debug2ts lea2eow lea2ts val2ts use2eow use2pro use2ts ide2eow ide2mtb ide2ts int2ts cre2ts ana2comp ana2pro ana2ts eco2mtb eco2pro eco2ts rep2reu rep2ts awa2ts ler2eow}] (factor-mat) {Tool Maturity};
	\node[factor,below of=factor-mat,actions ocg={}{reset-b lea2eow lea2ts}{df2comp df2eow df2exp df2mtb df2pro df2ts bx2comp bx2eow bx2exp bx2mtb bx2pro inc2comp inc2eow inc2exp map2comp map2eow map2exp map2mtb map2reu man2comp man2eow man2exp nav2comp nav2eow nav2exp gpl2comp gpl2eow gpl2exp gpl2mtb gpl2pro gpl2reu gpl2ts trv2comp trv2eow trv2exp trv2pro pm2comp pm2exp pm2pro rm2reu trc2comp trc2eow trc2exp trc2pro debug2comp debug2ts mat2ts val2ts use2eow use2pro use2ts ide2eow ide2mtb ide2ts int2ts cre2ts ana2comp ana2pro ana2ts eco2mtb eco2pro eco2ts rep2reu rep2ts awa2ts ler2eow}] (factor-lea) {Tool Learnability};
	\node[factor,below of=factor-lea,actions ocg={}{reset-b val2ts}{df2comp df2eow df2exp df2mtb df2pro df2ts bx2comp bx2eow bx2exp bx2mtb bx2pro inc2comp inc2eow inc2exp map2comp map2eow map2exp map2mtb map2reu man2comp man2eow man2exp nav2comp nav2eow nav2exp gpl2comp gpl2eow gpl2exp gpl2mtb gpl2pro gpl2reu gpl2ts trv2comp trv2eow trv2exp trv2pro pm2comp pm2exp pm2pro rm2reu trc2comp trc2eow trc2exp trc2pro debug2comp debug2ts mat2ts lea2eow lea2ts use2eow use2pro use2ts ide2eow ide2mtb ide2ts int2ts cre2ts ana2comp ana2pro ana2ts eco2mtb eco2pro eco2ts rep2reu rep2ts awa2ts ler2eow}] (factor-val) {Validation Tooling};
	\node[factor,below of=factor-val,actions ocg={}{reset-b use2eow use2pro use2ts}{df2comp df2eow df2exp df2mtb df2pro df2ts bx2comp bx2eow bx2exp bx2mtb bx2pro inc2comp inc2eow inc2exp map2comp map2eow map2exp map2mtb map2reu man2comp man2eow man2exp nav2comp nav2eow nav2exp gpl2comp gpl2eow gpl2exp gpl2mtb gpl2pro gpl2reu gpl2ts trv2comp trv2eow trv2exp trv2pro pm2comp pm2exp pm2pro rm2reu trc2comp trc2eow trc2exp trc2pro debug2comp debug2ts mat2ts lea2eow lea2ts val2ts ide2eow ide2mtb ide2ts int2ts cre2ts ana2comp ana2pro ana2ts eco2mtb eco2pro eco2ts rep2reu rep2ts awa2ts ler2eow}] (factor-use) {Tool Usability};
	\node[factor,below of=factor-use,actions ocg={}{reset-b ide2eow ide2mtb ide2ts}{df2comp df2eow df2exp df2mtb df2pro df2ts bx2comp bx2eow bx2exp bx2mtb bx2pro inc2comp inc2eow inc2exp map2comp map2eow map2exp map2mtb map2reu man2comp man2eow man2exp nav2comp nav2eow nav2exp gpl2comp gpl2eow gpl2exp gpl2mtb gpl2pro gpl2reu gpl2ts trv2comp trv2eow trv2exp trv2pro pm2comp pm2exp pm2pro rm2reu trc2comp trc2eow trc2exp trc2pro debug2comp debug2ts mat2ts lea2eow lea2ts val2ts use2eow use2pro use2ts int2ts cre2ts ana2comp ana2pro ana2ts eco2mtb eco2pro eco2ts rep2reu rep2ts awa2ts ler2eow}] (factor-ide) {IDE Tooling};
	\node[factor,below of=factor-ide,actions ocg={}{reset-b int2ts}{df2comp df2eow df2exp df2mtb df2pro df2ts bx2comp bx2eow bx2exp bx2mtb bx2pro inc2comp inc2eow inc2exp map2comp map2eow map2exp map2mtb map2reu man2comp man2eow man2exp nav2comp nav2eow nav2exp gpl2comp gpl2eow gpl2exp gpl2mtb gpl2pro gpl2reu gpl2ts trv2comp trv2eow trv2exp trv2pro pm2comp pm2exp pm2pro rm2reu trc2comp trc2eow trc2exp trc2pro debug2comp debug2ts mat2ts lea2eow lea2ts val2ts use2eow use2pro use2ts ide2eow ide2mtb ide2ts cre2ts ana2comp ana2pro ana2ts eco2mtb eco2pro eco2ts rep2reu rep2ts awa2ts ler2eow}] (factor-int) {Tool Interoperability};
	\node[factor,below of=factor-int,actions ocg={}{reset-b cre2ts}{df2comp df2eow df2exp df2mtb df2pro df2ts bx2comp bx2eow bx2exp bx2mtb bx2pro inc2comp inc2eow inc2exp map2comp map2eow map2exp map2mtb map2reu man2comp man2eow man2exp nav2comp nav2eow nav2exp gpl2comp gpl2eow gpl2exp gpl2mtb gpl2pro gpl2reu gpl2ts trv2comp trv2eow trv2exp trv2pro pm2comp pm2exp pm2pro rm2reu trc2comp trc2eow trc2exp trc2pro debug2comp debug2ts mat2ts lea2eow lea2ts val2ts use2eow use2pro use2ts ide2eow ide2mtb ide2ts int2ts ana2comp ana2pro ana2ts eco2mtb eco2pro eco2ts rep2reu rep2ts awa2ts ler2eow}] (factor-cre) {Tool Creation Effort};
	\node[factor,below of=factor-cre,actions ocg={}{reset-b ana2comp ana2pro ana2ts}{df2comp df2eow df2exp df2mtb df2pro df2ts bx2comp bx2eow bx2exp bx2mtb bx2pro inc2comp inc2eow inc2exp map2comp map2eow map2exp map2mtb map2reu man2comp man2eow man2exp nav2comp nav2eow nav2exp gpl2comp gpl2eow gpl2exp gpl2mtb gpl2pro gpl2reu gpl2ts trv2comp trv2eow trv2exp trv2pro pm2comp pm2exp pm2pro rm2reu trc2comp trc2eow trc2exp trc2pro debug2comp debug2ts mat2ts lea2eow lea2ts val2ts use2eow use2pro use2ts ide2eow ide2mtb ide2ts int2ts cre2ts eco2mtb eco2pro eco2ts rep2reu rep2ts awa2ts ler2eow}] (factor-ana) {Analysis Tooling};
	\node[factor,below of=factor-ana,actions ocg={}{reset-b eco2mtb eco2prob eco2ts}{df2comp df2eow df2exp df2mtb df2pro df2ts bx2comp bx2eow bx2exp bx2mtb bx2pro inc2comp inc2eow inc2exp map2comp map2eow map2exp map2mtb map2reu man2comp man2eow man2exp nav2comp nav2eow nav2exp gpl2comp gpl2eow gpl2exp gpl2mtb gpl2pro gpl2reu gpl2ts trv2comp trv2eow trv2exp trv2pro pm2comp pm2exp pm2pro rm2reu trc2comp trc2eow trc2exp trc2pro debug2comp debug2ts mat2ts lea2eow lea2ts val2ts use2eow use2pro use2ts ide2eow ide2mtb ide2ts int2ts cre2ts ana2comp ana2pro ana2ts rep2reu rep2ts awa2ts ler2eow}] (factor-eco) {Ecosystem};
	\node[factor,below of=factor-eco,actions ocg={}{reset-b rep2reu rep2ts}{df2comp df2eow df2exp df2mtb df2pro df2ts bx2comp bx2eow bx2exp bx2mtb bx2pro inc2comp inc2eow inc2exp map2comp map2eow map2exp map2mtb map2reu man2comp man2eow man2exp nav2comp nav2eow nav2exp gpl2comp gpl2eow gpl2exp gpl2mtb gpl2pro gpl2reu gpl2ts trv2comp trv2eow trv2exp trv2pro pm2comp pm2exp pm2pro rm2reu trc2comp trc2eow trc2exp trc2pro debug2comp debug2ts mat2ts lea2eow lea2ts val2ts use2eow use2pro use2ts ide2eow ide2mtb ide2ts int2ts cre2ts ana2comp ana2pro ana2ts eco2mtb eco2pro eco2ts awa2ts ler2eow}] (factor-rep) {Code Repositories};
	\node[factor,below of=factor-rep,actions ocg={}{reset-b awa2ts}{df2comp df2eow df2exp df2mtb df2pro df2ts bx2comp bx2eow bx2exp bx2mtb bx2pro inc2comp inc2eow inc2exp map2comp map2eow map2exp map2mtb map2reu man2comp man2eow man2exp nav2comp nav2eow nav2exp gpl2comp gpl2eow gpl2exp gpl2mtb gpl2pro gpl2reu gpl2ts trv2comp trv2eow trv2exp trv2pro pm2comp pm2exp pm2pro rm2reu trc2comp trc2eow trc2exp trc2pro debug2comp debug2ts mat2ts lea2eow lea2ts val2ts use2eow use2pro use2ts ide2eow ide2mtb ide2ts int2ts cre2ts ana2comp ana2pro ana2ts eco2mtb eco2pro eco2ts rep2reu rep2ts ler2eow}] (factor-awa) {Tooling Awareness};
	
	\node[property,actions ocg={}{reset-b gpl2comp dom2comp bx2comp inc2comp map2comp trc2comp trv2comp pm2comp nav2comp man2comp df2comp debug2comp ana2comp}{df2eow df2exp df2mtb df2pro df2ts bx2eow bx2exp bx2mtb bx2pro inc2eow inc2exp map2eow map2exp map2mtb map2reu man2eow man2exp nav2eow nav2exp gpl2eow gpl2exp gpl2mtb gpl2pro gpl2reu gpl2ts trv2eow trv2exp trv2pro pm2exp pm2pro rm2reu trc2eow trc2exp trc2pro debug2ts mat2ts lea2eow lea2ts val2ts use2eow use2pro use2ts ide2eow ide2mtb ide2ts int2ts cre2ts ana2pro ana2ts eco2mtb eco2pro eco2ts rep2reu rep2ts awa2ts ler2eow}] (prop-comp) at (7.5,-0.5) {Comprehensibility};
	\begin{pgfonlayer}{background}
		\node[mod-factor-box,below of=prop-comp] (mod-prop-comp) {\parbox{8em}{\scriptsize\begin{itemize}[label=$\bullet$]
					\item Choice of MTL
					\item Language Skills
					\item I/O Semantic gap
					\item Size
		\end{itemize}}};
	\end{pgfonlayer}
	
	\node[property,below of=prop-comp,actions ocg={}{reset-b df2eow bx2eow inc2eow man2eow nav2eow gpl2eow trv2eow trc2eow lea2eow use2eow ide2eow ler2eow}{df2comp df2exp df2mtb df2pro df2ts bx2comp bx2exp bx2mtb bx2pro inc2comp inc2exp map2comp map2exp map2mtb map2reu man2comp man2exp nav2comp nav2exp gpl2comp gpl2exp gpl2mtb gpl2pro gpl2reu gpl2ts trv2comp trv2exp trv2pro pm2comp pm2exp pm2pro rm2reu trc2comp trc2exp trc2pro debug2comp debug2ts mat2ts lea2ts val2ts use2pro use2ts ide2mtb ide2ts int2ts cre2ts ana2comp ana2pro ana2ts eco2mtb eco2pro eco2ts rep2reu rep2ts awa2ts}] (prop-eow) {Ease of Writing};
	\begin{pgfonlayer}{background}
		\node[mod-factor-box,below of=prop-eow] (mod-prop-eow) {\parbox{8em}{\scriptsize\begin{itemize}[label=$\bullet$,topsep=0pt]
					\item Choice of MTL
					\item Language Skills
					\item User Experience
					\item (Meta-) Models
					\item I/O Semantic gap
					\item Size
		\end{itemize}}};
	\end{pgfonlayer}
	
	\node[property,below of=prop-eow,actions ocg={}{reset-b df2exp bx2exp inc2exp map2exp man2exp nav2exp gpl2exp trv2exp pm2exp trc2exp}{df2comp df2eow df2mtb df2pro df2ts bx2comp bx2eow bx2mtb bx2pro inc2comp inc2eow map2comp map2eow map2mtb map2reu man2comp man2eow nav2comp nav2eow gpl2comp gpl2eow gpl2mtb gpl2pro gpl2reu gpl2ts trv2comp trv2eow trv2pro pm2comp pm2pro rm2reu trc2comp trc2eow trc2pro debug2comp debug2ts mat2ts lea2eow lea2ts val2ts use2eow use2pro use2ts ide2eow ide2mtb ide2ts int2ts cre2ts ana2comp ana2pro ana2ts eco2mtb eco2pro eco2ts rep2reu rep2ts awa2ts ler2eow}] (prop-ex) {Expressiveness};
	\begin{pgfonlayer}{background}
	\node[mod-factor-box,below of=prop-ex] (mod-prop-ex) {\parbox{8em}{\scriptsize\begin{itemize}[label=$\bullet$,itemsep=0.25em,topsep=0.25em]
				\item Choice of MTL
	\end{itemize}}};
	\end{pgfonlayer}
	
	\node[property,below of=prop-ex,actions ocg={}{reset-b df2ts gpl2ts debug2ts mat2ts lea2ts val2ts use2ts ide2ts int2ts cre2ts ana2ts eco2ts rep2ts awa2ts}{df2comp df2eow df2exp df2mtb df2pro bx2comp bx2eow bx2exp bx2mtb bx2pro inc2comp inc2eow inc2exp map2comp map2eow map2exp map2mtb map2reu man2comp man2eow man2exp nav2comp nav2eow nav2exp gpl2comp gpl2eow gpl2exp gpl2mtb gpl2pro gpl2reu trv2comp trv2eow trv2exp trv2pro pm2comp pm2exp pm2pro rm2reu trc2comp trc2eow trc2exp trc2pro debug2comp lea2eow use2eow use2pro ide2eow ide2mtb ana2comp ana2pro eco2mtb eco2pro rep2reu ler2eow}] (prop-ts) {Tool Support};
	\begin{pgfonlayer}{background}
	\node[mod-factor-box,below of=prop-ts] (mod-prop-ts) {\parbox{8em}{\scriptsize\begin{itemize}[label=$\bullet$,topsep=0pt]
				\item Choice of MTL
	\end{itemize}}};
	\end{pgfonlayer}
	
	\node[property,below of=prop-ts,actions ocg={}{reset-b df2mtb bx2mtb map2mtb gpl2mtb ide2mtb eco2mtb}{df2comp df2eow df2exp df2pro df2ts bx2comp bx2eow bx2exp bx2pro inc2comp inc2eow inc2exp map2comp map2eow map2exp map2reu man2comp man2eow man2exp nav2comp nav2eow nav2exp gpl2comp gpl2eow gpl2exp gpl2pro gpl2reu gpl2ts trv2comp trv2eow trv2exp trv2pro pm2comp pm2exp pm2pro rm2reu trc2comp trc2eow trc2exp trc2pro debug2comp debug2ts mat2ts lea2eow lea2ts val2ts use2eow use2pro use2ts ide2eow ide2ts int2ts cre2ts ana2comp ana2pro ana2ts eco2pro eco2ts rep2reu rep2ts awa2ts ler2eow}] (prop-mtb) {Maintainability};
	\begin{pgfonlayer}{background}
		\node[mod-factor-box,below of=prop-mtb] (mod-prop-mtb) {\parbox{8em}{\scriptsize\begin{itemize}[label=$\bullet$]
					\item Choice of MTL
					\item Language Skills
					\item User Experience
		\end{itemize}}};
	\end{pgfonlayer}
	
	\node[property,below of=prop-mtb,actions ocg={}{reset-b df2pro bx2pro gpl2pro trv2pro pm2pro trc2pro use2pro ana2pro eco2pro}{df2comp df2eow df2exp df2mtb df2ts bx2comp bx2eow bx2exp bx2mtb inc2comp inc2eow inc2exp map2comp map2eow map2exp map2mtb map2reu man2comp man2eow man2exp nav2comp nav2eow nav2exp gpl2comp gpl2eow gpl2exp gpl2mtb gpl2reu gpl2ts trv2comp trv2eow trv2exp pm2comp pm2exp rm2reu trc2comp trc2eow trc2exp debug2comp debug2ts mat2ts lea2eow lea2ts val2ts use2eow use2ts ide2eow ide2mtb ide2ts int2ts cre2ts ana2comp ana2ts eco2mtb eco2ts rep2reu rep2ts awa2ts ler2eow}] (prop-pro) {Productivity};
	\begin{pgfonlayer}{background}
		\node[mod-factor-box,below of=prop-pro] (mod-prop-pro) {\parbox{8em}{\scriptsize\begin{itemize}[label=$\bullet$]
					\item Choice of MTL
					\item Language Skills
					\item I/O Semantic gap
		\end{itemize}}};
	\end{pgfonlayer}
	
	\node[property,below of=prop-pro,actions ocg={}{reset-b map2reu gpl2reu rm2reu rep2reu}{df2comp df2eow df2exp df2mtb df2pro df2ts bx2comp bx2eow bx2exp bx2mtb bx2pro inc2comp inc2eow inc2exp map2comp map2eow map2exp map2mtb man2comp man2eow man2exp nav2comp nav2eow nav2exp gpl2comp gpl2eow gpl2exp gpl2mtb gpl2pro gpl2ts trv2comp trv2eow trv2exp trv2pro pm2comp pm2exp pm2pro trc2comp trc2eow trc2exp trc2pro debug2comp debug2ts mat2ts lea2eow lea2ts val2ts use2eow use2pro use2ts ide2eow ide2mtb ide2ts int2ts cre2ts ana2comp ana2pro ana2ts eco2mtb eco2pro eco2ts rep2ts awa2ts ler2eow}] (prop-reu) {Reusability};
	\begin{pgfonlayer}{background}
		\node[mod-factor-box,below of=prop-reu] (mod-prop-reu) {\parbox{8em}{\scriptsize\begin{itemize}[label=$\bullet$]
					\item Choice of MTL
					\item Language Skills
		\end{itemize}}};
	\end{pgfonlayer}
	
	\begin{scope}[ocg={name=egal,ref=df2comp,status=visible}]
	\path[l] (factor-df.east) -- (prop-comp.west);
	\end{scope}
	\begin{scope}[ocg={name=egal,ref=df2eow,status=visible}]
	\path[l] (factor-df.east) -- (prop-eow.west);
	\end{scope}
	\begin{scope}[ocg={name=egal,ref=df2exp,status=visible}]
	\path[l] (factor-df.east) -- (prop-ex.west);
	\end{scope}
	\begin{scope}[ocg={name=egal,ref=df2mtb,status=visible}]
	\path[l] (factor-df.east) -- (prop-mtb.west);
	\end{scope}
	\begin{scope}[ocg={name=egal,ref=df2pro,status=visible}]
	\path[l] (factor-df.east) -- (prop-pro.west);
	\end{scope}
	\begin{scope}[ocg={name=egal,ref=df2ts,status=visible}]
	\path[l] (factor-df.east) -- (prop-ts.west);
	\end{scope}
	
	\begin{scope}[ocg={name=egal,ref=bx2comp,status=visible}]
	\path[l] (factor-bx.east) -- (prop-comp.west);
	\end{scope}
	\begin{scope}[ocg={name=egal,ref=bx2eow,status=visible}]
	\path[l] (factor-bx.east) -- (prop-eow.west);
	\end{scope}
	\begin{scope}[ocg={name=egal,ref=bx2exp,status=visible}]
	\path[l] (factor-bx.east) -- (prop-ex.west);
	\end{scope}
	\begin{scope}[ocg={name=egal,ref=bx2mtb,status=visible}]
	\path[l] (factor-bx.east) -- (prop-mtb.west);
	\end{scope}
	\begin{scope}[ocg={name=egal,ref=bx2pro,status=visible}]
	\path[l] (factor-bx.east) -- (prop-pro.west);
	\end{scope}

	\begin{scope}[ocg={name=egal,ref=inc2comp,status=visible}]
	\path[l] (factor-inc.east) -- (prop-comp.west);
	\end{scope}
	\begin{scope}[ocg={name=egal,ref=inc2eow,status=visible}]
	\path[l] (factor-inc.east) -- (prop-eow.west);
	\end{scope}
	\begin{scope}[ocg={name=egal,ref=inc2exp,status=visible}]
	\path[l] (factor-inc.east) -- (prop-ex.west);
	\end{scope}

	\begin{scope}[ocg={name=egal,ref=map2comp,status=visible}]
	\path[l] (factor-map.east) -- (prop-comp.west);
	\end{scope}
	\begin{scope}[ocg={name=egal,ref=map2eow,status=visible}]
	\path[l] (factor-map.east) -- (prop-eow.west);
	\end{scope}
	\begin{scope}[ocg={name=egal,ref=map2exp,status=visible}]
	\path[l] (factor-map.east) -- (prop-ex.west);
	\end{scope}
	\begin{scope}[ocg={name=egal,ref=map2mtb,status=visible}]
	\path[l] (factor-map.east) -- (prop-mtb.west);
	\end{scope}
	\begin{scope}[ocg={name=egal,ref=map2reu,status=visible}]
	\path[l] (factor-map.east) -- (prop-reu.west);
	\end{scope}

	\begin{scope}[ocg={name=egal,ref=man2comp,status=visible}]
	\path[l] (factor-man.east) -- (prop-comp.west);
	\end{scope}
	\begin{scope}[ocg={name=egal,ref=man2eow,status=visible}]
	\path[l] (factor-man.east) -- (prop-eow.west);
	\end{scope}
	\begin{scope}[ocg={name=egal,ref=man2exp,status=visible}]
	\path[l] (factor-man.east) -- (prop-ex.west);
	\end{scope}

	\begin{scope}[ocg={name=egal,ref=nav2comp,status=visible}]
	\path[l] (factor-nav.east) -- (prop-comp.west);
	\end{scope}
	\begin{scope}[ocg={name=egal,ref=nav2eow,status=visible}]
	\path[l] (factor-nav.east) -- (prop-eow.west);
	\end{scope}
	\begin{scope}[ocg={name=egal,ref=nav2exp,status=visible}]
	\path[l] (factor-nav.east) -- (prop-ex.west);
	\end{scope}

	\begin{scope}[ocg={name=egal,ref=gpl2comp,status=visible}]
	\path[l] (factor-gpl.east) -- (prop-comp.west);
	\end{scope}
	\begin{scope}[ocg={name=egal,ref=gpl2eow,status=visible}]
	\path[l] (factor-gpl.east) -- (prop-eow.west);
	\end{scope}
	\begin{scope}[ocg={name=egal,ref=gpl2exp,status=visible}]
	\path[l] (factor-gpl.east) -- (prop-ex.west);
	\end{scope}
	\begin{scope}[ocg={name=egal,ref=gpl2mtb,status=visible}]
	\path[l] (factor-gpl.east) -- (prop-mtb.west);
	\end{scope}
	\begin{scope}[ocg={name=egal,ref=gpl2pro,status=visible}]
	\path[l] (factor-gpl.east) -- (prop-pro.west);
	\end{scope}
	\begin{scope}[ocg={name=egal,ref=gpl2reu,status=visible}]
	\path[l] (factor-gpl.east) -- (prop-reu.west);
	\end{scope}
	\begin{scope}[ocg={name=egal,ref=gpl2ts,status=visible}]
	\path[l] (factor-gpl.east) -- (prop-ts.west);
	\end{scope}

	\begin{scope}[ocg={name=egal,ref=trv2comp,status=visible}]
	\path[l] (factor-trv.east) -- (prop-comp.west);
	\end{scope}
	\begin{scope}[ocg={name=egal,ref=trv2eow,status=visible}]
	\path[l] (factor-trv.east) -- (prop-eow.west);
	\end{scope}
	\begin{scope}[ocg={name=egal,ref=trv2exp,status=visible}]
	\path[l] (factor-trv.east) -- (prop-ex.west);
	\end{scope}
	\begin{scope}[ocg={name=egal,ref=trv2pro,status=visible}]
	\path[l] (factor-trv.east) -- (prop-pro.west);
	\end{scope}

	\begin{scope}[ocg={name=egal,ref=pm2comp,status=visible}]
	\path[l] (factor-pm.east) -- (prop-comp.west);
	\end{scope}
	\begin{scope}[ocg={name=egal,ref=pm2exp,status=visible}]
	\path[l] (factor-pm.east) -- (prop-ex.west);
	\end{scope}
	\begin{scope}[ocg={name=egal,ref=pm2pro,status=visible}]
	\path[l] (factor-pm.east) -- (prop-pro.west);
	\end{scope}

	\begin{scope}[ocg={name=egal,ref=rm2reu,status=visible}]
		\path[l] (factor-rm.east) -- (prop-reu.west);
	\end{scope}

	\begin{scope}[ocg={name=egal,ref=trc2comp,status=visible}]
		\path[l] (factor-trc.east) -- (prop-comp.west);
	\end{scope}
	\begin{scope}[ocg={name=egal,ref=trc2eow,status=visible}]
		\path[l] (factor-trc.east) -- (prop-eow.west);
	\end{scope}
	\begin{scope}[ocg={name=egal,ref=trc2exp,status=visible}]
		\path[l] (factor-trc.east) -- (prop-ex.west);
	\end{scope}
	\begin{scope}[ocg={name=egal,ref=trc2pro,status=visible}]
		\path[l] (factor-trc.east) -- (prop-pro.west);
	\end{scope}

	\begin{scope}[ocg={name=egal,ref=ler2eow,status=visible}]
		\path[l] (factor-ler.east) -- (prop-eow.west);
	\end{scope}

	\begin{scope}[ocg={name=egal,ref=debug2comp,status=visible}]
	\path[l] (factor-debug.west) -- (prop-comp.east);
	\end{scope}
	\begin{scope}[ocg={name=egal,ref=debug2ts,status=visible}]
	\path[l] (factor-debug.west) -- (prop-ts.east);
	\end{scope}

	\begin{scope}[ocg={name=egal,ref=mat2ts,status=visible}]
	\path[l] (factor-mat.west) -- (prop-ts.east);
	\end{scope}

	\begin{scope}[ocg={name=egal,ref=lea2eow,status=visible}]
	\path[l] (factor-lea.west) -- (prop-eow.east);
	\end{scope}
	\begin{scope}[ocg={name=egal,ref=lea2ts,status=visible}]
	\path[l] (factor-lea.west) -- (prop-ts.east);
	\end{scope}

	\begin{scope}[ocg={name=egal,ref=val2ts,status=visible}]
	\path[l] (factor-val.west) -- (prop-ts.east);
	\end{scope}

	\begin{scope}[ocg={name=egal,ref=use2eow,status=visible}]
	\path[l] (factor-use.west) -- (prop-eow.east);
	\end{scope}
	\begin{scope}[ocg={name=egal,ref=use2pro,status=visible}]
	\path[l] (factor-use.west) -- (prop-pro.east);
	\end{scope}
	\begin{scope}[ocg={name=egal,ref=use2ts,status=visible}]
	\path[l] (factor-use.west) -- (prop-ts.east);
	\end{scope}

	\begin{scope}[ocg={name=egal,ref=ide2eow,status=visible}]
	\path[l] (factor-ide.west) -- (prop-eow.east);
	\end{scope}
	\begin{scope}[ocg={name=egal,ref=ide2mtb,status=visible}]
	\path[l] (factor-ide.west) -- (prop-mtb.east);
	\end{scope}
	\begin{scope}[ocg={name=egal,ref=ide2ts,status=visible}]
	\path[l] (factor-ide.west) -- (prop-ts.east);
	\end{scope}

	\begin{scope}[ocg={name=egal,ref=int2ts,status=visible}]
	\path[l] (factor-int.west) -- (prop-ts.east);
	\end{scope}

	\begin{scope}[ocg={name=egal,ref=cre2ts,status=visible}]
	\path[l] (factor-cre.west) -- (prop-ts.east);
	\end{scope}
	
	\begin{scope}[ocg={name=egal,ref=ana2comp,status=visible}]
	\path[l] (factor-ana.west) -- (prop-comp.east);
	\end{scope}
	\begin{scope}[ocg={name=egal,ref=ana2pro,status=visible}]
	\path[l] (factor-ana.west) -- (prop-pro.east);
	\end{scope}
	\begin{scope}[ocg={name=egal,ref=ana2ts,status=visible}]
	\path[l] (factor-ana.west) -- (prop-ts.east);
	\end{scope}

	\begin{scope}[ocg={name=egal,ref=eco2mtb,status=visible}]
	\path[l] (factor-eco.west) -- (prop-mtb.east);
	\end{scope}
	\begin{scope}[ocg={name=egal,ref=eco2pro,status=visible}]
	\path[l] (factor-eco.west) -- (prop-pro.east);
	\end{scope}
	\begin{scope}[ocg={name=egal,ref=eco2ts,status=visible}]
	\path[l] (factor-eco.west) -- (prop-ts.east);
	\end{scope}

	\begin{scope}[ocg={name=egal,ref=rep2reu,status=visible}]
	\path[l] (factor-rep.west) -- (prop-reu.east);
	\end{scope}
	\begin{scope}[ocg={name=egal,ref=rep2ts,status=visible}]
	\path[l] (factor-rep.west) -- (prop-ts.east);
	\end{scope}

	\begin{scope}[ocg={name=egal,ref=awa2ts,status=visible}]
	\path[l] (factor-awa.west) -- (prop-ts.east);
	\end{scope}

	\begin{scope}[ocg={name=egal,ref=reset-b,status=invisible}]
	\node[rectangle,draw,text=red,xshift=3em,yshift=-3em,right of=factor-ler,actions ocg={}{df2comp df2eow df2exp df2mtb df2pro df2ts bx2comp bx2eow bx2exp bx2mtb bx2pro inc2comp inc2eow inc2exp map2comp map2eow map2exp map2mtb map2reu man2comp man2eow man2exp nav2comp nav2eow nav2exp gpl2comp gpl2eow gpl2exp gpl2mtb gpl2pro gpl2reu gpl2ts trv2comp trv2eow trv2exp trv2pro pm2comp pm2exp pm2pro rm2reu trc2comp trc2eow trc2exp trc2pro debug2comp debug2ts mat2ts lea2eow lea2ts val2ts use2eow use2pro use2ts ide2eow ide2mtb ide2ts int2ts cre2ts ana2comp ana2pro ana2ts eco2mtb eco2pro eco2ts rep2reu rep2ts awa2ts ler2eow}{reset-b}] (reset-b) {Reset};
	\end{scope}
\end{tikzpicture}
	\caption{Structure model of influence and moderation effects of factors on MTL properties {\footnotesize (Due to its size, the model has been made interactive using standard PDF features. Clicking a factor will show only influences of the factor. Clicking a MTL property will show only influences on the property. This view can be reset by using the reset button that appears when using the interactive features. To use the interactive features please open the PDF in \textit{Adobe Reader} or \textit{okular}. Other PDF viewers might work but have not been tested.)}}
	\label{fig:structure_model}
\end{figure*}
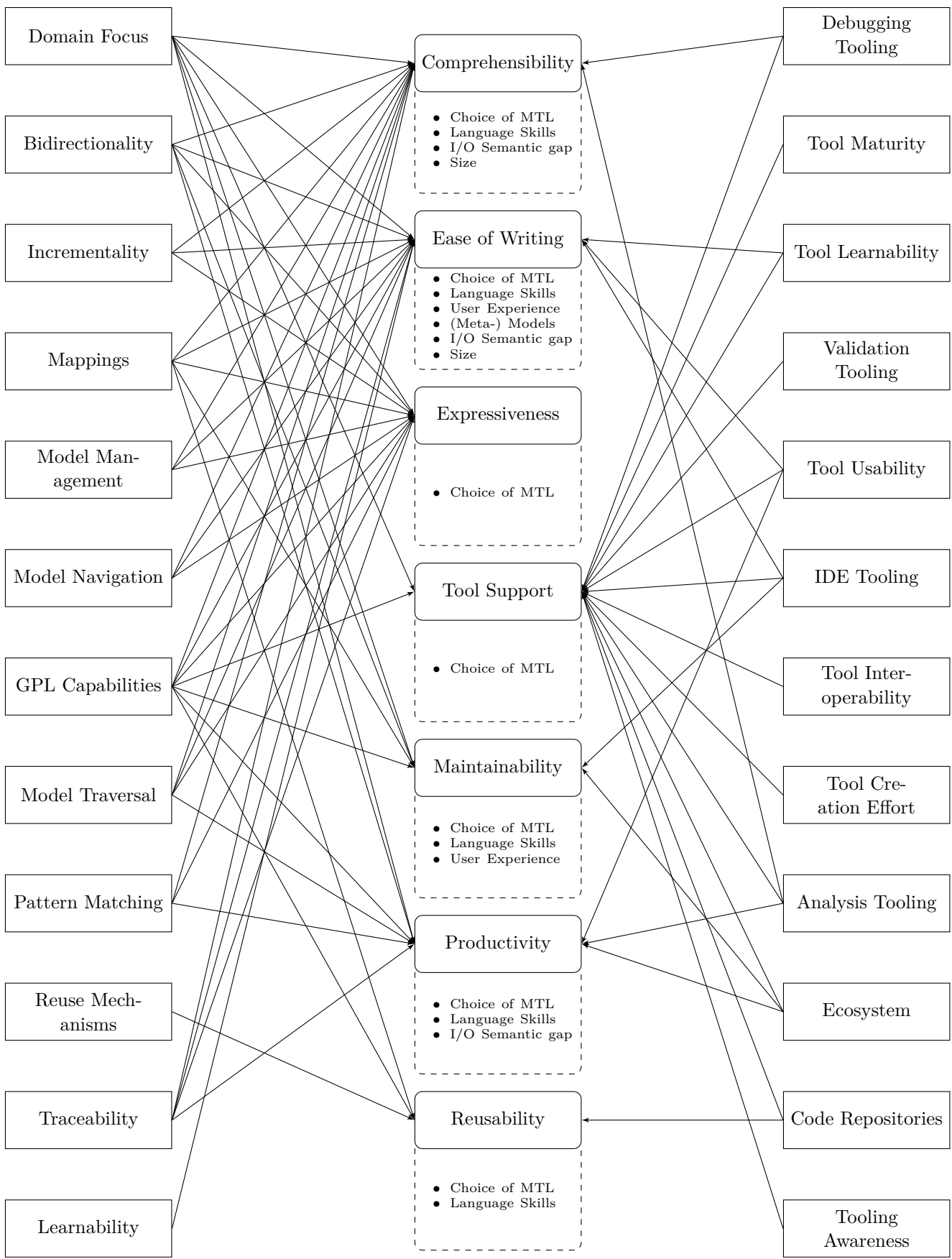

\subsection{The Effects of MTL Capabilities}
\label{sec:findings:effects_mtl_caps}

Capabilities of model transformation languages that go beyond what general purpose languages can offer, are regarded as opportunities for better support in development of transformations.
The advantage often boils down to not having to manually implement the functionality in question when it is required.
It also helps reduce clutter in transformation code, putting the mapping of input and output at the centre of attention.
Moreover, they aid developers in handling problems specific to the transformation domain, such as synchronisations and the relationship of input and output values.

This does however come with its own set of limitations.
Model transformation languages favour a different way of problem solving that is well suited to the problem at hand, but is unfamiliar for the common programmer.
This is amplified by an education that is heavily focused on imperative programming and lacks deeper exposure to logical and functional programming.
Knowledge and understanding of functional concepts would help developers when using query languages such as OCL, while logical concepts often find application in graph based transformation languages.
The domain specific mechanisms in model transformation languages also make generalisations harder.
This is highlighted in the discussions regarding reusability.
Interviewees commonly referred to transformations as conceptionally hard to reuse because of their specificity that makes them applicable only to the use case for which they were developed.

\subsection{Tooling Impact on Properties other than Tool Support}
\label{sec:findings:tooling}

Tooling, or the lack thereof, is a main factor that influences how people perceive the quality and availability of usable model transformation languages.
However, our interviews show that tooling also facilitates many other properties.
This is because tools are not developed as an end in themselves.
Tools are intended to support developers in their efforts to develop and maintain their code.

As a result, the quality of available tools is a major factor that impacts all aspects of a MTL.
\textit{``Basically all the good aids you see in a Java environment should be there even better in a MTL tool, because model transformation is so much more abstract and more relevant that you should be having tools that are again more abstract and more relevant.''} (\textbf{P28}).
Problems in the area of \textit{Usability}, \textit{Maturity} and \textit{Interoperability} of tools have also been reported on in empirical studies on MDE in general~\parencite{Whittle2013,Mohagheghi2013}.

Herein also lies the biggest problem for model transformation languages.
The quality of tools is inadequate.
While there do exist good and useable tools, they are far and between, only exist for certain languages and are not integrated with each other.
This greatly diminishes the potential of model transformation languages because, compared to general purpose languages, developing with them can often be scattered over multiple separate workflows and tools.
There do exist many tools, but most of them are prototypical in nature and only available for individual languages.
This makes it hard to fully utilise the capabilities of a MTL when suitable tools only exist in theory.

The lack of good tools can be attributed mainly to the amount of work required to develop them and the comparatively small community.
Moreover, there are no large commercial vendors, that could put in the required resources to develop tools of a commercially viable quality.

\subsection{The Importance of Moderating Factors}

The saying \textit{``Use the right tool for the job''} also applies to the context of model transformations.
One of the most important things to note is that depending on the context, such as \textit{Use Case} and \textit{developer Skills}, the right language to use can differ greatly.
This was highlighted time and time again in our interviews.

Interviewees insisted that the combination of use case and the concrete implementation of a language feature significantly change how well a feature supports properties such as \textit{Comprehensibility} or \textit{Productivity}, i.e. the influence of factors is moderated by `contextual' factors.
For one, the implementation of a feature in a language might not fit well for the problem that needs to be solved.
Or the feature is not required at all and thus could impose effort on developers that is seen as unnecessary.
In our opinion, this stems from the use cases language developers intended the language for.
For example, a language such as Henshin is intended for cases where patterns of model elements need to be matched and manipulated.
In such cases, the features provided by Henshin can bring significant advantages over implementing the intended transformation in general purpose languages.
Other use cases, where these features are not required, bring no advantage.
They can even have negative effects as the language design around them might hinder users from developing a straight forward solution.

The skills and background knowledge of users is relevant, as it can greatly influence how comfortable people are in using a language. 
This in turn reflects on how well they can perform.
This is problematic for the adoption of model transformation languages, as programmers tend to be trained in imperative, general purpose languages.
As a result, a gentle learning curve is essential and the initial costs of learning need to bear fruit in adequate time.
The choice of using an MTL is therefore a long term investment that is not necessarily suited for only a single project.

The considerations around \textit{Use Case}, \textit{Skills} and the \textit{Choice of MTL} are not novel, but they are rarely discussed explicitly.
This is concerning because almost any decision process will come back to these three factors and their sub-factors, as seen in the fact that the influence on each MTL property in \Cref{fig:structure_model} is moderated by at least one of them.
They provide organisational considerations that come into play before transformation development begins.
Moreover, organisational concerns have already been identified as relevant factors for general MDE adoption~\parencite{Whittle2013,Hutchinson2011a,Hutchinson2011}.
As such they have to be at the centre of attention of researchers and language developers too.

\section{Actionable Results}
\label{sec:discussion}

\newcounter{actionCounter}

\makeatletter
\newcommand\action[1]{\stepcounter{actionCounter}\textbf{(\alph{actionCounter}) #1}}
\makeatother

In this section we present and discuss actionable results that arise from the responses made by our interviewees and analysis thereof.
Results will largely focus on actions that can be taken by researchers, because they make up the largest portion of our interview participants.

\subsection{Evaluation and Development of MTL Capabilities}

Our interviewees mentioned a large number of model transformation language capabilities and reasoned about their implications for the investigated properties of MTL.
We believe the detailed results of the interviews can form a basis for further research into two key aspects:

\begin{enumerate}[label=(\Roman*)]
	\item backing up the expert opinions with empirical data
	\item improving existing model transformation languages
\end{enumerate}

\subsubsection{Evaluation of MTL Capabilities and Properties}
\label{sec:discussion:eval}

In our interviews, experts voiced many opinions on how and why factors influence the various MTL aspects examined in our study.
The opinions were always based on personal experiences, experiences of colleagues and reasoning.
We therefore believe, that our results provide a good insight into the communities sentiment and show that there exists consensus between the experts in many aspects.
Model transformation language capabilities are considered largely beneficial, except for certain edge cases.
However, empirical data to support this consensus is still missing.
The lack of empirical studies into the topic of model transformation languages has already been highlighted in our preceding study \parencite{Goetz2020}.

We are firmly convinced that researchers within the community need to carry out extensive empirical studies, to back up the expert opinions and to explore the exact limits that interviewees hinted at.

\radded{\ref{review-study-types}}{
We envision two main types of studies, experiments and case studies.
Setting up experiments that consider real-world examples with a large number of suitable participants would be optimal but is hard to achieve.
Introducing large transformation examples in experiments is very time-consuming and requires that all participants are experts in the languages used.
In addition, recruiting appropriate participants is a generally difficult in software engineering studies~\parencite{wholin2021credible}.
We recommend using existing transformations as the object of study in experiments instead.
This enables the analysis of complex systems to generate quantitative data without involving human subjects.

If the assessments and experiences of developers are to be the central object of study, we recommend to set up case studies.
This allows researchers to study effects in complex, real-world settings over a longer period of time.
This is important because the exact effects of, for example, the use of a certain language feature often only become apparent to developers after a long period of use.
Case studies of research projects or even industrial transformation systems can thus be used to obtain detailed information on the impact of the applied technologies. 
}

\modified{To design such studies, our results can form an important basis.}

\action{Empirical factor evaluation}.
How and under which circumstances the factors we have identified affect MTL properties needs to be comprehensively evaluated.
Here we envision both qualitative and quantitative studies that focus on the impact of a single factor or a group of related factors.
These could, for example, make comparisons between cases where a factor does or does not apply.
The results of such studies can help language developers make decisions about features to include in their MTLs.

Our interviews provide extensive context that should be taken into account in the study design and interpretation of results.
For example, our interviews show that the semantic gap between input and output defines a relevant context that needs to be considered.
For this reason, when investigating the advantages and disadvantages of mappings, transformations involving models with different levels of semantic gaps between input and output have to be used, to be able to fully evaluate all relevant use cases.
Some transformations need to contain complicated selection conditions or complex calculations for attributes while others need to have less complicated expressions.
Researchers can then evaluate how well mappings in a language fit the different scenarios to aid in providing a clear picture of their advantages and disadvantages.

\action{Empirical MTL property evaluation}.
What advantages or disadvantages MTLs really have is still up for debate.
We believe that the credibility of research efforts on MTL can be greatly improved with studies that provide empirical substantiation to the speculated properties.
\modified{Advances like those made by \textcite{Hebig2018} are rare and further ones, based on real world examples, must be carried out.}

Our results can also make a valuable contribution to such studies.
The factors we have identified as influencing a property can be taken into account in studies from the outset.
They can be used to formulate null hypotheses on why a MTL is superior or inferior to a GPL when considering one specific property.

For example, a study that is interested in investigating the \textit{Comprehensibility} of MTLs compared to GPLs can find a number of factors in our results that need to be taken into account.
Such factors include tracing mechanisms, mappings or pattern matching capabilities.
Researchers can consciously decide which of them are relevant for the transformations used in the study and what impact their presence or absence has on the study results.
Based on these considerations hypotheses can be formed.

\rmodified{\ref{review-unclear}}{A recent study we conducted provides an example of how these considerations can be used to expand the body of empirical studies on this topic~\parencite{Hoeppner2021}
By focusing the investigation on \textit{Mappings}, \textit{Model Navigation} and \textit{Tracing} we were able to present clear and focused results for comparing and explaining differences in the expressiveness of transformation code written in ATL and Java.
We concentrated our analysis on these factors because they all influence Expressiveness according to our interviews.}

Such considerations should of course be part of any proper study, but our results provide a basis that can be useful in ensuring that no relevant factors are overlooked.

\action{Influence Quantification}.
Lastly, the results of this study should be quantified.
The design of the reported study makes quantification of the importance of factors and their influence strengths impossible.
However, such quantification is necessary to prioritise which factors to focus on first, both for assessment and for improvement.
We intend to design and execute such a study as future work to this study.

We can use structural equation modelling methods~\parencite{Weiber2021} to quantify the factors and their influences because we already have a structure model.
We plan to use an online survey to query users of MTLs from research and industry about the amount they use different language features, their perception of qualitative properties of their transformations and demographic data surrounding use-case, skills \& experience and used languages.
The responses are used as input for universal structure modelling (USM)~\parencite{buckler2008identifying} based on the structural equation model developed from the interview responses.

USM is used to estimate the influence and moderation weights of all variables within the structure model.
We can therefore use it to produce quantified data on the influence and moderation effects of identified factors.

We are confident that the approach of using a survey to quantify interview results, can complement the current results, because several of the authors have had positive experiences applying it~\parencite{Liebel2018,Juhnke2020}.

\subsubsection{Improving MTL Capabilities}

To improve current model transformation languages the criticisms articulated by interviewees can be used as starting points for enhancements and innovation.
There are several aspects that are considered to be problematic by our interviewees.

\action{Improve reuse mechanism adoption}.
Reuse mechanisms in model transformation languages are one aspect where interviewees saw potential for improvement (see \Cref{sec:results:mtl_caps:reuse}).
Languages that do not currently possess mature reuse mechanisms can adopt them to become more usable.
For the adoption of mature reuse mechanisms in MTLs we see the languages developers as responsible.

\action{Reuse mechanism innovation}.
Innovation towards transformation specific reuse mechanisms, as has been requested by some participants (see \Cref{sec:results:mtl_caps:reuse}), should also be advanced.
This topic was discussed at length during the interviews on the statement \textit{``Having written several transformations, we have identified that current MTLs are too low a level of abstraction for succinctly expressing transformations between DSLs, because they demonstrate several recurring patterns that have to be reimplemented each time.''} in Question Set 3.

Interviewees pointed out a need for reuse mechanisms that allow transformations to adapt to differing inputs and outputs.
It would be conceivable to define transformation rules, or parts of them, independently of concrete model types, similar to generics in GPLs.
This would allow development of generic transformation `templates' of common transformation patterns.
One pattern, for example, could be finding and manipulating specific model structures, like cliques, independent of the \textit{concrete} model elements involved.
Such templates can then be reused and adapted in all transformations where the pattern is required.

We believe, that innovating such new transformation specific reuse mechanisms is a community wide effort that needs to be taken on in order to make them more widely usable.

\action{Improving MDSE education}.
The \textit{\modified{L}earnability} of MTLs has also been a point of criticism.
We believe, that more effort needs to be put into the transfer of knowledge for MDSE and its techniques like model transformations and MTLs.
This believe is supported by the findings of \textcite{Hutchinson2011}.
They also identified the lack of MDSE knowledge as a limiting factor for the adoption of the approach.

People need to come into contact with the principles earlier so that the inhibition threshold to apply them is lower.
This was also remarked by interviewees when discussing the \textit{Learnability} (see \Cref{sec:results:mtl_caps:learn}).
More focus needs to be given to modelling and modelling techniques in software engineering courses.
This is especially important since the skill of users has been said to be a largely impactful factor upon which many of the advantages from other MTL capabilities rely.
Furthermore, there exist studies such as the one by \textcite{Dieste2017}, which detected a connection between the experience of developers with a language and their productivity as well as the code quality of the resulting programs.

To achieve this, we believe, that the researchers from the community, in their role as higher education teachers and university staff, need to become active.
They should advocate for teaching the concepts of MDSE and the advantages/disadvantages in undergraduate studies in computer science study programmes.
This view is shared by \textcite{Samiee2018}.
Particularly, it should be taught that models can be used for more than documentation purposes, e.g., code generation, simulations early in the development cycle, test case generation.
These other uses are widely and successfully employed in the domain of cyber-physical systems according to \textcite{DBLP:journals/software/BucchiaroneCLPT21}.
Hence, it might be beneficial to include industrial modelling tools like Matlab/Simulink/Stateflow from this domain in addition to standard UML tools in undergraduate courses.
Furthermore, we successfully used simulation frameworks for autonomous cars, like Carla (see \textcite{Dosovitskiy17}), in the past as targets for student projects when teaching courses on the development of modeling languages and model transformations. For example, the students devised a state machine language and code generator targeting the simulation framework to develop an automatic parking functionality. Model transformations were developed to flatten hierarchical state machines to non-hierarchical state machines prior to code generation.

\action{Increase knowledge retention}.
It is also difficult to get to grips with the subject matter in general, as information on it is much harder to obtain than on general purpose programming (see \Cref{sec:results:tooling:aware}).
This starts with the fact that, we found websites on MTLs to often be outdated or unappealing and lack good tutorials and comprehensible documentation.
These points need to be fixed, by the language developers, to provide potential users with better resources to combat the perceived steepness of the learning curve.
More active community involvement is also conceivable here.
Users of MTLs could invest time in creating documentation and keeping it up-to-date.
The possibility of this working and producing good results can be seen in examples such as the arch-linux wiki\footnote{\url{wiki.archlinux.org}}.

\action{Improve community outwards presentation}.
The model transformation community is small.
In our opinion this leads to less innovation and poses the danger of entrenched practices.
The problem is not limited to small communities as seen by, for example, the risk averse movie industry or low innovation automotive industry.
An improved outwards presentation of the technology of model transformations can help alleviate the problem of limited human resources.
The current hype surrounding low-code-platforms can be used to inspire young and aspiring researchers to contribute to its underlying concepts such as model transformations.

\action{Improve industry outreach and cooperation}.
We think it is also paramount to pursue industry cooperation to gauge industrial needs in order to facilitate more industrial adoption of MTLs.
Here ambitious studies are required that attempt to provide the community with clear requirements specific domains of industry have for MDE and transformation languages, as well as to show for which domains application is reasonable at all.
There exist some field studies by \textcite{Staron2006,10.1007/978-3-540-69100-6_31, Mohagheghi2013} but they are far and in between and do not focus on the transformation languages involved.
The research community can attempt to organize solutions for these requirements based on such field study and industry research.
However, for such industry cooperation to be possible, a focused community outreach is required.
There are notable advancements in this direction e.g. MDENet\footnote{\url{community.mde-network.org}}, but they are still in their infancy and require more involvement by the research community.

\action{Provide \textit{representative} model transformation languages}.
To provide reasonable evidence that model transformation languages can be competitive against GPLs there also needs to be heavy focus on providing less prototypical and more pragmatic and useable transformation languages (see \Cref{sec:findings:tooling}).
To that end only a few selected languages should be attempted to be made production ready, potentially through further industry cooperation.
MTLs could be integrated into commercial modelling tools in order to be able to process models programmatically in the tool.

Alternatively, few modern standardised MTLs could be promoted by the community.
Since such a decision has far-reaching effects, a central, community wide respected body is needed.
The OMG could possibly take action for this as they are already deciding on community impacting standards.

The QVT standard was an ambitious push in this direction.
However, we believe that the initiative needs a fresh take, given the findings of the last 20 years of research.
This idea is supported by several interviewees who considered QVT to be bloated and outdated.
Especially in the areas of bidirectional and incremental transformations we see huge potential.
Furthermore, relying more on declarative approaches for defining uni-directional transformations should also be considered.
This trend can also be observed in the field of GPLs with the introduction of more and more functional concepts into them.

Innovation in prototypical languages should then be thoroughly evaluated for its usefulness before adoption into one of the flagship languages.
It is not the task of research to produce industry ready languages, but setting up the environment and using these languages should not be more complicated than for any general purpose programming language.

\action{Research legacy integration}.
The integration of MTLs into existing legacy systems has been remarked as a huge entry barrier for industry adoption (see \Cref{sec:results:domain_focus,sec:results:tooling:eco}).
We believe this stems from a lack of techniques that facilitate gradual integration of modelling technology into existing systems and infrastructure.
This is highlighted by the fact that basic literature such as that by \textcite{brambilla2017model} does not contain any suggestions to this end.
To combat this, we propose a dedicated branch of MDE research focused on developing tools and processes to integrate model driven techniques into legacy systems.

We envision distinct guidelines and processes on how to integrate transformations and transformation concepts into existing systems.
There should be terms of reference as to which types of system components lend themselves to the use of model transformations.
Furthermore, descriptions of which transformations and which transformation languages are suitable for which type of use case are also required.
Having such guides can reduce the barrier of entry, because they provide a clear course of action when trying to (gradually) adopt the paradigm.

This also includes accessible GPL bindings for applying model transformation concepts.
They can be used to gradually replace system components that can benefit from the use of transformations.
This can be done without the overhead of integrating a new language and intermediate models.
One example for this is DresdenOCL, a OCL dialect that can be used on Java code~\parencite{demuth2009model}.

\subsection{Steps Towards Solving the Tooling Problem}
\label{sec:discussion:tools}

From our interviews, we have to conclude that the biggest weak point of model transformation languages is their \textit{Tool Support}.

The two biggest tooling gaps that we were able to identify are:
\begin{enumerate}[label=(\Roman*)]
	\item many necessary tools do not exist
	\item existing tools lack user-friendliness and are not compatible with each other
\end{enumerate}
We hope that our work can be a starting point in counteracting these drawbacks.

\action{Provide essential tooling}.
In our view, tooling of flagship model transformation languages needs to be extended to include all the essential tools mentioned in the interviews to make MTLs production and industry ready.
This includes useable \textit{Editors}, \textit{Debuggers} and \textit{Validation} or \textit{Analysis} tools.
At best all such tools for a language should be useable within one IDE.
One way language developers can help with this task is by implementing the \textit{Language Server Protocol} (LSP)~\parencite{lsp} or its graphical counterpart GLSP~\parencite{glsp} for their MTL.
This would greatly improve the ability of tool developers to create and distribute tooling.

\action{Develop transformation specific debugging}.
As mentioned by our interviewees, for debuggers there is a need for model transformation specific techniques.
\textcite{troya2022model} showed that there are numerous advances in this area like by \textcite{10.1007/978-3-642-04425-0_59,10.1007/978-3-540-75209-7_40,8904583} but none of them have led to well rounded debuggers yet.
Further effort by researchers active in this area is therefore required.
They should strive to develop their approaches to a point where they can be productively used to demonstrate their usefulness for a productive transformation development.

\action{Improve tool usability}.
Most importantly, a lot of effort needs to be put into improving the usability of MTL tools.
Our interviews have shown, that unusable tools are the most off putting factor that hampers wider adoption.
To combat this, we believe usability studies to be essential.
Studies to identify usability issues in the likeness of what is proposed by \textcite{pietron2018study} can be used to gain insights into where problems originate from and how to improve them.
Such studies have already been successfully utilised for other MDE related tooling \parencite{8904612}.
We therefore need more researchers from the community to get involved in designing and conducting usability studies for tooling surrounding MTLs.

We think the results of usability studies can also provide useful lessons learned for tool developers to make tools more usable from the beginning.
The overall goal must be to find out what needs to be changed or improved in MTL tools to make their adoption significant.
Industrial efforts to provide proper tool support can then be based on these results and the existing, usable, tools.
This adoption is necessary because, in our view, the human resources required for providing adequate long-term support for the tools can only be provided by commercially operating companies.
Such long term support is necessary so that model transformation languages, and their accompanying tools, can gain a foothold in the fast-moving industrial world.
The industrialisation of MTL tooling was also proposed during an open community discussion detailed by \textcite{Burgueno2019}.

The goal should be to provide well rounded, all-in-one solutions that integrate all necessary tooling in one place, to make development as seamless as possible.
The appropriateness of this has been shown by Jonkers et al.~\parencite{jonkers2006bootstrapping}.

\action{Limit-test internal MTLs}.
A different approach that should be further explored is the attempt to thoroughly embed an internal model transformation language in a main stream GPL as done by \textcite{Hinkel2019a}.
The advantage of this approach is the ability to inherit tooling of the host language~ \parencite{Hinkel2019a} and it allows general purpose developers to apply their rich pool of experience.
However, there are some drawbacks to this approach, as discussed in \Cref{sec:results}.
The amount of tooling that can be properly integrated is limited and it is more difficult to develop transformation specific tooling for internal languages as it is hard to extract the required information from the code.
For this reason, we think, the required tools should be known at design time and the language has to be designed to expose all the required information while not imposing this as an additional burden on developers.
Researchers that plan to develop an internal model transformation language should therefore thoroughly asses the tool requirements for the use case for which they intend to develop their language.
%
%
%
%
\section{Threats to validity}
\label{sec:threats}

Our interview study was carefully designed, and followed reputable guidelines for preparation, conduction and analysis.
Nonetheless there are some threats to validity that need to be discussed to provide a complete picture of our study and its results.

\subsection{Internal Validity}

Internal validity describes the extent to which a casual conclusion based on the study is warranted.
The validity is threatened by manual errors and biases of the involved researchers throughout the study process.

Errors could have been introduced during the transcription phase and during the analysis of the data since both steps were conducted by a single author at a time.

To prevent transcription errors, all transcripts were re-examined after completion to ensure consistency between the transcripts and audio recordings.

To minimize possible confirmation biases introduced during analysis and categorisation of interviewee statements, random samples were checked by other authors to find possible discrepancies between the authors assessments on statements.
In cases where such discrepancies were encountered, thorough discussions between all authors were conducted to find a consensus that was then applied to all transcripts containing similar considerations.

Lastly there is the potential of misinterpretation of interviewees responses during analysis.
While we carefully stuck to interpret statements literally during coding, there are words and phrases that have overloaded meanings.
During the interviews, it would always be necessary to ask exactly what meaning interviewees used, but this was not always possible.
Therefore the threat could not be mitigated completely as contextual information was required to interpret interviewees responses in some cases.

\subsection{External Validity}
\label{sec:threats:external}

External validity describes the extent to which the results of a study can be generalised.
In our interview study this validity is threatened by our interview participant assortment
,which is a result of our sampling and selection method.

We utilise convenience sampling interviewing any and all people that respond to our emails.
This can limit how representative the final group of interviewees is of the target population.
The issue here is that we do not know much about the makeup of the target population.
It is therefore difficult to assess how much the group of participants deviates from a representative set.

Using research publications as the starting point for participant selection also introduces a bias towards users from research.
This can be clearly seen in \Cref{fig:participant_types}.
There is an apparent lack of participants from industry which limits the applicability of our results to industrial cases.
This threat is somewhat mitigated by the fact that half of all participants do have at least some contact with industry, either through research projects in conjunction with industry or by having worked in industry.

Another threat to external validity relates to model to text (M2T) transformations.
Only a few of our participants stated to have experience in applying M2T transformations.
This is a result of how the initial set of potential participants was constructed.
The search terms used in the SLR miss terms that relate to M2T such as `code generation' or `model to text'.
This limitation was opted into to avoid having to differentiate between the two transformation approaches during analysis.
Moreover, the consensus during discussions was that we were talking about model to model transformations.
As such, our results can not be applied to the field of model to text languages.

Lastly, there is the threat of participation bias.
Participants may disproportionately posses a trait that reduces the generalisability of their responses.
People that view model transformation languages positively might be more inclined to participate than critics.
We can not preclude this threat, but, the amount of critique we were able to elicit from the interviews suggests the effects from this bias to be weak.
Other impacts of this bias are discussed in \Cref{sec:threats:conclusion}.

\subsection{Construct Validity}

Construct validity describes the extent to which the right method was applied to find answers for the research question.
This validity is threatened by an inappropriate method that allows for errors.

Prior to conducting our research much work went into designing a proper framework to use.
Here we relied on reputable existing guidelines for both the interview and analysis parts of this work.
We used open ended questions to facilitate an open space for participants to bring forth any and all their opinions and considerations for the topic at hand.
The statements used as guidance can however present a potential threat since their wording could introduce an unconscious bias in our interviewees.
To combat this we selected broad statements as well as used both a negative and a positive statement for each discussed property.
However, there is a chance that these measures were not fully sufficient.

Lastly, it can not be excluded that some relevant factors have not been raised during our interviews.
We have interviewed a large number of people, but this threat cannot be overcome because of the study design and the open nature of our research question.

\subsection{Conclusion Validity}
\label{sec:threats:conclusion}

Conclusion validity describes the extent to which our results stem from the investigated variables and are reproducible.
Here, the biases of our participants represent the biggest threat.

It is safe to assume that people who do research on a subject are more likely to see it in a positive light and less likely to find anything negative about it.
As such there is the possibility that too little negative impact factors were considered and presented.
However, we found that the people we interviewed were also able to deal with the topic in a very critical way.
We therefore conclude that the statements may have been somewhat more positively loaded, but that the results themselves are meaningful.

\section{Related Work}
\label{sec:rw}

To the best of our knowledge, there exists no other interview study that focuses on influence factors on the advantages and disadvantages of model transformation languages.
Nonetheless there exist several works that can be related to our study.
The related work is divided into empirical studies on model transformation languages, empirical studies on model transformations in general and interview studies on MDE.

\subsection{Empirical studies on model transformation languages}

A structured literature review we conducted~\parencite{Goetz2020} forms the basis for the work presented in this paper.
The goal of the reported literature review was to extract and categorize claims about the advantages and disadvantages of model transformation languages as well as to learn and report on the current state of evaluation thereof.
The authors searched over 4000 publications to extract a total of 58 publications that directly claim properties of model transformation languages.
In total the authors found 137 claims and categorized them into 15 properties.
From their work the authors conclude that while many advantages and disadvantages are claimed little to no studies have been executed to verify them.
They also point out a general lack of context and background information on the claimed properties that hinders evaluation and prompts scepticism.

\textcite{Burgueno2019} report on a online survey as well as a subsequent open discussion at the 12th edition of the International Conference on Model Transformations (ICMT'2019) about the future of model transformation languages.
Their goal for the survey was to identify reasons as to why developers decided for or against the use of model transformation languages and what their opinion on the future of these languages was.
At ICMT'2019 where the results of the survey were presented they then moderated an open discussion on the same topic.
The results of the study indicate that MTLs have fallen in popularity compared to at the beginning of the decade which they attribute to technical issues, tooling issues, social issues and the fact that general purpose languages have assimilated ideas from MTLs making GPLs a more viable option for defining model transformations.
While their methodology differed from our interview study, the results of both studies support each other.
However the results of our study are more detailed and provide a larger body of background knowledge that is relevant for future studies on the subject.

The notion of general purpose programming languages as alternatives to MTLs for writing model transformations has been explored by \textcite{Hebig2018} and by us~\parencite{Goetz2021}.
Hebig et al~\parencite{Hebig2018} report on a controlled experiment where student participants had to complete three tasks involved in the development of model transformations.
One task was to comprehend an existing transformation, one task involved modifying an existing transformation and one task required the participants to develop a transformation from scratch.
The authors compare how the use of ATL, QVT-O and the general purpose language Xtend affect the outcome of the three tasks.
Their results show no clear evidence of an advantage when using a MTL compared to a GPL but concede the narrow conditions under which the observation was made.
The study provides a rare example of empirical evaluation of MTLs of which we suggest that more be made.
The narrow conditions the authors struggled with could be alleviated by follow-up studies that draw from our results for defining their boundaries.

In a recent study by us~\parencite{Goetz2021} we put the value of model transformation language into a historical perspective and drew from the preliminary results of the interview study for the study setup.
We compare the complexity of a set of 10 model transformations written in ATL with their counterparts written in Java SE5, which was current around 2006 when ATL was first introduced, and Java SE14.
The Java transformations were translated from the ATL modules based on a predefined translation schema.
The findings support the assumptions from Burgueno et al.~\parencite{Burgueno2019} in part.
While we found that newer Java features such as Streams allow for a significant reduction in cyclomatic complexity and lines of code the relative amount of complexity of aspects that ATL can hide stays the same between the two Java versions.

\textcite{Gerpheide2016} use an exploratory study with expert interviews, a literature review and introspection to formalize a quality model for the QVTo model transformation standard by the OMG.
They validate their quality model using a survey and afterwards use the quality model to identify tool support need of transformation developers.
In a final step the authors design and evaluate a code test coverage tool for QVTo.
Their study is similar to ours in that they also relied on expert interviews for their goal.
The end goal of the study however differs from ours as they used the interviews to design a quality model for QvTo while we used it to formulate influence factors on quality attributes of model transformation languages

Lastly there are two study templates for evaluating model transformation languages which have yet to be used for executing actual studies.
\textcite{Kramer2016} present a template for a controlled experiment to evaluate the comprehensibility of model transformation languages.
Their approach suggests the use of a paper-based questionnaire to let participants prove their ability to understand what a transformation code snippet does.
The influence of the language in which the code is written on comprehension speed and quality is then measured by comparing the average number of correct answers and the average time spent to fill out the questionnaires.
\textcite{Strueber2016} propose a controlled experiment for comparing the benefits and drawbacks of the reusability mechanisms \textit{rule refinement} and \textit{variability-based rules}.
They suggest that the value of the reusability of an approach can be measured by looking at the comprehensibility of the two mechanisms as well their changeability, which is measured through bug-fixing and modification tasks.
The results of studies executed based on both study templates could draw from our results for their final design and would provide valuable empirical data, a gap we identified in this and the preceding literature review.

\subsection{Empirical studies on model transformations}

\textcite{tehrani2016requirements} executed an interview based study on requirements engineering for model transformation development.
Their goal was to identify and understand the contexts and manner in which model transformations are applied as well as how requirements for them are established.
To this end they interviewed 5 industry experts.
From the interviews the authors found that out of 7 transformation projects only a single project was developed in an already existing project while all other projects were created from scratch.
Their findings are relevant to our work since participants in our study agreed that it is hard to integrate MTLs in existing infrastructures.
Whether the fact that MTLs are hard to integrate was an influence factor for the projects considered in the interview study by them is however not clear.

\textcite{Groner2020} utilize an exploratory mixed method study consisting of a survey and subsequent interviews with a selection of the survey participants to try and evaluate how developers deal with performance issues in their model transformations.
They also asses the causes and solutions that developers experienced.
The survey results show that over half of all developers have experienced performance issues in their transformations.
While the interviews allowed the authors to identify and categorize performance causes and solutions into 3 categories: \textit{Engine related}, \textit{Transformation definition related} and \textit{Model related}.
From the interviews they were also able to identify that tools such as useable profilers and static analyses would help developers in managing performance issues.
The results of their study highlight that some of the factors identified by us are also relevant for other MTL properties not directly investigated in our study.



\subsection{Interview studies on model driven software engineering}
\label{sec:rw:MDE}

There are numerous publications and several groups of researchers that have carried out large scale, in-depth empirical studies on model driven engineering as a whole.
We focus on a selection of those that have relation to our study in terms of findings.

Whittle, Hutchinson, Rouncefield et al. used questionnaires~\parencite{Hutchinson2011,Hutchinson2014} and interviews~\parencite{Whittle2013,Hutchinson2011,Hutchinson2011a,Hutchinson2014} to elicit positive and negative consequences of the usage of MDE in industrial settings.
Apart from technical factors related to tooling they also found organisational and social factors that impact the adoption and efficacy of MDE.
Several of their findings for MDE in general coincide with results from our study.
Related to tooling they too found the factors of \textit{Interoperability}, \textit{Maturity} and \textit{Usability} to be influential.
Moreover, on the organisational side, the small amount of people that are knowledgable in MDE techniques and the problem of integrating into existing infrastructure are also results Whittle et al. found.
Lastly, developers being more interested in using techniques that help build their CV was identified by them as a limiting factor too.

\textcite{Staron2006} analyse data collected from a case study of MDE adoption at two companies where one company withdrew from adopting MDE while the other was in the process of adoption.
Their findings suggest that legacy code was a main influence factor on whether a cost efficient MDE adoption was possible.
This observation is consistent with our findings that integrating MTLs into existing infrastructures has a negative impact on the \textit{Productivity} that can be achieved with MTLs.

The research group surrounding Mohagheghi also carried out multiple empirical studies on MDE, focusing on factors for and consequences of adoption thereof.
They use surveys and interviews at several companies~\parencite{Mohagheghi2013,Mohagheghi2013a} as well as a literature review~\parencite{10.1007/978-3-540-69100-6_31} for this purpose.
In addition to mature tooling, factors identified by the authors are usefulness, ease of use and compatibility with existing tools.
Similar to statements by our interviewees, they also found that MDE is seen as a long term investment.
It is not well suited for single projects.

Lastly, \textcite{Akdur2018} report on a large online survey of people from the domain of embedded systems industry.
They too found tools surrounding MDE to be a major factor.
Another interesting finding by them was that UML models are by far the most commonly used models.
This is of relevance to our results since one of our interviewees pointed out, that the makeup of some UML models can have detrimental effects on the usefulness of MTLs.

The results of all presented research groups show, that many of the factors we identified for MTLs also apply to MDE in general which provides additional confidence in our results and shows that advancements in these areas would have a high impact.
\section{Conclusion}
\label{sec:conclusion}

There are many claims about the advantages and disadvantages of model transformation languages.
In this paper, we presented and argued the detailed factors that play a role for such claims.
Based on interviews with 56 participants from research and industry we present a \textbf{structure model} of relevant factors for the \textit{Ease of writing}, \textit{Expressiveness}, \textit{Comprehensibility}, \textit{Tool Support}, \textit{Productivity}, \textit{Reuse} and \textit{Maintainability} of model transformation languages.
For each factor we detail which properties they influence and \textbf{how} they influence them.
We have identified two types of factors.
There are factors that have a \textbf{direct impact} on said properties, e.g. different capabilities of model transformation languages like automatic trace handling.
And there are factors that define a \textbf{context} whose characteristics \textit{moderate} the the impact of the former factors, e.g. the \textit{Skills} of developers.

Based on the interview results we suggest a number of tangible actions that need to be taken in order to convey the viability of model transformation languages and MDSE.
For one, empirical studies need to be executed to provide proper substantiation to claimed properties.
We also need to see more innovation for transformation specific reuse, legacy integration and need to improve outreach and presentation to both industry and academia.
Lastly, efforts must be made to improve tool support and especially tool usability for MTLs.

For all of the suggested actions, our results can provide detailed data to draw from.
\appendix
\section{Interview Questions}
\label{apdx:questions}

\subsection*{Demographic Questions}
\begin{itemize}
	\item In what context have you used model transformation languages? Research, industrial projects or other?
	\item How much experience do you have in using model transformation languages? Rough estimate in years is sufficient.
	\item What model transformation languages have you used to date? 
\end{itemize}

\subsection*{Question Set 1}
\subsubsection*{Ease of \modified{W}riting}
\begin{flushleft}
	\textbf{The use of MTLs increases the ease of writing model transformations.}
	\begin{itemize}
		\item On a scale of 1 to 5, with 1 being strongly disagree and 5 being strongly agree, how would you rate your agreement with the statement?
		\item What is the reasoning behind your answer?
	\end{itemize}
	
	\textbf{Model transformation languages ease development efforts by offering succinct syntax to query from and map model elements between different modelling domains.}
	\begin{itemize}
		\item On a scale of 1 to 5, with 1 being strongly disagree and 5 being strongly agree, how would you rate your agreement with the statement?
		\item What is the reasoning behind your answer?
	\end{itemize}
	
	\textbf{Model transformation languages require specific skills to be able to write model transformations.}
	\begin{itemize}
		\item On a scale of 1 to 5, with 1 being strongly disagree and 5 being strongly agree, how would you rate your agreement with the statement?
		\item What is the reasoning behind your answer?
	\end{itemize}
\end{flushleft}

\subsubsection*{Comprehensibility}
\begin{flushleft}
	\textbf{The use of MTLs increases the comprehensibility of model transformations.}
	\begin{itemize}
		\item On a scale of 1 to 5, with 1 being strongly disagree and 5 being strongly agree, how would you rate your agreement with the statement?
		\item What is the reasoning behind your answer?
	\end{itemize}
	
	\textbf{Model transformation languages incorporate high-level abstractions that make them more understandable than GPLs.}
	\begin{itemize}
		\item On a scale of 1 to 5, with 1 being strongly disagree and 5 being strongly agree, how would you rate your agreement with the statement?
		\item What is the reasoning behind your answer?
	\end{itemize}
	
	\textbf{Most MTLs lack convenient facilities for understanding the transformation logic.}
	\begin{itemize}
		\item On a scale of 1 to 5, with 1 being strongly disagree and 5 being strongly agree, how would you rate your agreement with the statement?
		\item What is the reasoning behind your answer?
	\end{itemize}
\end{flushleft}

\subsection*{Question Set 2}
\subsubsection*{Tool \modified{S}upport}
\begin{flushleft}
	\textbf{There is sufficient tool support for the use of MTLs for writing model transformations.}
	\begin{itemize}
		\item On a scale of 1 to 5, with 1 being strongly disagree and 5 being strongly agree, how would you rate your agreement with the statement?
		\item What is the reasoning behind your answer?
	\end{itemize}
	
	\textbf{Tool support for external transformation languages is potentially more powerful than for internal MTL or GPL because it can be tailored to the DSL.}
	\begin{itemize}
		\item On a scale of 1 to 5, with 1 being strongly disagree and 5 being strongly agree, how would you rate your agreement with the statement?
		\item What is the reasoning behind your answer?
	\end{itemize}
	
	\textbf{Model transformation languages lack tool support.}
	\begin{itemize}
		\item On a scale of 1 to 5, with 1 being strongly disagree and 5 being strongly agree, how would you rate your agreement with the statement?
		\item What is the reasoning behind your answer?
	\end{itemize}
\end{flushleft}

\subsubsection*{Productivity}
\begin{flushleft}
	\textbf{The use of MTLs increases the productivity of writing model transformations.}
	\begin{itemize}
		\item On a scale of 1 to 5, with 1 being strongly disagree and 5 being strongly agree, how would you rate your agreement with the statement?
		\item What is the reasoning behind your answer?
	\end{itemize}
	
	\textbf{Model transformation languages, being DSLs, improve the productivity.}
	\begin{itemize}
		\item On a scale of 1 to 5, with 1 being strongly disagree and 5 being strongly agree, how would you rate your agreement with the statement?
		\item What is the reasoning behind your answer?
	\end{itemize}
	
	\textbf{Productivity of GPL development might be higher since expert users for GPLs are easier to hire.}
	\begin{itemize}
		\item On a scale of 1 to 5, with 1 being strongly disagree and 5 being strongly agree, how would you rate your agreement with the statement?
		\item What is the reasoning behind your answer?
	\end{itemize}
\end{flushleft}

\subsection*{Question Set 3}
\subsubsection*{Reuseability \& Maintainability}
\begin{flushleft}
\textbf{The use of MTLs increases the reusability and maintainability of model transformations.}
\begin{itemize}
	\item On a scale of 1 to 5, with 1 being strongly disagree and 5 being strongly agree, how would you rate your agreement with the statement?
	\item What is the reasoning behind your answer?
\end{itemize}

\textbf{Bidirectional model transformations have an advantage in maintainability.}
\begin{itemize}
	\item On a scale of 1 to 5, with 1 being strongly disagree and 5 being strongly agree, how would you rate your agreement with the statement?
	\item What is the reasoning behind your answer?
\end{itemize}

\textbf{Model transformation languages lack sophisticated reuse mechanisms.}
\begin{itemize}
	\item On a scale of 1 to 5, with 1 being strongly disagree and 5 being strongly agree, how would you rate your agreement with the statement?
	\item What is the reasoning behind your answer?
\end{itemize}
\end{flushleft}

\subsubsection*{Expressiveness}
\begin{flushleft}
	\textbf{The use of MTLs increases the expressiveness of model transformations.}
	\begin{itemize}
		\item On a scale of 1 to 5, with 1 being strongly disagree and 5 being strongly agree, how would you rate your agreement with the statement?
		\item What is the reasoning behind your answer?
	\end{itemize}
	
	\textbf{Model transformation languages hide transformation complexity and burden from the user.}
	\begin{itemize}
		\item On a scale of 1 to 5, with 1 being strongly disagree and 5 being strongly agree, how would you rate your agreement with the statement?
		\item What is the reasoning behind your answer?
	\end{itemize}
	
	\textbf{Having written several transformations we have identified that current MTLs are too low a level of abstraction for succinctly expressing transformations between DSLs because they demonstrate several recurring patterns that have to be reimplemented each time.}
	\begin{itemize}
		\item On a scale of 1 to 5, with 1 being strongly disagree and 5 being strongly agree, how would you rate your agreement with the statement?
		\item What is the reasoning behind your answer?
	\end{itemize}
\end{flushleft}
\section{Mail Templates}
\label{apdx:mails}

\subsection*{Mail Template}

Dear \$\{Author Name\},

I'm a PhD student with Matthias Tichy at Ulm University.
We recently conducted an SLR about the advantages and disadvantages of model transformation languages
as claimed in literature. Our results have been published in the software and systems modelling journal
here \url{http://dx.doi.org/10.1007/s10270-020-00815-4}.
One of our main takeaways from the study was that a large portion of claims about model
transformation languages is never substantiated.
One main reason for this, we believe, is implicit knowledge authors tend to omit for different reasons.

Since you are an author of one of the publications we considered during our SLR it would be great to talk
to you about your experiences and stance with regard to model transformation languages
and the claims we extracted from literature. We would need max. 30 minutes of your time.
The interview would be conducted by me via an online conferencing system.

In order to organize the interview dates I would like to ask you to
chose a suitable date, under the following link \url{https://terminplaner4.dfn.de/F1mIEEwSSkTwh8XA}.
Please note that the times are given in UTC.
The password for the poll is "claims". Your response will not be visible to anyone other than myself.
If  none of the dates is suitable for you, you are welcome to contact me to find another date for the interview.

Before your interview I would like to ask you to agree to the data \modified{protection } agreement
under the following link \url{https://pmx.informatik.uni-ulm.de/limesurvey/index.php/924713?lang=en}.
I have also attached a copy of how we handle the interview data to this mail.

Best regards

Stefan Götz

\subsection*{Reminder Mail Template}

Dear \$\{Author Name\},

If you already filled out our organization poll please ignore this mail. 

I wanted to remind you to maybe take part in our interview study about the implicit knowledge of users with regards to advantages and disadvantages of model transformation languages.
It would be great to talk to you about your experiences and stance with regard to model transformation languages and the claims we extracted from literature. We would need max. 30 minutes of your time.

In order to organize the interview dates I would like to ask you to chose a suitable date, under the following link \url{https://terminplaner4.dfn.de/F1mIEEwSSkTwh8XA}.
Please note that the times are given in UTC. Please also note that you need to press the SAVE button at the right hand side of the poll.
The password for the poll is "claims". Your response will not be visible to anyone other than myself.
If  none of the dates is suitable for you, you are welcome to contact me to find another date for the interview.

Before your interview I would like to ask you to agree to the data \modified{protection } agreement under the following link \url{https://pmx.informatik.uni-ulm.de/limesurvey/index.php/924713?lang=en}. I have also attached a copy of how we handle the interview data to this mail.

Best regards

Stefan Götz
\section{Demographics}
\label{apdx:demographics}
See~\Cref{tbl:demographicsOverview}.

\onecolumn
\begin{xltabular} {\textwidth} {@{}l|l|r|X@{}}
	\caption{Overview over the interviewee demographic data}~\label{tbl:demographicsOverview}\\
	\toprule
	
	\textbf{PID} & \textbf{Background} & \textbf{Experience in years} & \textbf{Language types used for writing transformations}\\
	\midrule
	\midrule
	\endfirsthead
	
	\multicolumn{2}{c}%
	{\tablename\ \thetable{} -- continued from previous page} \\
	\toprule
	\textbf{PID} & \textbf{Background} & \textbf{Experience in years} & \textbf{Language types used for writing transformations}\\
	\midrule
	\midrule
	\endhead

	\textbf{P1} & Research & >10 & GPLs\\
	\midrule
	\textbf{P2} & Research & 10-15 & dedicated MTLs\\
	\midrule
	\textbf{P3} & Research & 8 & dedicated MTLs\\
	\midrule
	\textbf{P4} & Research & 7 & dedicated MTLs \& internal MTLs\\
	\midrule
	\textbf{P5} & Research & >5 & dedicated MTLs \& GPLs\\
	\midrule
	\textbf{P6} & Research \& Industry Projects & 13 & dedicated MTLs \& GPLs\\
	\midrule
	\textbf{P7} & Research \& Industry Projects & 10 & dedicated MTLs \& GPLs\\
	\midrule
	\textbf{P8} & Research \& Industry Projects & 18 & dedicated MTLs \& GPLs\\
	\midrule
	\textbf{P9} & Industry & 20 & dedicated MTLs\\
	\midrule
	\textbf{P10} & Research & 4 & dedicated MTLs \& GPLs\\
	\midrule
	\textbf{P11} & Research & 5-6 & dedicated MTLs\\
	\midrule
	\textbf{P12} & Research \& Industry Projects & 8 & dedicated MTLs\\
	\midrule
	\textbf{P13} & Industry with History in Research & 6 & dedicated MTLs \& internal MTLs\\
	\midrule
	\textbf{P14} & Research \& Industry Projects & 15 & dedicated MTLs \& internal MTLs \& GPLs\\
	\midrule
	\textbf{P15} & Research \& Industry Projects & 5 & dedicated MTLs \& GPLs\\
	\midrule
	\textbf{P16} & Research & 7 & dedicated MTLs \& GPLs\\
	\midrule
	\textbf{P17} & Research \& Industry Projects & 18 & dedicated MTLs \& GPLs\\
	\midrule
	\textbf{P18} & Research \& Industry Projects & 10 & dedicated MTLs \& GPLs\\
	\midrule
	\textbf{P19} & Research & 7 & dedicated MTLs\\
	\midrule
	\textbf{P20} & Research \& Industry Projects & 3 & dedicated MTLs \& GPLs\\
	\midrule
	\textbf{P21} & Research \& Industry Projects & 15 & dedicated MTLs \& GPLs\\
	\midrule
	\textbf{P22} & Research \& Industry Projects & 8 & dedicated MTLs \& GPLs\\
	\midrule
	\textbf{P23} & Research & 13 & dedicated MTLs\\
	\midrule
	\textbf{P24} & Research \& Industry Projects & 15 & dedicated MTLs \& GPLs\\
	\midrule
	\textbf{P25} & Research & 8 & dedicated MTLs\\
	\midrule
	\textbf{P26} & Industry & >10 & dedicated MTLs\\
	\midrule
	\textbf{P27} & Industry with History in Research & 10-12 & dedicated MTLs \& GPLs\\
	\midrule
	\textbf{P28} & Research & 15 & dedicated MTLs \& GPLs\\
	\midrule
	\textbf{P29} & Research \& Industry Projects & 12 & dedicated MTLs\\
	\midrule
	\textbf{P30} & Research \& Industry Projects & 17 & dedicated MTLs \& GPLs\\
	\midrule
	\textbf{P31} & Research & 8 & dedicated MTLs\\
	\midrule
	\textbf{P32} & Research \& Industry Projects & 15 & dedicated MTLs \& GPLs\\
	\midrule
	\textbf{P33} & Research & 5-6 & dedicated MTLs \& GPLs\\
	\midrule
	\textbf{P34} & Research & 5-6 & GPLs\\
	\midrule
	\textbf{P35} & Research & 10 & dedicated MTLs\\
	\midrule
	\textbf{P36} & Research & 10 & dedicated MTLs \& GPLs\\
	\midrule
	\textbf{P37} & Research \& Industry Projects & 10-11 & dedicated MTLs\\
	\midrule
	\textbf{P38} & Research & 4-5 & dedicated MTLs\\
	\midrule
	\textbf{P39} & Industry & 28 & dedicated MTLs \& GPLs\\
	\midrule
	\textbf{P40} & Research & 9 & dedicated MTLs\\
	\midrule
	\textbf{P41} & Research & 7-8 & dedicated MTLs\\
	\midrule
	\textbf{P42} & Industry with History in Research & 13 & dedicated MTLs \& internal MTLs \& GPLs\\
	\midrule
	\textbf{P43} & Research \& Industry Projects & 8-10 & dedicated MTLs\\
	\midrule
	\textbf{P44} & Research \& Industry Projects & 10 & dedicated MTLs\\
	\midrule
	\textbf{P45} & Research & 1-2 & dedicated MTLs\\
	\midrule
	\textbf{P46} & Research \& Industry Projects & 9 & dedicated MTLs\\
	\midrule
	\textbf{P47} & Research & 4 & dedicated MTLs\\
	\midrule
	\textbf{P48} & Research & 7-8 & dedicated MTLs \& internal MTLs\\
	\midrule
	\textbf{P49} & Research \& Industry Projects & 10 & dedicated MTLs \& GPLs\\
	\midrule
	\textbf{P50} & Research & 20 & dedicated MTLs\\
	\midrule
	\textbf{P51} & Research \& Industry Projects & 3 & dedicated MTLs\\
	\midrule
	\textbf{P52} & Research & 13-14 & dedicated MTLs\\
	\midrule
	\textbf{P53} & Research & 12 & dedicated MTLs \& GPLs\\
	\midrule
	\textbf{P54} & Research & 7 & dedicated MTLs\\
	\midrule
	\textbf{P55} & Research \& Industry Projects & 16 & dedicated MTLs \& GPLs\\
	\midrule
	\textbf{P56} & Research & 16 & dedicated MTLs\\
	\bottomrule
\end{xltabular}
\section{Data Privacy Agreement}
\label{apdx:consent_form}



**General information, declaration of consent**

**General information about the interviews about claims about model transformation languages of the Institute for Software Engineering and Programming Languages of Ulm University.**

At the Institute for Software Engineering and Programming Languages of Ulm University model transformation languages are being examined. This includes claims about the advantages and disadvantages and evidence thereof. Within one work package of the doctoral thesis of Stefan Götz, researchers and practitioners are being surveyed about their opinions on certain claims.

The goal of this work package is to gain a deeper understanding of the reasoning people use for believing certain claims about model transformation languages.

## 1 Procedure

- If you decide to participate in our interview study, please fill out the poll at https://terminplaner4.dfn.de/F1mIEEwSSkTwh8XA so we can set up a date for the interview.
- We will contact you about one week in advance of the chosen date to arrange the interview.
- The interview will take place using an online conferencing tool hosted at Ulm University or via Skype depending on your preference.
- Please fill out the consent form at https://pmx.informatik.uni-ulm.de/limesurvey/index.php/924713 before the interview.
- At the beginning of the interview we will ask you questions about your experience level with regards to model transformation languages such as the languages you have used.
- During the interview we will show you different claims from literature about model transformation languages for which we would like you to tell us if you agree with them or not and your reasoning behind the decision.
- **An audio recording of the interview will be made if you consent to it.**
- The audio recordings will be transcribed after the interview and deleted as soon as transcribing has been completed.

## 2 Conditions of participation:
- You have some for of experience with model transformation languages.

Created based on a template by ZENDAS    Page 1 of 3    Status of the template: 05.04.2019

## 3 Handling the data in the research project

1. Your interview will be recorded on audio and notes will be taken. The procedure described in 2-5 is followed for anonymizing your interview.
2. The non-anonymous raw data (first name, last name, e-mail address, notes and the audio recording of your interview) is only shared between the project partners (Stefan Götz, Yves Haas and Matthias Tichy from Ulm University) for transcribing and analyzing the answers.
3. We will anonymize your interview by transcribing it and delete the information about your first name, last name, e-mail and the audio recording of your interview as soon as possible and not later than 30 October 2020. Which means that individuals cannot be deduced from their interview.
4. For scientific publication, anonymized answers to this interview (transcribed interview and notes) will be further processed, e.g., coded or aggregated. Special risks for your person are not apparent with the processed results because individual persons cannot be inferred.
5. We strongly believe in open data to allow replication of our results as well as enabling further research. The anonymized answers (unprocessed and processed) to this interview (transcribed interview and notes ) will be made available online to the public to accompany publications and stored in open data repositories. Please note that such published data can no longer be completely deleted and can be accessed and used by any person. Special risks for your person are not apparent with the anonymized answers, because individual persons cannot be inferred.

**Contact person**

If you have any questions, concerns or doubts, please do not hesitate to contact us:

Stefan Götz
Ulm University
Institute of Software Engineering and Programming Languages
89081 Ulm
E-Mail: stefan.goetz@uni-ulm.de

## Declaration of consent

..................................................................................................................
(Name, Surname of the participating person)

I have read the general information on the interview study and agree to participate in the research project and the associated data processing.

(You can also give the following consent:)

☐ I have read the general information on the interview, which is conducted as part of a research project. I consent to participate in the interview and I consent to the related data processing as described in 1-4.

☐ I hereby consent to the publication of my anonymized answers (unprocessed and processed) to this interview (transcribed interview, notes and linked questionnaire) as described in 5.

I am aware that the consents are voluntary and can be refused without disadvantages (even individually) or revoked at any time without giving reasons. I am aware that in case of revocation, the legality of the processing carried out on the basis of the consent until revocation is not affected. I understand that I can simply contact the contact person named in the information for a revocation and that no disadvantages arise from the refusal of consent or its revocation.
I was informed and provided with the information on the collection of personal data during the interview study. I have also received a copy of this consent form.

Created based on a template by ZENDAS    Page 3 of 3    Status of the template: 05.04.2019


\section{Quotations}
\label{apdx:quotations}

\begin{xltabular} {\textwidth} {@{}l|l|l|X@{}}
	\caption{Selection of quotations from interview participants for specific factors}~\label{tbl:quotations}\\
	\toprule
	
	\textbf{Factor} & \textbf{QID} & \textbf{PID} & \textbf{Quotation}\\
	\midrule
	\midrule
	\endfirsthead
	
	\multicolumn{4}{l}%
	{\tablename\ \thetable{} -- continued from previous page} \\
	\toprule
	\textbf{Factor} & \textbf{QID} & \textbf{PID} & \textbf{Quotation}\\
	\midrule
	\midrule
	\endhead
	\multirow{3}{7em}{GPL Capabilities} & \newtag{\textit{$Q_{gpl}1$}}{que:1} & \textbf{P42} & \textit{``[General purpose languages are] very good at constructing objects and filling in their fields [...] and computing 'simple' expressions.''}\\
	\cmidrule{2-4}
	 & \newtag{\textit{$Q_{gpl}2$}}{que:2} & \textbf{P14} & \textit{``I think that in the end you have more tools for development. And I feel more productive.''} \\
	\cmidrule{2-4}
	 & \newtag{\textit{$Q_{gpl}3$}}{que:3} & \textbf{P8} & \textit{``[...] when you reach the maintenance phase, maybe the [original] developers are gone. And you have an [MTL] program that might be more difficult to understand for [new] developers''} \\
	\midrule
		\multirow{5}{7em}{Domain Focus} & \newtag{\textit{$Q_{df}1$}}{que:4} & \textbf{P6} & \textit{``What is better by using MTLs instead of GPLs is the fact that you are on the same abstraction level of the modelling language. You are basically treating apples with apples.''}\\
	\cmidrule{2-4}
	& \newtag{\textit{$Q_{df}2$}}{que:5} & \textbf{P19} & \textit{``[...] you are gonna cut away all those unneeded code and complexity and focus on your problem.''} \\
	\cmidrule{2-4}
	& \newtag{\textit{$Q_{df}3$}}{quedf:2} & \textbf{P23} & \textit{``Once you have things like rules and helpers and things like left hand side and right hand side and all these patterns then [it is] easier to create things like meta-rules to take rules from one version to another version [...]''} \\
	\cmidrule{2-4}
	& \newtag{\textit{$Q_{df}4$}}{que:6} & \textbf{P13} & \textit{``To do this [tool support for analysing rule dependencies] [...] you have to resolve parameter dependencies and I immediately run into Turing completeness. And I don't have that with an external language [...]''} \\
	\cmidrule{2-4}
	& \newtag{\textit{$Q_{df}5$}}{que:7} & \textbf{P6} & \textit{``They have existing infrastructure and people and everything that is based on established languages which is hard to change.''} \\
	\midrule
	\multirow{4}{7em}{Bidirectionality} & \newtag{\textit{$Q_{bx}1$}}{que:8} & \textbf{P42} & \textit{``in a general purpose programming language you would have to add a bit of clutter, a bit of distraction, from the real heart of the matter''} \\
	\cmidrule{2-4}
	& \newtag{\textit{$Q_{bx}2$}}{que:9} & \textbf{P40} & \textit{``So either you write your own program to create unidirectional transformations in both directions or you write both directions by hand and that has the disadvantage that if, in the future, something changes in the transformation, then you have to rework both directions''} \\
	\cmidrule{2-4}
	& \newtag{\textit{$Q_{bx}3$}}{que:10} & \textbf{P41} & ``\textit{[...] That makes it harder for them to see whether something is correct or not and to master the complexity of these transformations.}'' \\
	\cmidrule{2-4}
	& \newtag{\textit{$Q_{bx}4$}}{que:11} & \textbf{P11} & \textit{``And as soon as I am at bidirectional transformations and there is somehow a loss of information. [...] And then [the question is] how difficult it is to access e.g. context elements that I have already created and need again later, because I want to refer to them.''} \\
	\midrule
	\multirow{2}{7em}{Incrementality} & \newtag{\textit{$Q_{inc}1$}}{que:12} & \textbf{P56} & \textit{``Declarative MTLs may have different computation paradigms which may be unfamiliar for developers used to imperative languages''} \\
	\cmidrule{2-4}
	& \newtag{\textit{$Q_{inc}2$}}{que:13} & \textbf{P42} & \textit{``[...] do not try to do it manually, because you will definitely have bugs,[...] because there will be some specific kind of change trajectory that you have missed, [...] this is a super hard problem.''} \\
	\midrule
	\multirow{8}{7em}{Mappings} & \newtag{\textit{$Q_{map}1$}}{que:14} & \textbf{P24} & ``\textit{They hide those dimensions that reflect how graph-wise it would be computationally complex to interpret the problem to transform one model into another}'' \\
	\cmidrule{2-4}
	& \newtag{\textit{$Q_{map}2$}}{que:15} & \textbf{P25} & \textit{``So it restricts you in the way you can work and that makes it easier because that is what you need to do.''} \\
	\cmidrule{2-4}
	& \newtag{\textit{$Q_{map}3$}}{que:16} & \textbf{P55} & \textit{``This means that you can write the rules independently of the execution sequence, you can define them more declaratively and, at least in my experience, you can still manage to define these rule blocks in a comprehensible way for large transformations.''} \\
	\cmidrule{2-4}
	& \newtag{\textit{$Q_{map}4$}}{que:17} & \textbf{P5} & \textit{``I mentioned language engineering because a lot of the transformation difficulties are understanding the syntactical and semantic differences between two domain specific languages.''} \\
	\cmidrule{2-4}
	& \newtag{\textit{$Q_{map}5$}}{que:18} & \textbf{P30} & \textit{``[...] Hidden mechanisms or built in mechanisms may be more difficult to understand [thus] learning the language may be a bit more difficult.''} \\
	\cmidrule{2-4}
	& \newtag{\textit{$Q_{map}6$}}{que:19} & \textbf{P38} & \textit{``[...] you have a formal correspondence between the two models. And if you can transform in both directions, then you can practically keep both models, between which you want to transform back and forth, synchronous.''} \\
	\cmidrule{2-4}
	& \newtag{\textit{$Q_{map}7$}}{que:20} & \textbf{P16} & \textit{``Whereas when you need to do some more elaborate business logic or when you need to hook some external services or other sources of information into your transformation  then I am saying that MTLs can start to be a little bit of a limit''} \\
	\cmidrule{2-4}
	& \newtag{\textit{$Q_{map}8$}}{que:21} & \textbf{P3} & \textit{``If I want to reuse this model transformation just changing 2 words in ATL [is enough], if I wanted to do the same in Java instead of changing something in 2 places I have to do it in 5 or 6.''} \\
	\midrule
	\multirow{3}{7em}{Traceability} & \newtag{\textit{$Q_{trc}1$}}{que:22} & \textbf{P22} & \textit{``So that is something you often have to do manually in a GPL. So you have to maintain the trace information yourself and kind of re-implement that.''} \\
	\cmidrule{2-4}
	& \newtag{\textit{$Q_{trc}2$}}{que:23} & \textbf{P32} & \textit{``You have to know what a trace is. [...] And at some point, at the latest when you do something more complex, you need this stuff. ''} \\
	\cmidrule{2-4}
	& \newtag{\textit{$Q_{trc}3$}}{que:24} & \textbf{P31} & \textit{``[...] a model transformation rule only [contains] the domain transformation, so which domain object of the source domain is mapped to an object and how the object is mapped to the target domain. And that is what someone who tries to understand the model transformation is trying to get [...] out of the source code.''} \\
	\midrule
	\multirow{1}{7em}{Automatic Traversal} & \newtag{\textit{$Q_{trv}1$}}{que:25} & \textbf{P49} & \textit{``That means abstracting away from the order of traversal and then also knowing in which context this thing came up, that is a bit of a double-edged sword for me, [...] it has the potential to mask serious errors.''} \\
	\midrule
	\multirow{1}{7em}{Pattern-Matching} & \newtag{\textit{$Q_{pm}1$}}{que:26} & \textbf{P14} & \textit{``[...] all the complexity of pattern matching is in the engine, but if you try to implement a mapping then all the complexity of keeping the traces you have to do that manually.''} \\
	\midrule
	\multirow{2}{7em}{Model Navigation} & \newtag{\textit{$Q_{nav}1$}}{que:27} & \textbf{P41} & \textit{``[...] you don't have to worry about the efficiency of the procedure, just figures out the optimal way of kind of traversing it. For me that is the biggest thing actually, they go and get me the data, if we can hide that from the user, that is great.''} \\
	\cmidrule{2-4}
	& \newtag{\textit{$Q_{nav}2$}}{que:28} & \textbf{P11} & \textit{``[...] I do not have to iterate over the model. I only say, I need this or that.''} \\
	\midrule
	\multirow{1}{7em}{Model Management} & \newtag{\textit{$Q_{man}1$}}{que:29} & \textbf{P2} & \textit{``[...] this technical level, how I access a model, [...]I get the elements out. That gets abstracted away.''} \\
	\midrule
	\multirow{3}{7em}{Reuse Mechanism} & \newtag{\textit{$Q_{rm}1$}}{que:30} & \textbf{P51} & \textit{``[...] we usually use object oriented programming languages and those already have some pretty strong tools for reusability in the appropriate contexts. So i think the bar here, that would we want model transformation languages to jump over, is to provide something more targeted towards modeling [...]''} \\
	\cmidrule{2-4}
	& \newtag{\textit{$Q_{rm}2$}}{que:31} & \textbf{P30} & \textit{``[...] for ATL there are things like module superimposition, and other kinds, we have helper libraries.''} \\
	\cmidrule{2-4}
	& \newtag{\textit{$Q_{rm}3$}}{que:32} & \textbf{P27} & \textit{``[...] in the case of VIATRA one of the main goals of the pattern language we are using there is to allow reusing previously defined patterns. Basically any pattern can be included. So there is a lot of stuff you can do to reuse the element.''} \\
	\midrule
	\multirow{2}{7em}{Learnability} & \newtag{\textit{$Q_{ler}1$}}{que:33} & \textbf{P23} & \textit{``So the learning curve is pretty steep when trying to use MTLs. You need to learn a lot of stuff before you can use them properly.''} \\
	\cmidrule{2-4}
	& \newtag{\textit{$Q_{ler}2$}}{que:34} & \textbf{P6} & \textit{``You can take 10 Java developers and out of them probably 2 would understand what a MTL is. They don’t have experience in modelling. Not because they are dumb, because they are not used to that.''} \\
	\midrule
	\multirow{1}{7em}{Debugging Tooling} & \newtag{\textit{$Q_{db}1$}}{que:35} & \textbf{P51} & \textit{``Well I think one of the other important points would be to [be] able to prove properties of transformations or check properties of transformations, [...] but we don't really have that for model transformations.''} \\
	\midrule
	\multirow{3}{7em}{Ecosystem} & \newtag{\textit{$Q_{eco}1$}}{que:36} & \textbf{P49} & \textit{``people from industry have a hard time when they are required to use multiple languages.''} \\
	\cmidrule{2-4}
	& \newtag{\textit{$Q_{eco}2$}}{que:37} & \textbf{P49} & \textit{``It is often on a technical level that the integration into the overall ecosystem of tools you have is not so great.''} \\
	\cmidrule{2-4}
	& \newtag{\textit{$Q_{eco}3$}}{que:38} & \textbf{P31} & \textit{``Something I see as a problem with some model transformation languages, which limit the applicability, is the coupling to Eclipse. This is what will cause us as a research community big problems some day [...].''} \\
	\midrule
	\multirow{1}{7em}{Interoperability} & \newtag{\textit{$Q_{int}1$}}{que:39} & \textbf{P36} & \textit{``But the technologies, to combine them, it is difficult [...]''} \\
	\midrule
	\multirow{1}{7em}{Tooling Awareness} & \newtag{\textit{$Q_{awa}1$}}{que:40} & \textbf{P35} & \textit{``And, I think, it is hard for new users to see, for example, what, which tool to use. Or which technology you should work with.''} \\
	\midrule
	\multirow{2}{7em}{Tool Creation Effort} & \newtag{\textit{$Q_{tce}1$}}{que:41} & \textbf{P01} & \textit{``I am keenly aware of the cost to being able to develop a good programming language, the cost of maintaining it and the cost of adding debuggers and refactoring engines. It is enormous.''} \\
	\cmidrule{2-4}
	& \newtag{\textit{$Q_{tce}2$}}{que:42} & \textbf{P6} & \textit{``But it is definitely easier and faster to build the tool support and it allows you to do more advanced stuff. You can play around with your domain specific concepts in a lot of different ways.''} \\
	\midrule
	\multirow{1}{7em}{Tool Learnability} & \newtag{\textit{$Q_{tle}1$}}{que:43} & \textbf{P34} & \textit{``Because when i started to work with model transformation languages and to hear about them, [...] I do not think that [...] there was like initial go-to documentation.''} \\
	\midrule
	\multirow{2}{7em}{Tool Usability} & \newtag{\textit{$Q_{use}1$}}{que:44} & \textbf{P22} & \textit{``Basically all the good aids you see in a Java environment should be there even better in a MTL tool because model transformation is so much more abstract and more relevant that you should be having tools that are again more abstract and more relevant.''} \\
	\cmidrule{2-4}
	& \newtag{\textit{$Q_{use}2$}}{que:45} & \textbf{P48} & \textit{``There are quite a few corner cases, which are often not quite fixed and especially the usability is often very bad.''} \\
	\midrule
	\multirow{1}{7em}{Tool Maturity} & \newtag{\textit{$Q_{mat}1$}}{que:46} & \textbf{P23} & \textit{``Because [MTLs] have been around for like 30 years. And other languages and frameworks, they are created in 2-3 years, and they are good to go. And MTLs have been around for so long. And I think its mostly because industry has not taken it in. And it’s just a problem of manpower put into the languages.''} \\
	\midrule
	\multirow{1}{7em}{Validation Tooling} & \newtag{\textit{$Q_{val}1$}}{que:47} & \textbf{P8} & \textit{``For example I can not remember any tool that offers reasonable support for testing. In Java you have JUnit and other. In ATL there is nothing.''} \\
	\midrule
	\multirow{1}{7em}{Language Skills} & \newtag{\textit{$Q_{skl}1$}}{que:48} & \textbf{P1} & \textit{``[...] this is the way you have to think in terms of formulating your problem''} \\
	\cmidrule{2-4}
	& \newtag{\textit{$Q_{skl}2$}}{que:49} & \textbf{P12} & \textit{``And then you [need to] learn a language, the MTL.''} \\
	\midrule
	\multirow{2}{7em}{User Experience/ Knowledge} & \newtag{\textit{$Q_{exp}1$}}{que:50} & \textbf{P21} & \textit{``One of the reasons why Ada is virtually extinct is that developers preferred to have C++ on their CVs. Simply because there were more job postings with C++. And that develops a momentum of its own, which of course makes languages suffer. That also applies to DSLs.''} \\
	\cmidrule{2-4}
	& \newtag{\textit{$Q_{exp}2$}}{que:51} & \textbf{P06} & \textit{``Many MDSE courses are just given too late, when people are too acquainted with GPLs, and then its really hard for students to see the point of using models, modelling and MTLs, because it's comparable with languages and stuff they have already learned and worked with.''} \\
	\midrule
	\multirow{1}{7em}{Involved (meta-) models} & \newtag{\textit{$Q_{mod}1$}}{que:52} & \textbf{P28} & \textit{``As soon as you venture into eGenericType there is a lot of pain to be had and there is poor documentation.''} \\
	\midrule
	\multirow{1}{7em}{I/O Semantic gap} & \newtag{\textit{$Q_{gap}1$}}{que:53} & \textbf{P22} & \textit{``[...] as soon as I wanted to do something a bit more complex, I have often found that I was not able to express what I wanted to do easily and I had to resort to advanced features of the language in order to achieve what I want to do.''} \\
	\midrule
	\multirow{1}{7em}{Size} & \newtag{\textit{$Q_{siz}1$}}{que:54} & \textbf{P55} & \textit{``The size is a good point. I would reduce that now to rules. But if I have several rules that then build on each other, then it will probably be easier with an MTL. Especially if you have a lot of dependencies between the rules.''} \\
	\bottomrule
\end{xltabular}

\twocolumn

\section*{Declarations}
\textbf{Conflict of Interests} The authors have no competing interests to declare that are relevant to the content of this article.

\begingroup
\raggedright
\sloppy
\printbibliography
\endgroup


\end{document}